\documentclass[twocolumn,aps,prd,longbibliography]{revtex4-2}
\usepackage{dcolumn}
\usepackage{enumerate}
\usepackage{amsmath}
\usepackage{amssymb}
\usepackage{graphicx}
\usepackage{graphics}
\usepackage[dvipsnames]{xcolor}

\begin{document}
 
    \title{Non-axisymmetric modes of magnetorotational and possible hydrodynamical instabilities in the upcoming
     DRESDYN-MRI experiments – linear and nonlinear dynamics}
 
	\author{Ashish Mishra}
	\email{a.mishra@hzdr.de}
	\affiliation{Helmholtz-Zentrum Dresden-Rossendorf, Bautzner Landstr. 400, D-01328 Dresden, Germany}
	\affiliation{Center for Astronomy and Astrophysics, ER 3-2, TU Berlin, Hardenbergstr. 36, 10623 Berlin, Germany}
	\author{George Mamatsashvili}
	\affiliation{Helmholtz-Zentrum Dresden-Rossendorf, Bautzner Landstr. 400, D-01328 Dresden, Germany}
	\affiliation{Abastumani Astrophysical Observatory, Abastumani 0301, Georgia}
	
	\author{Frank Stefani}
	\affiliation{Helmholtz-Zentrum Dresden-Rossendorf, Bautzner Landstr. 400, D-01328 Dresden, Germany}
	
	\begin{abstract}
	The quest for an unambiguous detection of magnetorotational instability (MRI) in experiments is still ongoing despite recent promising results. To conclusively identify MRI in the laboratory, a large cylindrical Taylor-Couette experiment with liquid sodium is under construction within the DRESDYN project. Recently, we have analyzed the nonlinear dynamics and scaling properties of axisymmetric standard MRI with an axial background magnetic field in the context of the DRESDYN-MRI experiment. In this sequel paper, we investigate the linear and nonlinear dynamics of non-axisymmetric MRI in the same magnetized Taylor-Couette flow of liquid sodium. We show that the achievable highest Lundquist $Lu = 10$ and magnetic Reynolds $Rm = 40$ numbers in this experiment are large enough for the linear instability of non-axisymmetric modes with azimuthal wavenumber $|m|=1$, although the corresponding critical values of these numbers are usually higher than those for the axisymmetric mode. The structure of the ensuing nonlinear saturated state and its scaling properties with respect to Reynolds number $Re$ are analyzed, which are important for the DRESDYN-MRI experiment having very high $Re \gtrsim 10^6$. It is shown that for $Re \lesssim 4\times 10^4$, the non-axisymmetric MRI modes eventually decay, since the modified shear profile of the mean azimuthal velocity due to the nonlinear axisymmetric MRI appears to be stable against non-axisymmetric instabilities. By contrast, for larger $Re \gtrsim 4\times 10^4$, a rapid growth and saturation of the non-axisymmetric modes of nonmagnetic origin occurs, which are radially localized near the inner cylinder wall, forming a turbulent boundary layer. However, for all the parameters considered, the saturation amplitude of these non-axisymmetric modes is always a few orders smaller than that of the axisymmetric MRI mode. Therefore, the results of our previous axisymmetric study on the scaling properties of nonlinear MRI states also hold when non-axisymmetric modes are included.  
	\end{abstract}	
	\maketitle
	
\section{Introduction}
Since the inception of magnetorotational instability (MRI \cite{Velikhov_1959}) as the most likely mechanism transporting angular momentum in accretion disks \cite{Balbus_Hawley_1991}, there have been numerous attempts to capture MRI in the laboratory \cite{Sisan_etal2004,Nornberg_Ji_etal_2010_PhysRevLett,Roach_etal2012PhRvL, Hung_etal2019CmPhy, Wang_etal2022b_PRL}. 
The detection of a standard version of MRI (SMRI) with an imposed purely axial magnetic field in a experimental cylindrical Taylor-Couette (TC) device filled with a liquid metal is challenging. For the onset of SMRI both magnetic Reynolds ($Rm$) and Lundquist ($Lu$) numbers must be high enough $\sim 1-10$, which, because of very small magnetic Prandtl numbers $Pm \sim 10^{-6}-10^{-5}$ of liquid metals used in the experiments, is very difficult to achieve without destabilizing the flow itself. In spite of great progress over the past two decades on both analytical and numerical sides (see reviews \cite{Ruediger_etal_2018_PhysRepo, Ji_Goodman2023}), a clear-cut and definitive identification of SMRI in the laboratory is still missing \cite{Sisan_etal2004,Nornberg_Ji_etal_2010_PhysRevLett,Roach_etal2012PhRvL, Hung_etal2019CmPhy}. 
In this respect, recent claims made by the Princeton team on the detection of the axisymmetric and non-axisymmetric modes of MRI \cite{Wang_etal2022b_PRL, Wang_etal2022a_NatCom} are currently under scrutiny by other groups \cite{Ruediger2023}. In their experiment, both the axisymmetric and non-axisymmetric modes of the alleged MRI manifest in close proximity to each other and at $Lu$ and $Rm$ that both are much lower than what global linear stability analysis predicts \cite{Wang_etal2022a_NatCom,Ruediger2023}). 
	
By contrast, the helical (HMRI, \cite{Hollerbach_Rudiger_2005}) and azimuthal (AMRI, \cite{Hollerbach_Rudiger_2010}) versions of MRI have been characterized and identified conclusively in the liquid metal (GaInSn) experiment PROMISE \cite{Stefani_Gundrum_Gerbeth_etal_2006PhRvL, Stefani_Gerbeth_Gundrum_Etal_2009PhysRevE, Seilmayer_etal2014, Mishra_etal2021}. In spite of this success of the PROMISE experiment, its constructional restriction makes it incapable  of achieving $Rm \sim 1-10$ required for the onset of SMRI in a TC flow, which, given very small $Pm$, amounts to huge $Re \sim 10^6-10^7$. This has motivated the construction of a much bigger TC machine using liquid sodium as a working fluid within the DRESDYN project \cite{Stefani_etal2019}. The main advantage of this new, technologically advanced DRESDYN-MRI machine is its ability to achieve high enough $Lu \sim 10$ and $Rm \sim 40$ necessary for SMRI to set on and grow \cite{Mishra_etal_2022PhRvF}. The capabilities of the DRESDYN-MRI machine are not restricted only to SMRI, but should allow us to experimentally study also Tayler instability \cite{Seilmayer_etal2012PhRvL} and the recently discovered Super-HMRI at positive shear \cite{Mamatsashvili_Stefani_Hollerbach_Rudiger_2019}. 
Previously, we (Mishra et al. \cite{Mishra_etal_2022PhRvF}, hereafter Paper I) carried out a detailed linear analysis of axisymmetric SMRI for the parameter regimes of the upcoming DRESDYN-MRI experiment. It was shown that SMRI can, in principle, be detected for all Rayleigh-stable rotation profiles of interest, including the astrophysically important Keplerian profile, for those ranges of $Lu$ and $Rm$ accessible in the DRESDYN-MRI experiment. 

After linear analysis, of immediate interest is subsequent nonlinear saturation and evolution of SMRI, which have been extensively investigated since its rediscovery in both astrophysical \cite{Hawley_Balbus_1991ApJ_nonlinear, Balbus_Hawley_1998, Balbus2003} and laboratory \cite{Knobloch_Julien_2005PhFl,Liu_Goodman_Ji_NonlinearMRI_2006ApJ} settings. Recent liquid metal experiments on MRI in a magnetized TC devices and related numerical simulations have advanced our understanding of the nonlinear saturation and dynamics of SMRI \cite{Gellert_etal2012, Gissinger_Goodman_Ji_2012PhFl, Wei_etal2016, Choi_etal2019, Winarto_etal2020}. Early work by Liu et al. \cite{Liu_Goodman_Ji_NonlinearMRI_2006ApJ} performed numerical simulations of axisymmetric (with azimuthal wavenumber $m=0$) SMRI in a TC geometry without endcaps and observed a jet-like outflow at the mid-height of the cylinder, the reconnection layer, and different scalings for normalized perturbation torque at moderate and large magnetic Reynolds numbers, but still much lower Reynolds numbers than those found in experiments. Subsequent studies further explored the scaling for saturation amplitude of energy and angular momentum transport in a channel or TC geometry using weakly nonlinear analysis with periodic axial boundary conditions \cite{Umurhan_Menou_Regev_2007PhRvL, Umurhan_Regev_Menou_2007PhRvE, Clark_Oishi_2017ApJ_local_geometry, Clark_Oishi_2017ApJ_TC_geometry}. However, these scaling properties turned out to depend on the geometry of the TC setup (thin or wide gap, height of cylinders, etc.) in question. Recently, we (Mishra et al. \cite{Mishra_etal_nonlinear_2022}, hereafter Paper II) have studied in detail the fully nonlinear evolution and saturation properties of axisymmetric SMRI in a wide-gap TC flow without endcaps and analysed its scaling behavior over a wide range of $Lu$, $Rm$ and $Re$ accessible in the DRESDYN-MRI device. It was shown that the saturation of SMRI occurs via magnetic reconnection. The dependence of the saturated magnetic energy and torque on Reynolds number was also analyzed and shown to follow a power law scaling $Re^a$, where $-0.6\leq a \leq -0.5$, and $Re^b$, where $0.4 \leq b \leq 0.5$, respectively, at large $ Re \geq 4000$ and all sets of $(Lu, Rm)$, so that the exponents always satisfy the relationship $b-a\approx 1$. 

Relaxing the idealization of infinite cylinders, a number of studies \cite{Gissinger_Goodman_Ji_2012PhFl, Wei_etal2016, Choi_etal2019, Winarto_etal2020, Wang_etal2022b_PRL} focused on the saturation properties and nonlinear dynamics of SMRI in a more realistic setup of finite-height TC flow with top and bottom endcaps primarily in the context of the Princeton MRI experiment. These studies analyzed the dependence of the saturated amplitude on $Lu, Rm$ and $Re$ under the influence of insulating or conducting endcaps, and revealed different scalings with $Pm$ (or $Re$) in these two cases, taking into account  the complexity introduced by Ekman circulations near the top and bottom endcaps. 

In the DRESDYN-MRI experiments, the main parameters $Lu, Rm$ and $Re$ are high enough (see Table \ref{Table1:non_dimensional} below) for {\it non-axisymmetric} ($m \neq 0$) SMRI modes to grow also in parallel with axisymmetric ones, although with a smaller growth rate. For clear identification of different modes and the parameter regimes in which each mode can be excited, it is important to study the linear stability of non-axisymmetric modes too. In Papers I and II, we discussed the linear and nonlinear axisymmetric SMRI in the context of DRESDYN-MRI experiment. Extending this analysis, in this paper our aim is to investigate in detail the dynamics of non-axisymmetric SMRI modes. Although non-axisymmetric SMRI in TC flows has been investigated before both in ideal \cite{Ogilvie_Pringle1996, Khalzov2006, Goedbloed_Keppens2022, Ebrahimi2022} and non-ideal MHD primarily for liquid metals \cite{Rudiger_Schultz_Shalybkov_2003PhRvE, Shalybkov_Rudiger_Schultz_2002A&A_nonaxisMRI, Gellert_etal2012, Wang_etal2022a_NatCom}, its analysis in the parameter regimes specific to the DRESDYN-MRI experiments at very small $Pm \sim 10^{-6}-10^{-5}$ of liquid sodium and comparison with axisymmetric SMRI mode is still lacking. Motivated by this, in the present paper we first investigate the linear regime of non-axisymmetric SMRI and then its nonlinear development, saturation and scaling properties as a function of different system parameters.

The paper is organized as follows. The setup of the problem and the basic equations are described in Sec. \ref{sec_2_math_setting}. The formulation of the linear stability problem is given in Sec. \ref{linear_results} where the main results for non-axisymmetric SMRI and the comparison of the growth rates of axisymmetric and non-axisymmetric SMRI modes are also discussed. Nonlinear evolution and saturation properties of the non-axisymmetric modes are discussed in Sec. \ref{nonlin_results}. A summary and conclusions are in Sec. \ref{conclusion}.

\section{Physical model and main equations} \label{sec_2_math_setting}
	
\begin{figure}
\includegraphics[width=0.4\textwidth]{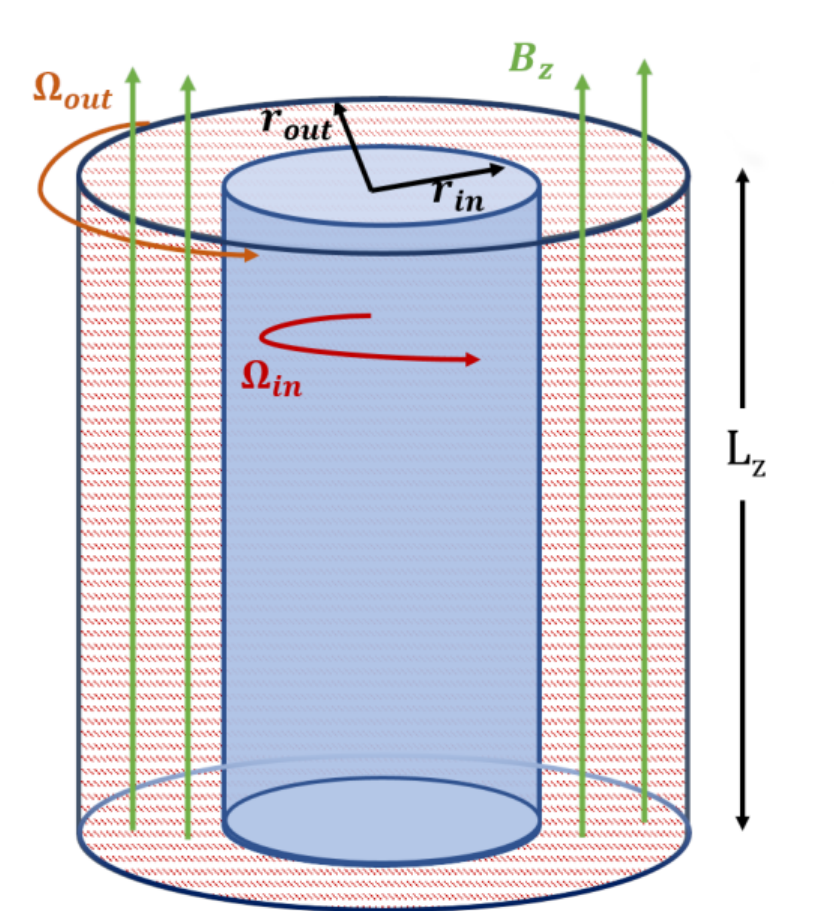}
\caption{A sketch of a cylindrical Taylor-Couette flow setup with an axial background magnetic field.}\label{fig:TC_sketch}
\end{figure}

As in papers I and II, we consider a cylindrical TC flow setup -- a basis for the DRESDYN-MRI experiment, which contains liquid sodium as a working fluid in the cylindrical coordinate system $(r,\phi,z)$ (Fig. \ref{fig:TC_sketch}). In this setup, the inner and outer cylinders with radii $r_{in}$ and $r_{out}$ rotate, respectively, with the angular velocities $\Omega_{in}$ and $\Omega_{out}$. In the DRESDYN-MRI machine, the ratio of the inner and outer cylinder radii is fixed to $r_{in}/r_{out}=0.5$ and the aspect ratio $L_z/r_{in}=10$ is large, where $L_z$ is the length of the cylinders (Table \ref{Table1:non_dimensional}). For simplicity, here we assume that the cylinders do not have endcaps and perturbations can extend along the cylinder axis. The ratio of the angular velocities of the outer and inner cylinders, $\mu=\Omega_{out}/\Omega_{in}$, can be varied (see Table \ref{Table1:non_dimensional}). A uniform axial magnetic field ${\bf B}_0=B_{0z}{\bf e}_z$ is imposed by a current-carrying solenoid surrounding the outer cylinder. This field is current-free between the cylinders and therefore does not exert any Lorentz force on the fluid. As a result, in the absence of endcaps, the equilibrium azimuthal flow ${\bf U}_0=(0,r\Omega(r),0)$ between the cylinders has a classical hydrodynamical TC profile of angular velocity 
\begin{equation} \label{TC_flow}
	\Omega(r) = C_1+\frac{C_2}{r^2}
\end{equation}
where the coefficients $C_1$ and $C_2$ are
\begin{equation*}
	C_1=\frac{\Omega_{out}r_{out}^2-\Omega_{in}r_{in}^2}{r_{out}^2-r_{in}^2}, \hspace{0.3cm} C_2=\frac{(\Omega_{in}-\Omega_{out})r_{in}^2r_{out}^2}{r_{out}^2-r_{in}^2}.
\end{equation*}
In laboratory TC experiments, this equilibrium profile is inevitably modified due to Ekman circulations (pumping) induced by the endcaps \cite{Hollerbach_Fournier2004,Szklarski2007,Stefani_Gerbeth_Gundrum_Etal_2009PhysRevE, Gissinger_Goodman_Ji_2012PhFl}. This effect for DRESDYN-MRI experiments will be discussed elsewhere. Here we only note that the DRESDYN-TC device has a large aspect ratio $L_z/r_{in}=10$ and split ring system at the endcaps designed such as to minimize Ekman pumping as much as possible \cite{Szklarski2007}.

The basic equations of non-ideal MHD governing the motion of an incompressible conducting fluid are
\begin{equation}\label{moment}
	\frac{\partial {\bf U}}{\partial t}+({\bf U}\cdot \nabla) {\bf U}=-\frac{1}{\rho}\nabla P + \frac{{\bf J}\times {\bf B}}{\rho} + \nu\nabla^2 {\bf U}, 
\end{equation}
\begin{equation}\label{induc}
	\frac{\partial {\bf B}}{\partial t}=\nabla\times \left( {\bf U}\times {\bf B}\right)+\eta\nabla^2{\bf B}, 
\end{equation}
\begin{equation}\label{div}
\nabla\cdot {\bf U}=0,~~~\nabla\cdot {\bf B}=0,
	\end{equation}
where $\rho$ is the constant density, ${\bf U}$ is the velocity, $P$ is the thermal pressure, ${\bf B}$ is the magnetic field, ${\bf J}=\mu_0^{-1}\nabla \times {\bf B}$ is the current density with $\mu_0$ being the magnetic permeability of vacuum. The fluid has constant kinematic viscosity $\nu$ and Ohmic resistivity $\eta$. 

The basic TC flow (\ref{TC_flow}) is hydrodynamically stable according to Rayleigh's criterion $\mu > r_{in}^2/r_{out}^2=0.25$, ensuring that all the linear instabilities in this flow are of magnetic nature. The values of $\mu$ are extended up to quasi-Keplerian rotation  when the cylinders' angular velocities are related through Kepler's law, $\Omega_{in, out}\propto r_{in, out}^{-3/2}$,  giving $\mu \approx 0.35$, which is important for astrophysical disks.

We non-dimensionalize time by $\Omega_{in}^{-1}$, angular velocity profile $\Omega(r)$ by $\Omega_{in}$, length by $r_{in}$, velocity ${\bf u}$ by $r_{in}\Omega_{in}$, pressure and kinetic energy density by $\rho\Omega^2r_{in}^2$ and magnetic field ${\bf B}$ by the background field $B_{0z}$. The main parameters are Reynolds number $Re = \Omega_{in} r_{in}^2/\nu$, magnetic Reynolds number $Rm = \Omega_{in}r_{in}^2/\eta$, magnetic Prandtl
number $Pm = \nu/\eta = Rm/Re$ and Lundquist number
$Lu = V_Ar_{in}/\eta$, where $V_A = B_{0z}/\sqrt{\rho\mu_0}$ is the Alfv\'en speed. Table \ref{Table1:non_dimensional} gives the ranges of these parameters typical of the DRESDYN-MRI experiment, which we use below.

\begin{table}
		\resizebox{\columnwidth}{!}{%
			\begin{tabular}[b]{ccc}			\hline
				\textbf{Dimensionless Parameter} & \textbf{Definition} & \textbf{Values} \\
				\hline
				$\mu$  & $\Omega_{out}/ \Omega_{in} $ & (0.25,0.35]\\
				Aspect ratio & $L_z/r_{in}$ & 10\\
				Reynolds number ($Re$) & $\Omega_{in} r_{in}^2/\nu$ & $\leq 7.72\times10^{6}$ \\
				Lundquist number ($Lu$) & $B_{0z}r_{in}/\eta\sqrt{\rho\mu_0}$ & $\leq 10$\\
				Magnetic Prandtl number ($Pm$) & $\nu/\eta $ & $7.77 \times 10^{-6}$\\
				Magnetic Reynolds number ($Rm$) & $Re\cdot Pm$ & $\leq 40$\\
				\hline
				
			\end{tabular}
		}
		\caption{Non-dimensional parameters of the DRESDYN-MRI experiment with liquid sodium at $T=130^{\circ}$C (see also Paper I).}\label{Table1:non_dimensional}
		
	\end{table}
	
\begin{figure*}
\centering		
\includegraphics[width=0.3\textwidth]{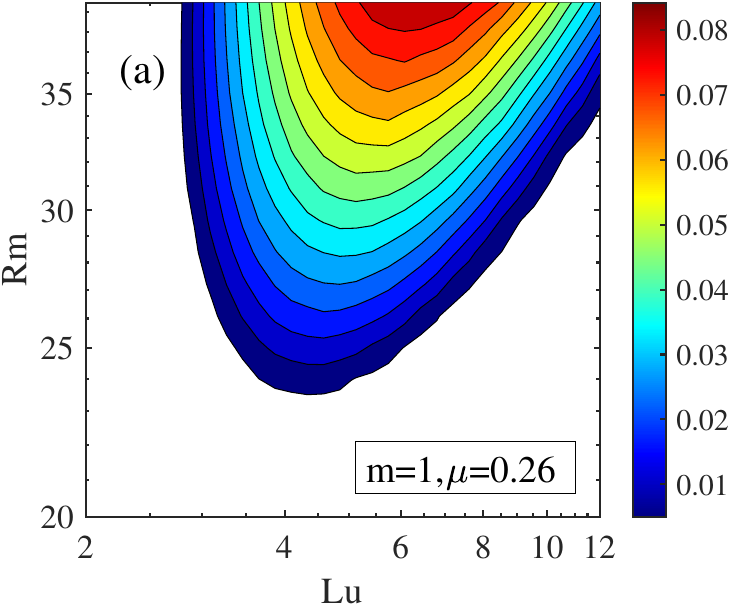}
\includegraphics[width=0.3\textwidth]{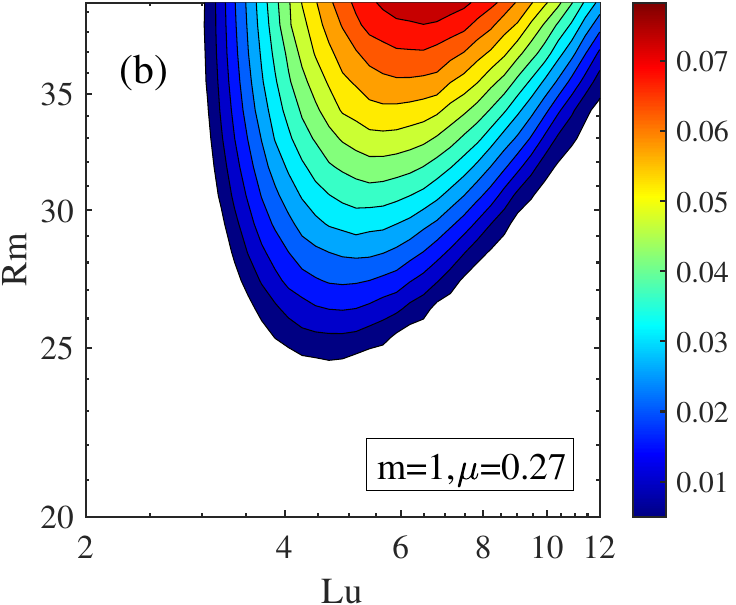}
\includegraphics[width=0.3\textwidth]{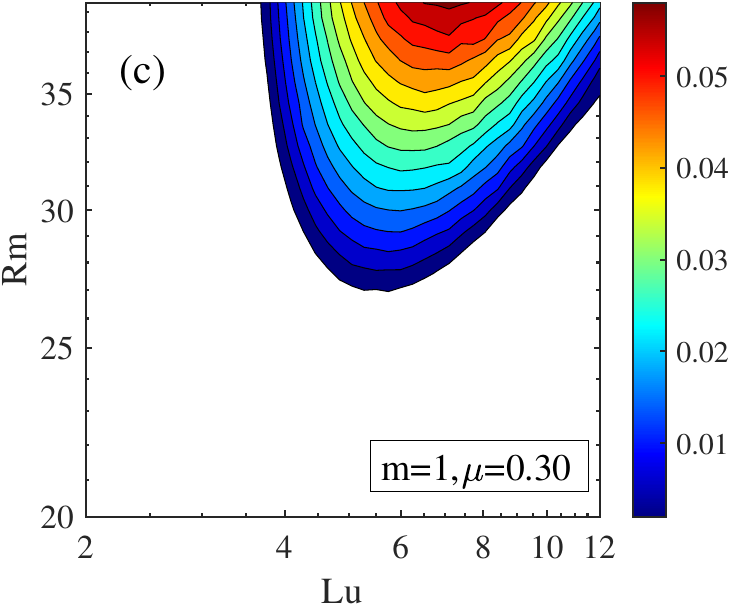}
\includegraphics[width=0.3\textwidth]{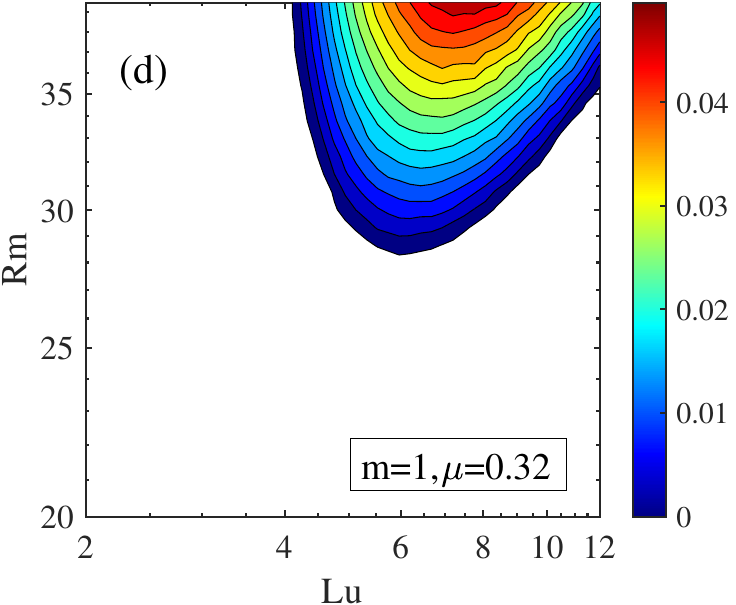}
\includegraphics[width=0.32\textwidth]{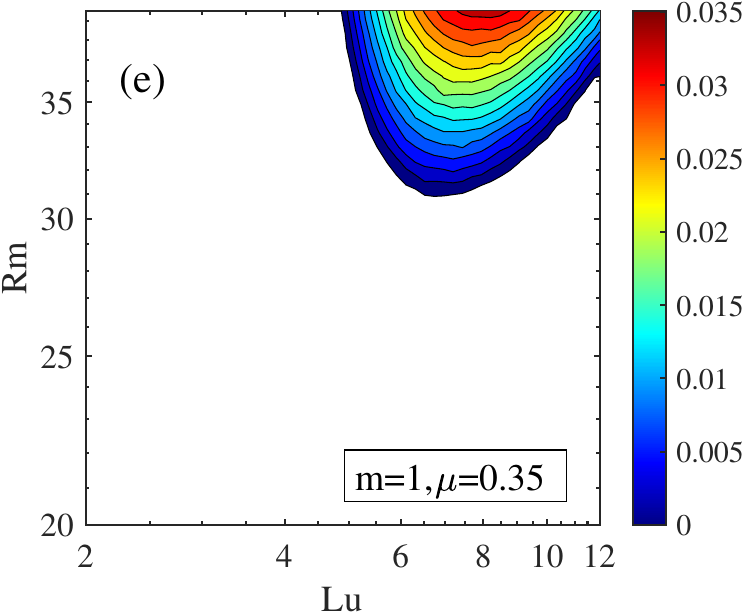}
\includegraphics[width=0.26\textwidth]{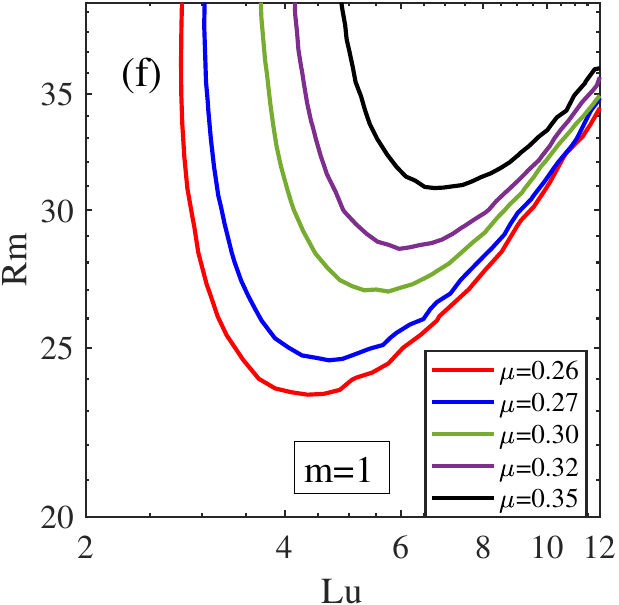}
\caption{Growth rate, ${\rm Re}(\gamma)>0$, of the non-axisymmetric $|m|=1$ mode of SMRI for $Pm = 10^{-5}$ and (a) $\mu=0.26$, (b) $\mu=0.27$, (c) $\mu=0.30$, (d) $\mu=0.32$, (e) $\mu=0.35$. (f) The marginal stability curves for the $|m|=1$ mode at these $\mu$.} \label{fig:Gr_m1_beta0_allMu}
\end{figure*}

\section{Linear analysis of non-axisymmetric SMRI} \label{linear_results}

The linear instability of this base TC flow ${\bf U}_0$ with the imposed constant axial magnetic field ${\bf B}_0$ to small perturbations, ${\bf u}={\bf U}-{\bf U}_0$, $p=P-P_0$, ${\bf b}={\bf B}-{\bf B}_0$ is studied. These perturbations are assumed to have a standard modal form $\propto \exp(\gamma t+ i m\phi+ik_z z)$, where $\gamma$ is the complex eigenvalue and $m$ and $k_z$ are, respectively, the azimuthal (integer) and axial wavenumbers. The flow is unstable when the real part of $\gamma$ is positive, ${\rm Re}(\gamma)>0$. Linearizing the main Eqs. (\ref{moment})-(\ref{div}), we obtain the system of non-dimensional perturbation equations as given in Paper I. Employing a similar spectral collocation method as in \cite{Hollerbach_Rudiger_2005, Hollerbach_Rudiger_2010}, we discretize the radial structure of the variables using the Chebyshev polynomials with typical resolution $N=120-140$ for a given pair of $(m, k_z)$. This resolution is taken to be higher than that used for axisymmetric $m=0$ SMRI modes in Paper I in order to resolve a thin region around co-rotation radius $r_c$, where $m\Omega(r_c)=-{\rm Im}(\gamma)$, in the structure of non-axisymmetric SMRI mode eigenfunctions (see below). Here we assume at least one full axial wavelength of non-axisymmetric $|m|\geq 1$ modes fits in the cylinder length $L_z$, so we set for the minimum axial wavenumber $k_{z, \rm min}=2\pi/L_z$ and maximize the growth rate over larger wavenumbers $k_z \geq k_{z,min}$. No-slip boundary conditions are used for the velocity and insulating  for the magnetic field. The linearized Eqs. (\ref{moment})-(\ref{div}) supplemented with these boundary conditions constitute a large matrix ($4N \times 4N$) eigenvalue problem for $\gamma$ the real part of which, when positive,  is the growth rate of the instability. 

In the linear analysis below, we focus on the non-axisymmetric $|m|=1$ SMRI mode \footnote{Because of symmetry, the results for the $m<0$ modes are the same as those for the $m>0$ modes, so without loss of generality everywhere below we use the absolute value $|m|$ when describing non-axisymmetric mode dynamics.}, which is the second most unstable mode after the axisymmetric $m=0$ SMRI mode, while higher $|m|\geq 2$ modes appear to be linearly stable in the considered ranges of the main parameters as given in Table \ref{Table1:non_dimensional}. On the other hand, in the more interesting fully nonlinear regime, we include also high-$m$ modes, because, as shown below, they can grow as a result of the modification (deviation) of the radial shear of the mean azimuthal velocity from the initial TC profile (\ref{TC_flow}) due to the nonlinear saturation of the axisymmetric SMRI mode.  For numerical reasons, we adopt $Pm=10^{-5}$ in the linear stability analysis below, which is about 1.3 times larger than the experimental value $Pm=7.77\times 10^{-6}$ (Table \ref{Table1:non_dimensional}). But since at small $Pm \ll 1$ the \textit{linear} dynamics and hence the growth rate of SMRI for a given $Lu$ and $Rm$ are essentially insensitive to $Pm$ (\cite{Rudiger_Schultz_Shalybkov_2003PhRvE}, see also Appendix \ref{Pm_dependence}), this difference in $Pm$ is not important.

Figure \ref{fig:Gr_m1_beta0_allMu} shows the growth rate for the unstable non-axisymmetric $|m|=1$ mode of SMRI, maximized over axial wavenumbers $k_{z} \geq k_{z,\rm min}$, in the $(Lu, Rm)$-plane at various $\mu$. It is seen that, similar to the axisymmetric case studied in Paper I, the instability region for the $|m|=1$ mode moves towards higher $Lu$ and $Rm$ with increasing $\mu$. This, in turn, increases the critical $Lu_c$ and  $Rm_c$ for the instability onset with $\mu$, as given in  Table \ref{Table2:critical_m0_m1}, but the critical $Lu_c\approx 6.7$ and $Rm_c\approx 30.9$ for the highest adopted $\mu=0.35$ is still well within the maximum achievable ranges of Lundquist and magnetic Reynolds numbers in the DRESDYN-MRI experiment (Table \ref{Table1:non_dimensional}). Notably, on comparison with Fig. 2 of Paper I, it is seen that the non-axisymmetric $|m|=1$ mode has about 2-3 times smaller growth rate than that of the axisymmetric $m=0$ mode for a fixed pair of $(Lu, Rm)$. At any rate, this non-axisymmetric mode for all considered $\mu\in(0.25, 0.35]$ appears to be unstable within the range of Lundquist and magnetic Reynolds numbers accessible in DRESDYN-TC machine and hence it can in principle be excited in the upcoming experiments together with the axisymmetric SMRI.

\begin{figure}
\includegraphics[width=\columnwidth]{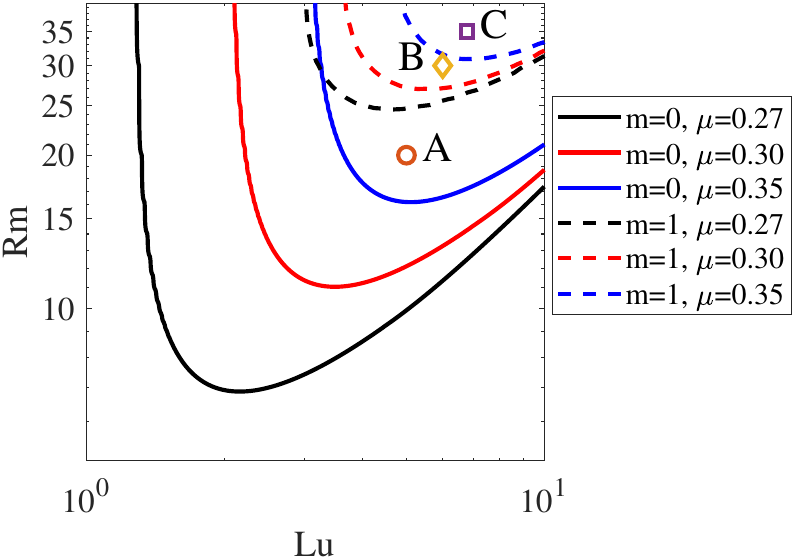}
\caption{Marginal stability curves for the axisymmetric $m=0$ (solid lines) and non-axisymmetric $|m|=1$ (dashed lines) SMRI modes at different $\mu=0.27~(black), 0.30~(red), 0.35~(blue)$ and $Pm=10^{-5}$. For all $\mu$, $m=0$ mode has much lower critical $Lu_c$ and $Rm_c$ than the $|m|=1$ mode. Points A, B, and C, which lie, respectively, outside, near and inside the marginal stability curve of the $|m|=1$ mode for the classical TC profile with quasi-Keplerian rotation $\mu=0.35$, are used for the nonlinear analysis below.} \label{fig:compare_m0_m1_mu}
	\end{figure}

In the DRESDYN-MRI experiment, large values of $Lu$ and $Rm$ are achievable where the axisymmetric and non-axisymmetric modes can co-exist in the flow. Given the importance of definitive detection of these modes, one should determine the parameter ranges over which these axisymmetric $m=0$ and non-axisymmetric $m=1$ SMRI modes can be uniquely identified. For comparison, in Fig. \ref{fig:compare_m0_m1_mu} we show the marginal stability curves for these modes together for different $\mu \in \{0.27, 0.30, 0.35\}$. As mentioned above, the instability region for both $m=0$ and $|m|=1$ modes in the $(Lu, Rm)$-plane becomes smaller as $\mu$ increases, because in this case the shear of the basic TC flow, which mainly drives SMRI, decreases. Evidently, the instability region of the axisymmetric $m=0$ mode is much larger than that of the non-axisymmetric $|m|=1$ mode and therefore the critical $Lu_{c}$ and  $Rm_{c}$ for $m=0$ mode are, respectively, about 1.3-2.87  and 1.9-4.85 times smaller than those of the non-axisymmetric $|m|=1$ mode for fixed $\mu$ (see Table \ref{Table2:critical_m0_m1} and Paper I). Hence, this gap  between the $m=0$ and $|m|=1$ modes' critical values at fixed $\mu$ can be used for the detection of axisymmetric SMRI mode distinct from non-axisymmetric one. The characteristic points A, B, and C in Fig. \ref{fig:compare_m0_m1_mu}, which lie, respectively, within the instability region of the $m=0$ SMRI mode but outside that of the $|m|=1$ mode, near the marginal stability of the latter mode, and within its instability region in the case of the quasi-Keplerian rotation $\mu=0.35$, are used as the reference points in the following nonlinear analysis.
	
	\begin{table}
		\resizebox{\columnwidth}{!}{%
			\begin{tabular}[b]{ccc}			\hline
				\textbf{$\mu$} & \hspace{8em} \textbf{$(Lu_c, Rm_c)$}  \\
				& \textbf{$m=0$}  & \textbf{$|m|=1$}\\
				\hline
				0.26 & (1.5, \, 4.8) & (4.3, 23.3)\\
				0.27 & (2.1, \, 6.9) & (4.7, 24.5)\\
				0.30 & (3.5, \,\, 11) & (5.7, 26.9) \\
				0.32 & (4.2, 13.2) & (5.9, 28.5)\\
				0.35 & (5.1, 16.2) & (6.7, 30.9)\\
				
				\hline
				
			\end{tabular}
		}
		\caption{Critical values for the onset of axisymmetric $m=0$ and non-axisymmetric $|m|=1$ SMRI modes for different $\mu$ obtained via 1D linear stability analysis.}\label{Table2:critical_m0_m1}
		
	\end{table}

\begin{figure*}
\centering
\includegraphics[width=\textwidth]{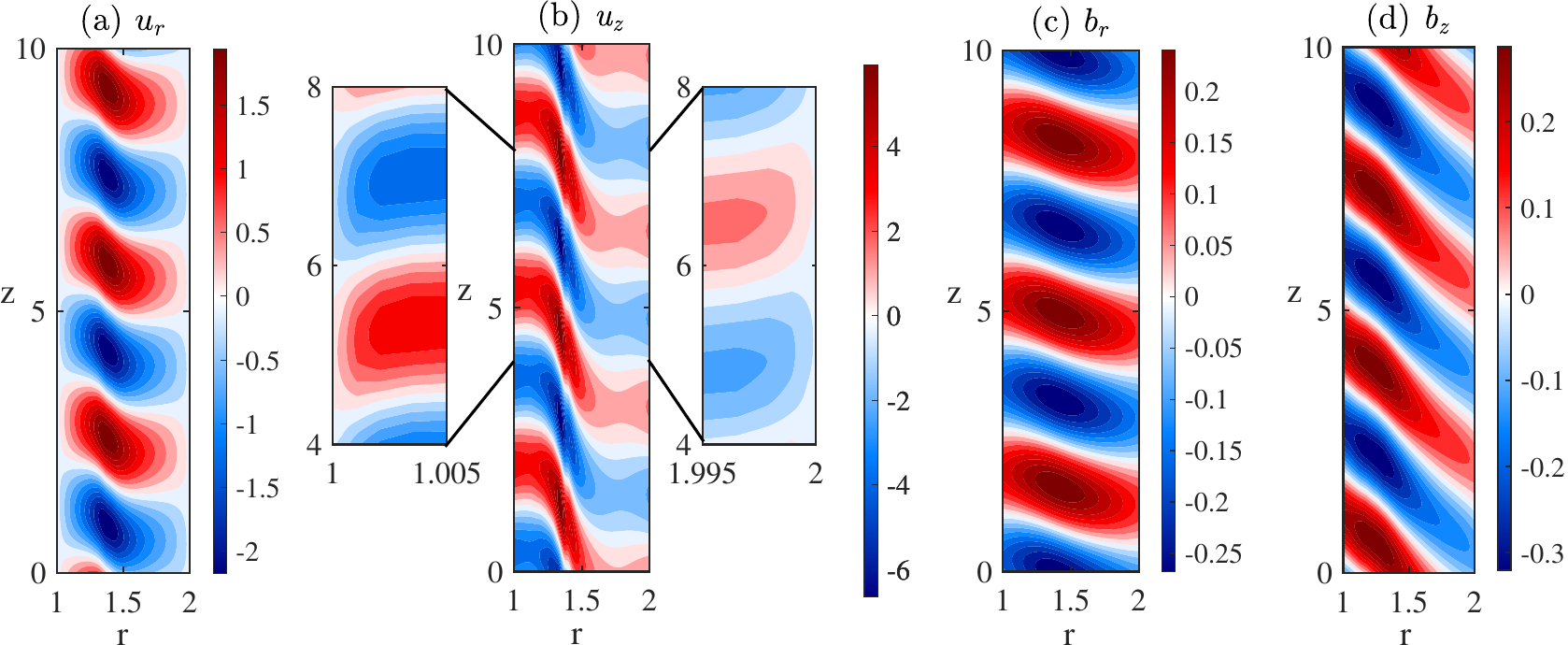}
\caption{Structure of the (a) radial velocity $u_r$, (b) axial velocity $u_z$,  (c) radial magnetic field $b_r$ and (d) axial magnetic field $b_z$ eigenfunctions in the $(r,z)$-plane for $m=1$ mode of SMRI at $\mu = 0.35, Lu=6.78, Rm=35, Pm = 10^{-5}$ (point C in Fig. \ref{fig:compare_m0_m1_mu}) and $k_z = 1.88$. A zoomed in axial velocity $u_z$ at the cylinder boundaries in panel (b) confirms the validity of the no-slip condition for velocity.} \label{fig:eigen_linearcode}
\end{figure*}
	
Figure \ref{fig:eigen_linearcode} shows the typical structure of the radial and axial velocities, $(u_r, u_z)$, and magnetic field, $(b_r,b_z)$, eigenfunctions in the $(r,z)$-plane for the non-axisymmetric $m=1$ mode at $\mu=0.35, Lu=6.78, Rm=35$ (point C in Fig. \ref{fig:compare_m0_m1_mu}) and the axial wavenumber $k_z=1.88$, which we have chosen in order to relate the linear analysis results with the nonlinear evolution presented below (see Appendix \ref{appendix_r_z_structures_growth_phase}). In the nonlinear analysis, the flow domain is periodic in the $z$-direction with a period of $L_z$ and hence the axial wavenumber is discrete $k_z=2\pi n_z/L_z$, where the integer $n_z=\pm 1, \pm 2,...$ (by contrast, in the linear analysis $k_z$ is a free parameter). The chosen $k_z=1.88=3\cdot 2\pi/L_z$, corresponding to $n_z=3$ wavelength within the cylinder length $L_z$, yields the maximum growth rate among the discrete $k_z$ values (see Fig. \ref{fig:diff_Pm} in Appendix \ref{Pm_dependence}).

The radial and axial velocity eigenfunctions reach their highest values and have a strong elongation of the cells in the axial direction within a thin layer around $r_c\approx 1.3$. This point represents the co-rotation radius at which the azimuthal phase velocity of the mode is equal to the flow velocity, i.e., $m\Omega(r_c)=-{\rm Im}(\gamma)$, and is located closer to the inner boundary. This is consistent with the results of Ref. \cite{Ebrahimi2022} that non-axisymmetric $m=1$ SMRI mode eigenfunctions are confined between two Alfv\'en resonance points located on either side of the co-rotation radius and reach higher values around that radius \footnote{In the present case with resistivity, the meaning of the Alfv\'en resonance points is, however, less clear.}. Since it is a higher-frequency mode, its co-rotation radius and hence the eigenfunctions tend to be located nearer the inner cylinder wall. Similarly, the radial and axial magnetic field eigenfunctions are also elongated near the co-rotation radius, but to a lesser degree, and reach higher values there.  In Appendix \ref{appendix_r_z_structures_growth_phase}, we compare these eigenfunctions obtained from the 1D linear stability code with the velocity and magnetic field structures at the early exponential growth stage in the nonlinear simulations.

 \section{Nonlinear evolution of non-axisymmetric modes}\label{nonlin_results}

From the above 1D linear stability analysis, we inferred that the axisymmetric and non-axisymmetric SMRI modes can co-exist in the same parameter regime with the latter typically having several times lower growth rates. In the experiments, a flow is allowed to reach a steady state before the measurements are taken. Thus, in order to uniquely identify different modes in the flow, it is necessary to first understand the nonlinear saturation properties of both axisymmetric and non-axisymmetric modes. In Paper II, we studied the nonlinear saturation and dynamics of \textit{axisymmetric} SMRI mode, since it is the most unstable one in the considered TC flow. It was shown to saturate via magnetic reconnection and the corresponding scaling relations of the total magnetic energy and torque of perturbations at the cylinders in the saturated state were derived with respect to Reynolds number $Re$. Importantly, these scalings have allowed us to extrapolate and estimate the expected magnitudes of velocity and magnetic field perturbations at those high $Re \gtrsim 10^6$ which are relevant for the DRESDYN-MRI experiment. In this section, focusing on the quasi-Keplerian rotation $\mu=0.35$, we investigate the nonlinear evolution and saturation of non-axisymmetric modes in comparison with axisymmetric ones, which is important for the preparation of the DRESDYN-MRI experiment and understanding its outcomes.

To solve the basic non-ideal MHD Eqs. (\ref{moment})-(\ref{div}), as in Paper II, we use the pseudo-spectral code described in \cite{Guseva_etal_2015NJPh}. This code employs a high-order finite-difference method for radial expansion, and Fourier expansions in the axial and azimuthal directions. The time-update is done using an implicit Crank–Nicolson scheme of second order. The nonlinear terms are calculated using the pseudo-spectral method with the 2/3-dealiasing rule. As in the linear analysis above, the boundary conditions are no-slip for velocity and insulating for magnetic field at the cylinder walls, which is consistent with the conditions present in the DRESDYN-MRI facility, and periodic in the axial $z$-direction. Additional details about this code and its validation can be found in \cite{Guseva_etal_2015NJPh}. The cylindrical flow domain is defined by the cylindrical coordinates $(r, \phi, z) \in [r_{in}, r_{out}] \times [0, 2\pi] \times [0, L_z]$, where, as before, the non-dimensional inner and outer radii $r_{in}=1$, $r_{out}=2$, and the cylinder length $L_z=10$. To achieve high resolution close to the cylinder walls, the Chebyshev collocation method is used to distribute points radially. We take $N_r=480$ finite difference points in the radial direction and a total of $N_z=480$ Fourier modes in the axial $z$-direction, that is, $|k_z|\leq 2\pi N_z/L_z$. The number of non-axisymmetric modes, $N_{\phi}$, is set to $20$, that is, the azimuthal wavenumber $|m|\leq N_{\phi}=20$.  In Appendix \ref{appendix_resolution_test}, we perform a resolution study for different $N_\phi$ and $N_z$ Fourier mode numbers to optimize them.  It was found that the adopted values $N_\phi=20$ and $N_z=480$ are in fact sufficient for the convergence of the azimuthal $m$- and axial $k_z$-spectra of energy and hence for the reliable representation of the non-axisymmetric mode dynamics in the present setup.  Still, future higher-resolution studies are needed to further explore the fully nonlinear dynamics  of non-axisymmetric modes, especially at larger $Re\gtrsim  10^6$ relevant to MRI-experiments than those considered here.  As in the linear analysis, the minimum axial wavenumber is set to $k_{z,\rm min}=2\pi/L_z$ to ensure at least one full axial wavelength fits in the domain length $L_z$. 

To qualitatively analyze the growth and saturation of non-axisymmetric modes, we define a radially integrated spectral magnetic energy density \cite{Mamatsashvili_etal2018},
\begin{equation}\label{rad_int_spectral_energy_density}
    \bar{\mathcal{E}}_{mag}(m,k_z)=\pi L_z \int_{r_{in}}^{r_{out}} (|\bar{b}_r|^2+|\bar{b}_\phi|^2+|\bar{b}_z|^2)\,rdr, 
\end{equation}
\noindent
where $\bar{b}_r,~\bar{b}_{\phi},~\bar{b}_z$ are the Fourier transforms of the perturbed magnetic field components with respect to $m$ and $k_z$ wavenumbers at a given radius $r$,
\begin{equation}\label{b_i_(r,m,k_z)}
    \bar{b}_i(r, m, k_z)=\frac{1}{2\pi L_z}\int_{0}^{2\pi} \int_{0}^{L_z} b_i(r, \phi, z)e^{-im\phi-ik_zz} \,d\phi dz
\end{equation}
with $i=r,\phi,z$. We define the azimuthal $m$-spectrum of the magnetic energy density as the sum of $\bar{\mathcal{E}}_{mag}(m,k_z)$ over all $k_z$,  $\mathcal{E}_{mag}(m)=\sum_{k_z} \bar{\mathcal{E}}_{mag}(m,k_z)$, which thus represents a total magnetic energy of modes with a given $m$. In a similar manner, we define the non-dimensional kinetic energy spectrum $\bar{\mathcal{E}}_{kin}(m,k_z)$ by replacing the magnetic field ${\bf b}$ with the velocity ${\bf u}$ in Eq. (\ref{rad_int_spectral_energy_density}) and hence its azimuthal component $\mathcal{E}_{kin}(m)=\sum_{k_z}\bar{\mathcal{E}}_{kin}(m,k_z)$.

\begin{figure*}
    \centering
    \includegraphics[width=0.3\textwidth]{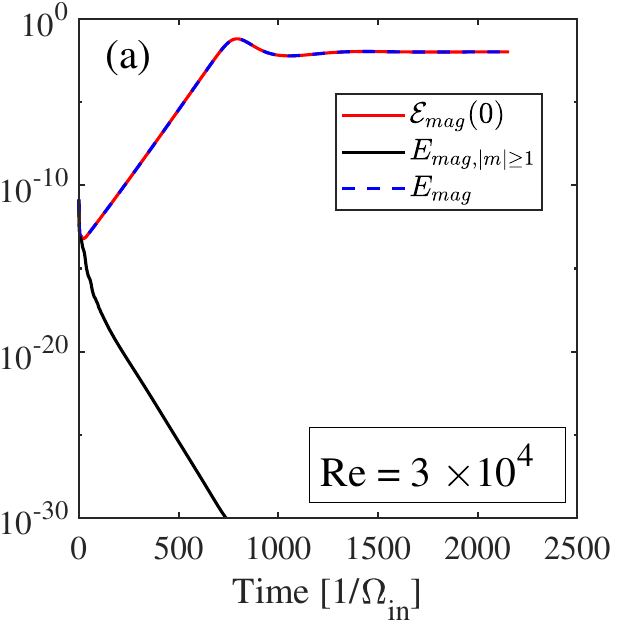}
    \includegraphics[width=0.3\textwidth]{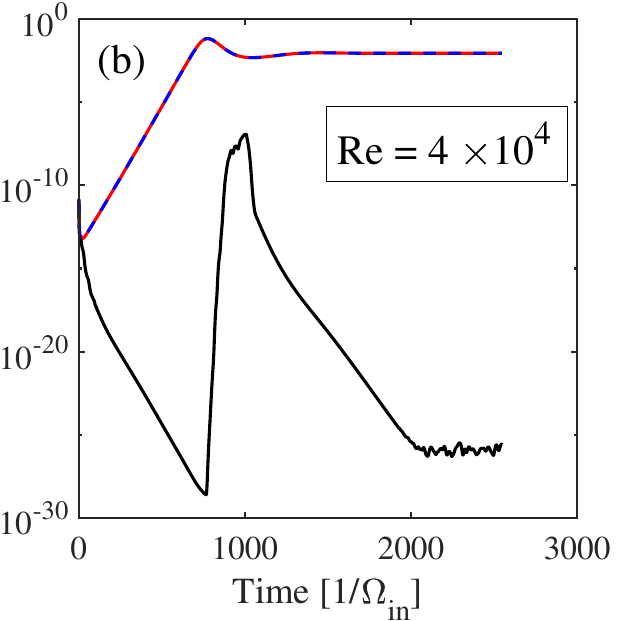}
    \includegraphics[width=0.3\textwidth]{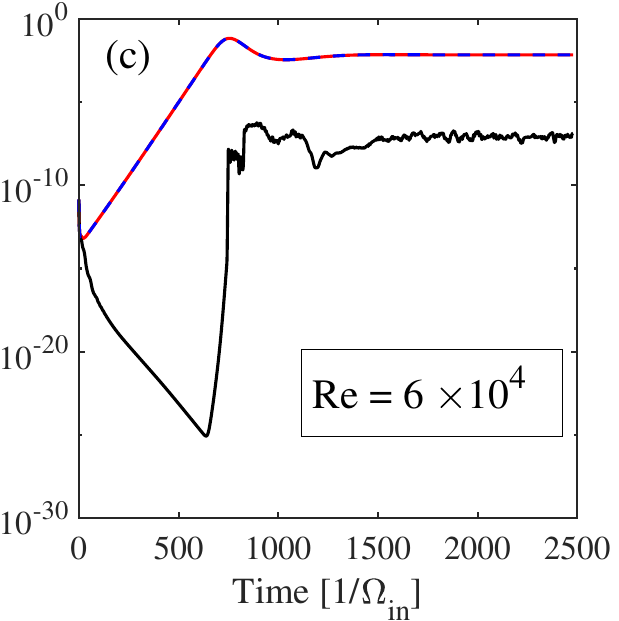}
    \caption{Evolution of the magnetic energy of axisymmetric $m=0$ modes, $\mathcal{E}_{mag}(0)$ (red), of all the non-axisymmetric $|m|\geq 1$ modes,  $E_{mag,|m|\geq 1}$ (black), and their sum $E_{mag}$ (dashed) for $Lu=5$, $Rm=20$ (point A in Fig. \ref{fig:compare_m0_m1_mu}) and (a) $Re = 3\times 10^4$, (b) $4\times 10^4$ and (c) $6\times 10^4$. The non-axisymmetric modes are linearly stable and hence initially decay for all $Re$ in the exponential growth phase of the axisymmetic SMRI mode. As seen in panels (b) and (c), a rapid growth of the non-axisymmetric modes occurs at large enough $Re$ (or small $Pm$) just when the dominant axisymmetric SMRI mode saturates and, as a result, the mean azimuthal flow profile changes (deviates) from the original TC flow (\ref{TC_flow}). At $Re=6\times 10^4$ in panel (c), the non-axisymmetric modes' total magnetic energy does not decay after the growth and settles down to a constant value, which is still several orders smaller than that of the axisymmetric mode.}
    \label{fig:Mag_energyRm20Lu5_m20}
\end{figure*}

\subsection{Magnetic energy -- evolution and radial dependence}

Figure \ref{fig:Mag_energyRm20Lu5_m20} shows the evolution of the magnetic energy for axisymmetric $m=0$ modes, $\mathcal{E}_{mag}(0)$, for all the non-axisymmetric $|m|\geq 1$ modes, $E_{mag, |m|\geq 1}\equiv \sum_{|m|\geq 1}\mathcal{E}_{mag}(m)$, and their sum $E_{mag}\equiv \mathcal{E}_{mag}(0)+E_{mag, |m|\geq 1}$, i.e., the total magnetic energy of perturbations at $Lu=5, Rm=20$ (point A in Fig. \ref{fig:compare_m0_m1_mu}) and different $Re = \{3, 4, 6\} \times 10^4$.  Note that because of numerical constraints,  these values of $Re$ adopted in the nonlinear analysis here (as in Paper II) are smaller than $Re \sim 10^6$ used in the above linear analysis and in the experiments, but are at least an order of magnitude higher than those typically used in previous simulations of the nonlinear (non-)axisymmetric SMRI in a magnetized Taylor-Couette flow \cite{Gissinger_Goodman_Ji_2012PhFl,Wei_etal2016,Choi_etal2019,Winarto_etal2020,Wang_etal2022a_NatCom, Wang_etal2022b_PRL}. In this regard, although the points A, B and C in Fig. \ref{fig:compare_m0_m1_mu} have been obtained for different Reynolds, or magnetic Prandtl numbers than those in the simulations, these points would not move in the $(Lu, Rm)$-plane (i.e., the corresponding growth rates would not change) with $Re$, or $Pm$ since, as mentioned above, the linear dynamics of SMRI is in fact insensitive to these two numbers for $Re \gg 1$, or $Pm \ll 1$ (here $Re \geq 10^4$ and  $Pm \leq 3.5\times 10^{-3}$). Therefore, we can use A, B and C points as the reference points in the nonlinear analysis.

It is seen in Fig. \ref{fig:compare_m0_m1_mu} that point A is way outside the linearly unstable regime of the $|m|=1$ mode in the presence of the original TC profile but still falls in the linearly unstable regime of $m=0$ SMRI mode. At $Re=3\times 10^4$, the magnetic energy of non-axisymmetric modes, $E_{mag,|m|\geq 1}$, decays over time while that of the axisymmetric mode, $\mathcal{E}_{mag}(0)$, grows and saturates [Fig. \ref{fig:Mag_energyRm20Lu5_m20}(a)]. For higher $Re=4\times 10^4$, $E_{mag,|m|\geq 1}$ initially decays during the exponential growth phase of the axisymmetric mode, but then it transiently increases very rapidly, reaching a peak, during the saturation of the latter mode [Fig. \ref{fig:Mag_energyRm20Lu5_m20}(b)]. However, after the axisymmetric mode has saturated, this peak is not sustained and the energy $E_{mag,|m|\geq 1}$ decays afterwards to a very small noise level. The non-axisymmetric mode energy reaches a similar peak for even higher $Re=6\times 10^4$, at which it eventually saturates at orders of magnitude higher levels, though still a few orders lower than the energy of the axisymmetric mode [Fig. \ref{fig:Mag_energyRm20Lu5_m20}(c)]. This implies that depending on $Re$, the non-axisymmetric modes, which would be stable for the original TC profile (\ref{TC_flow}), can nevertheless undergo rapid growth during the saturation process of the axisymmetric mode. A similar trend is seen at $Lu=6$, $Rm=30$ (point B in Fig. \ref{fig:compare_m0_m1_mu}), where the non-axisymmetric $|m|=1$ mode is marginally stable initially and becomes unstable in the course of saturation of the axisymmetric mode, as described in Appendix \ref{appendix_Lu6_Rm30_erergy_growth}. In all these cases, the axisymmetric mode is a dominant contributor to the total magnetic energy of perturbations.

\begin{figure*}
	\centering
	\includegraphics[width=0.32\textwidth]{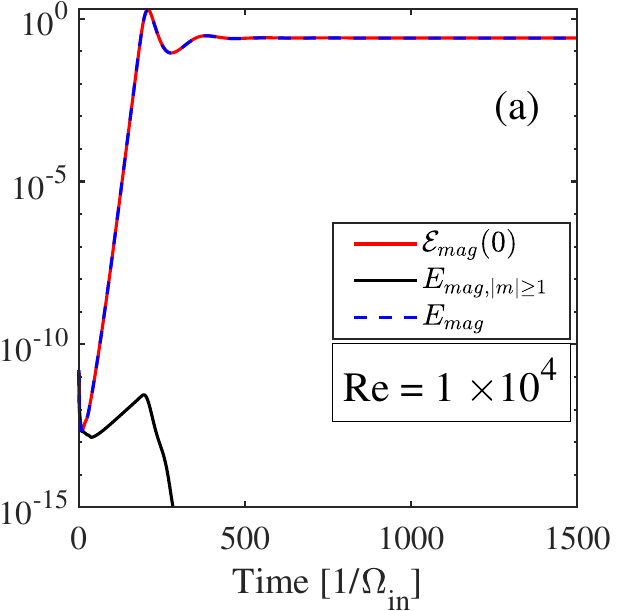}
	\includegraphics[width=0.3\textwidth]{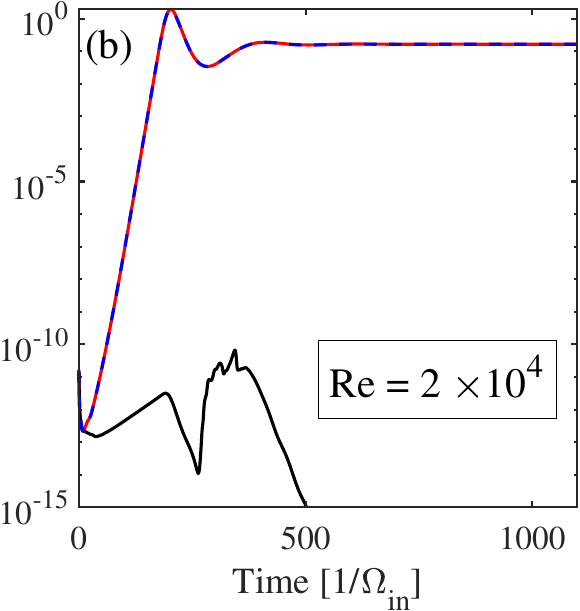}
	\includegraphics[width=0.31\textwidth]{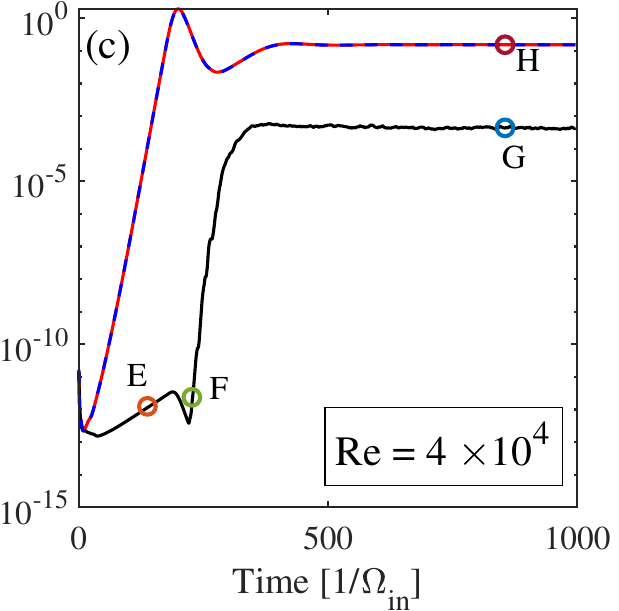}
	\caption{Same as in Fig. \ref{fig:Mag_energyRm20Lu5_m20} but for $Lu=6.78, Rm=35$ (point C in Fig. \ref{fig:compare_m0_m1_mu}) and (a) $Re=10^4$, (b) $2\times 10^4$ and (c) $4\times 10^4$. Note that the point C is inside the marginal stability curve of $|m|=1$ mode obtained from the original TC profile (\ref{TC_flow}) and therefore this mode grows for all $Re$ in the initial linear regime. As in Fig. \ref{fig:Mag_energyRm20Lu5_m20}, much steeper amplification of the non-axisymmetric mode energy occurs at higher $Re$ during the saturation of the axisymmetric mode as seen in panels (b) and (c), which in the latter case does not decay and saturates at a constant value. The time moments denoted by E, F, G and H in panel (c) will be used as the reference moments in subsection \ref{subsection_rz_and_meridional slices}.}
	\label{fig:Mag_energyRm35Lu678_m20}
\end{figure*}

The evolution of the magnetic energy of axisymmetric and all the non-axisymmetric modes for $Lu=6.78, Rm=35$ (point C in Fig. \ref{fig:compare_m0_m1_mu}) and $Re = \{1, 2, 4\}\times 10^4$ is shown in Fig. \ref{fig:Mag_energyRm35Lu678_m20}. Since the point C falls well within the linearly unstable regime of the non-axisymmetric $|m|=1$ mode, we see an initial exponential growth phase of this mode. For these parameters, the growth rate obtained from the nonlinear code is about $0.012$ which is close to that from the linear code $0.015$ for all $Re$ discussed in Fig. \ref{fig:Mag_energyRm35Lu678_m20}. Hence, it can be inferred that the unstable $|m|=1$ mode exhibiting growth at early times in the simulations is in fact the SMRI mode. For $Re=10^4$ and $2\times 10^4$, the non-axisymmetric modes do not, however, saturate after the exponential growth, but instead start to decrease during the saturation process of the axisymmetric SMRI mode [Figs. \ref{fig:Mag_energyRm35Lu678_m20}(a) and \ref{fig:Mag_energyRm35Lu678_m20}(b)]; at $Re=2\times 10^4$ this decrease is followed by some transient amplification, which reaching a maximum, falls again. By contrast, at larger $Re=4\times 10^4$, the magnetic energy of the non-axisymmetric modes, after an initial exponential growth phase, increases much more steeply during the saturation of the axisymmetric mode and eventually saturates itself to a certain value few orders smaller than that of the axisymmetric mode [Fig. \ref{fig:Mag_energyRm35Lu678_m20}(c)]. Thus,
Figs. \ref{fig:Mag_energyRm20Lu5_m20} and \ref{fig:Mag_energyRm35Lu678_m20} indicate that at high enough $Re$, irrespective of their initial linear stability or instability, non-axisymmetric modes exhibit rapid amplification in the nonlinear regime, when the dominant axisymmetric mode saturates. 

\begin{figure*}
\centering
\includegraphics[width=\textwidth]{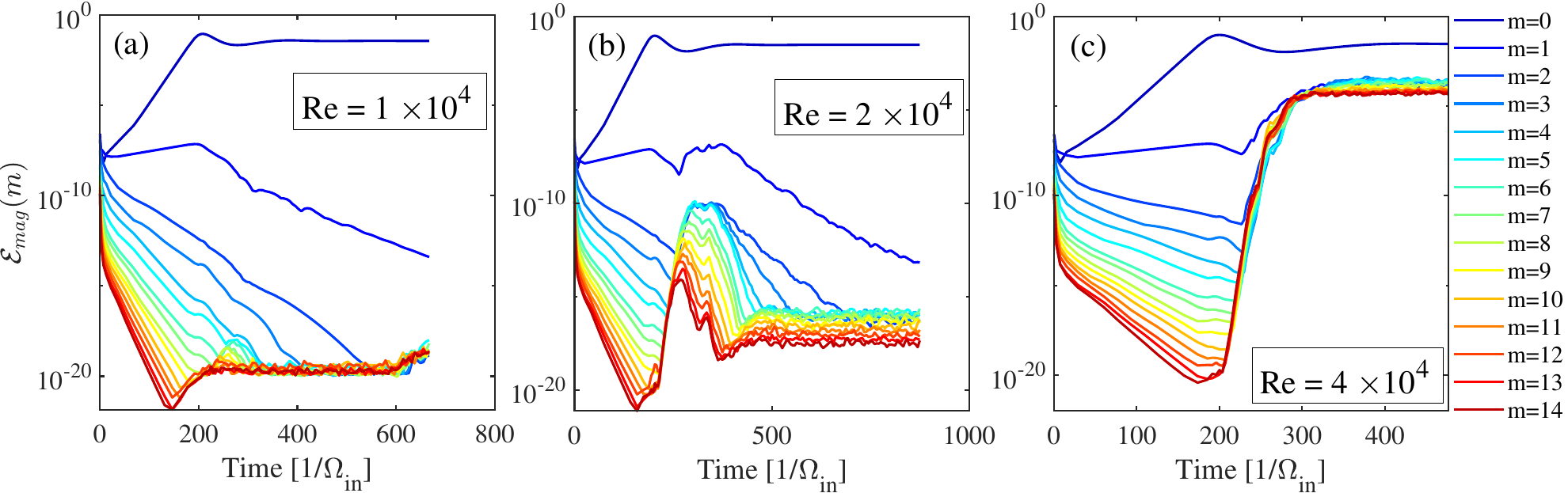}
\caption{Evolution of the azimuthal spectral magnetic energy density $\mathcal{E}_{mag}(m)$ for different $m \in [0,14]$ at the same $Lu=6.78$, $Rm=35$ and $Re$ as in Fig. \ref{fig:Mag_energyRm35Lu678_m20}. In the initial linear regime, only $|m|=1$ mode grows exponentially, because the point C lies in the unstable regime of this mode, while other non-axisymmetric $|m|>1$ modes are linearly stable for all considered $Re$. For the smallest $Re=10^4$ (a), all the non-axisymmetric $|m| \geq 1$ modes eventually decay in the saturated state. These modes undergo rapid growth at higher $Re$ during the saturation of the axisymmetric mode. For all $m$ this growth is only transient at $Re=2\times 10^4$ (b) and decays to a noise level, whereas it saturates at $Re=4\times 10^4$ (c) at about the same level though orders of magnitude smaller than that of the axisymmetric one.}
 \label{fig:Mode_energyRm35Lu678_m20}
\end{figure*}

In Fig. \ref{fig:Mode_energyRm35Lu678_m20}, we plot the evolution of $\mathcal{E}_{mag}$ for different $m\in[0,14]$ and the same parameters as in Fig. \ref{fig:Mag_energyRm35Lu678_m20}. It is evident that in the initial linear regime the exponential growth of the total magnetic energy of non-axisymmetric modes depicted in Fig. \ref{fig:Mag_energyRm35Lu678_m20} is essentially dominated by the $|m|=1$ mode while all other $|m|>1$ modes are stable and their energies decrease. For smaller $Re=10^4$, the $|m|=1$ mode also starts to decay to a very small noise level, as the axisymmetric $m=0$ mode undergoes the saturation phase and  modifies the mean azimuthal flow profile [Fig. \ref{fig:Mode_energyRm35Lu678_m20}(a)]. For larger $Re=2\times10^4$, all non-axisymmetric $|m| \geq 1$ modes exhibit transient growth nearly at the same time when the axisymmetric mode saturates [Fig. \ref{fig:Mode_energyRm35Lu678_m20}(b)]. This transient growth lasts the longer and reaches the higher peaks, the smaller $m$ is and then decays again to a very small level. For even higher $Re=4\times10^4$ transition to a nonlinear (turbulent) state finally takes place -- all $|m| \geq 1$ modes are excited, growing now rapidly during the same time of the axisymmetric mode saturation, and settle down to nearly similar amplitudes due to strong mutual nonlinear interaction at such high $Re$ [Fig. \ref{fig:Mode_energyRm35Lu678_m20}(c)]. (A similar behavior is observed also at point B of Fig. \ref{fig:compare_m0_m1_mu} and is discussed in Appendix \ref{appendix_Lu6_Rm30_erergy_growth}). Note that in all the cases considered above the saturation level of non-axisymmetric modes is few orders of magnitude smaller than that of the axisymmetric one. This indicates that the axisymmetric SMRI is always a dominant mode in the TC flow for the considered ranges of the main parameters $Lu$ and $Rm$ relevant to the DRESDYN-MRI experiment and as high $Re$ (i.e., as small $Pm$) as we can afford in our numerical simulations. 
 
\begin{figure*}
\centering
\includegraphics[width=0.8\textwidth]{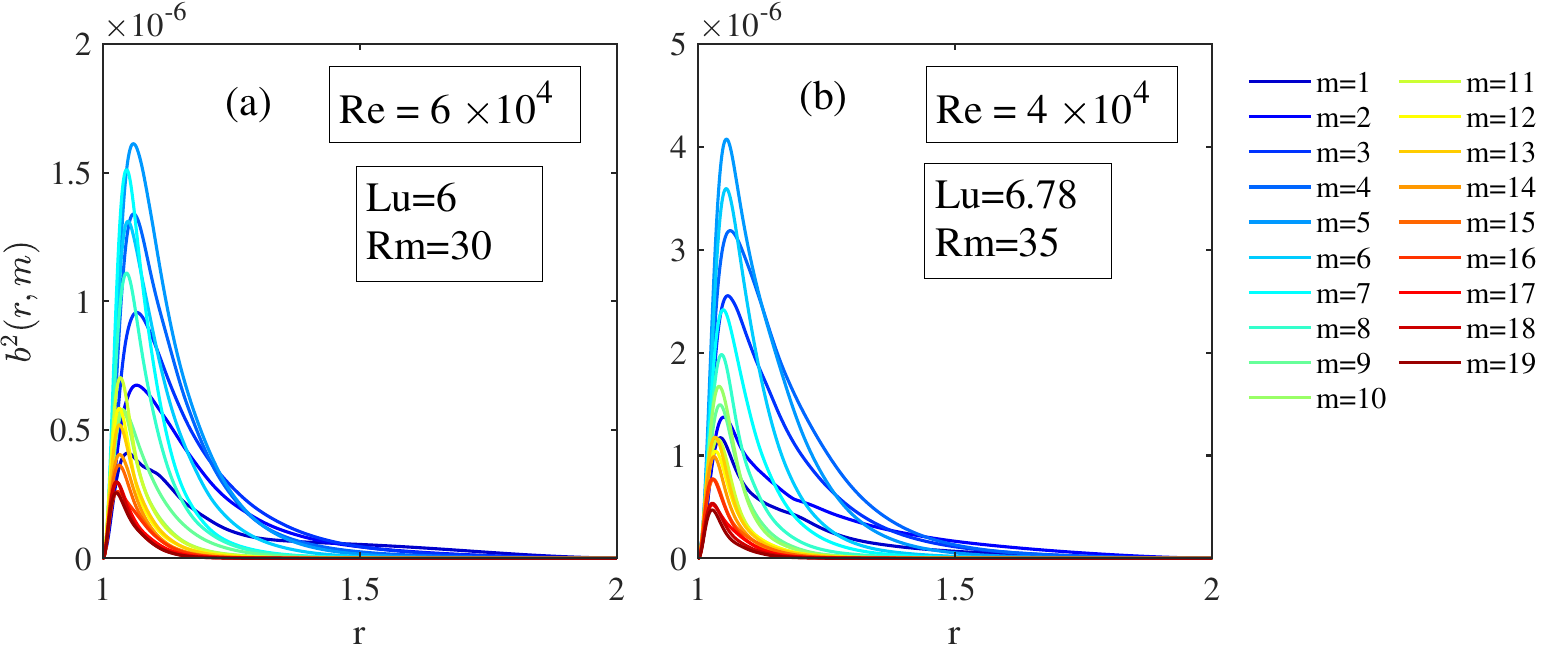}
\caption{Axially integrated magnetic energy $b^2(r,m)$ (see text) as a function of $r$ in the saturated state at different $m \in [1,19]$ for (a) $Lu=6, Rm=30, Re=6\times 10^4$ and (b) $Lu=6.78, Rm=35, Re=4\times 10^4$ (points B and C in Fig. \ref{fig:compare_m0_m1_mu}, respectively). From panels (a) and (b) it is clear that the larger-$m$ non-axisymmetric modes are more concentrated near the inner cylinder wall, indicating that they are mainly sustained by high shear $q$ values at this boundary.}\label{fig:Mode_energy_vs_r}
\end{figure*}

\begin{figure*}
\centering
\includegraphics[width=\textwidth]{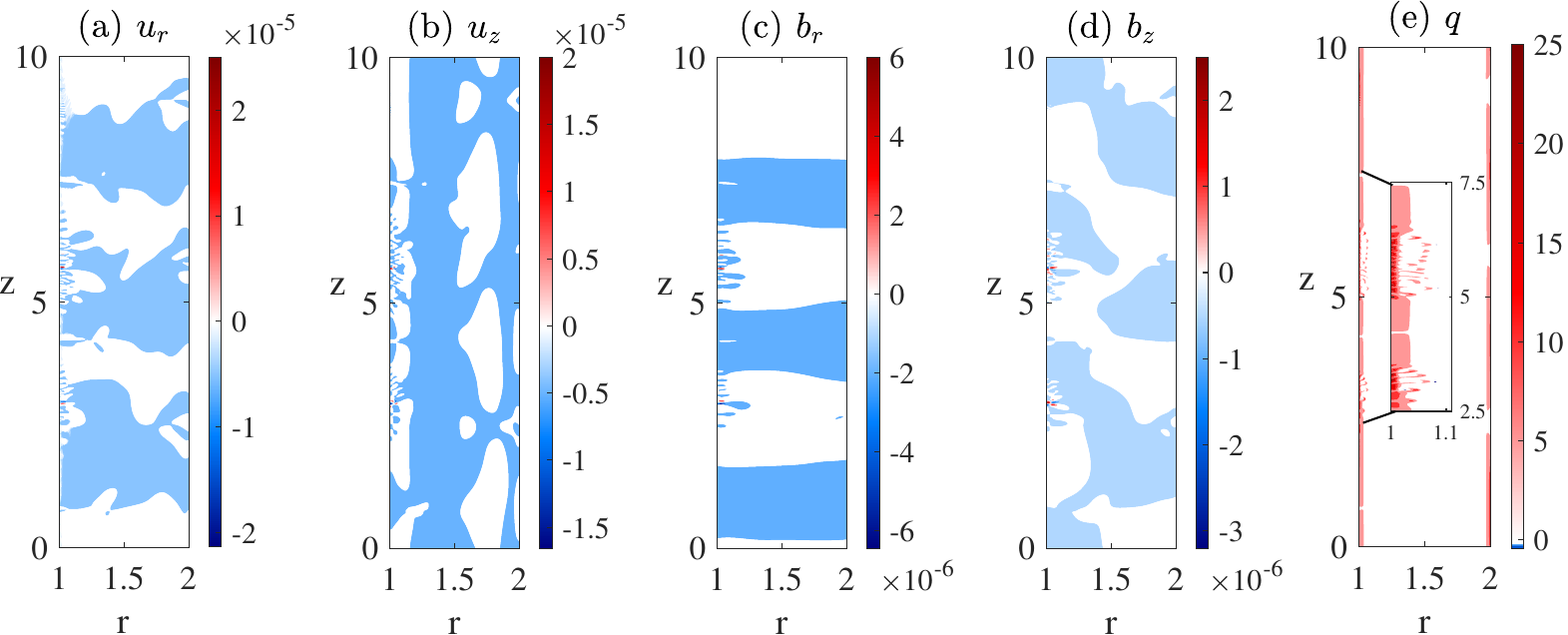}
\includegraphics[width=\textwidth]{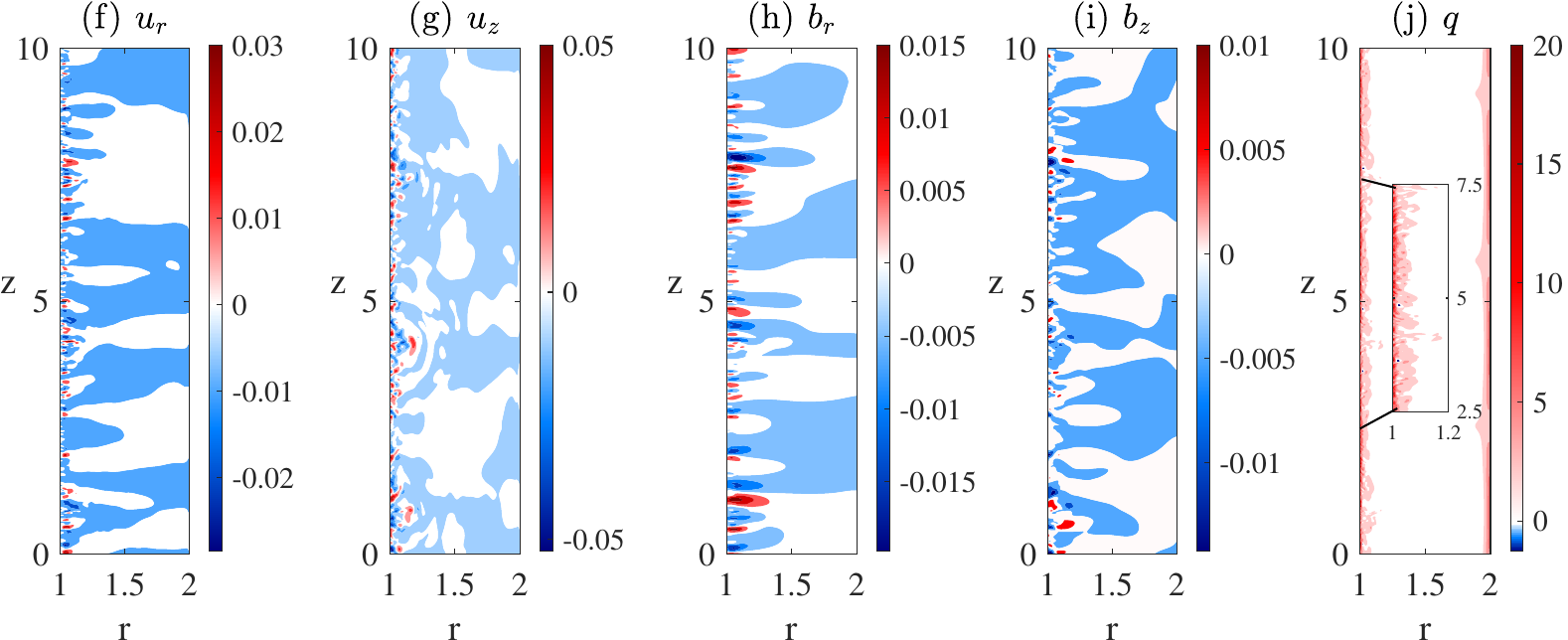}
\includegraphics[width=\textwidth]{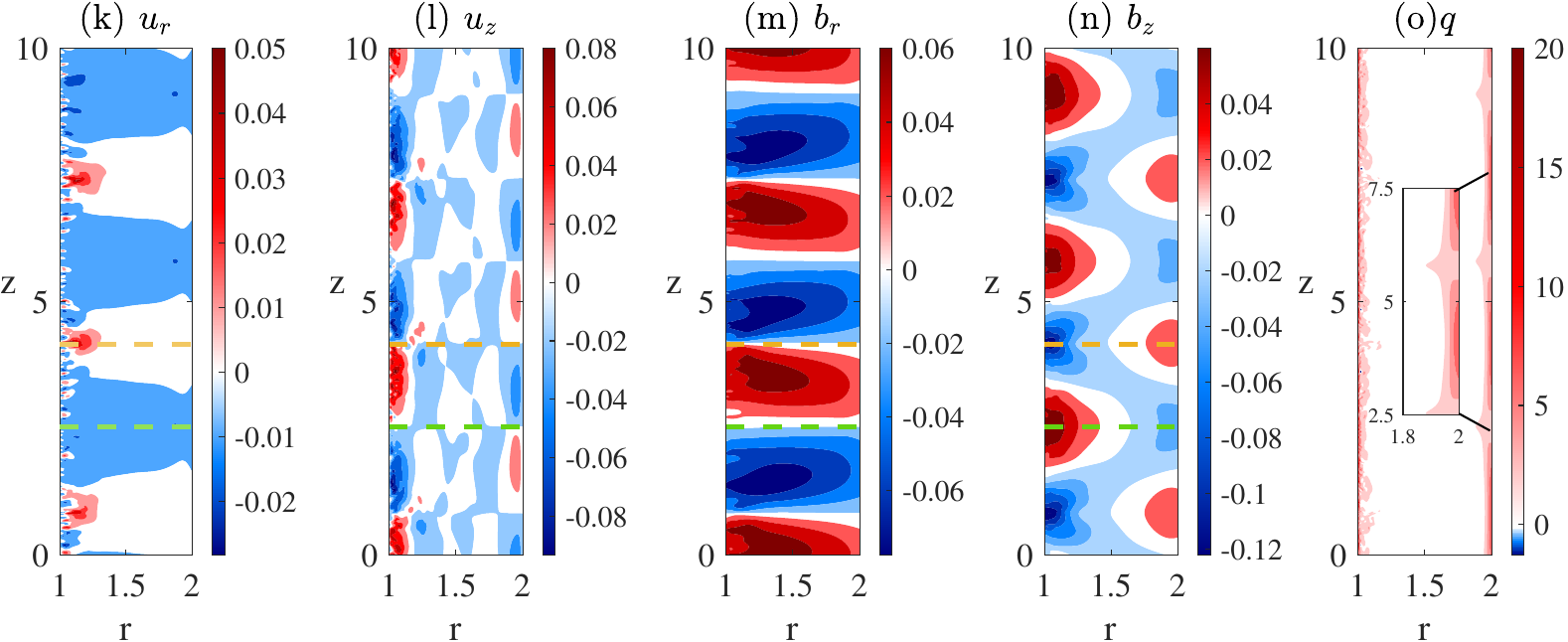}
\caption{Radial and axial velocity, ($u_r, \, u_z$), and magnetic field, ($b_r, \, b_z$), structures in the ($r,z$)-plane for $Lu=6.78$, $Rm=35$ and $Re=4\times10^4$ at different evolution times of the axisymmetric and non-axisymmetric modes corresponding to the time moments F, G and H in Fig. \ref{fig:Mag_energyRm35Lu678_m20}(c). Panels (a)-(d) and (f)-(i) show \textit{only} non-axisymmetric modes at moments F and G, respectively. (k)-(n) show all the modes in the saturated state at moment H. Rightmost panels (e), (j) and (o) show the distribution of the local shear parameter $q=-\partial {\rm ln} \, \Omega/\partial {\rm ln} \, r$ in the ($r,z$)-plane.
Orange and green dashed lines are plotted at $z=4.16$ and $z=2.52$ and analysed in Figs. \ref{fig:r_theta_plane_z600} and \ref{fig:r_theta_plane_z365}, respectively.}
\label{fig:r_z_slices}
\end{figure*}

To investigate the radial structure of the non-axisymmetric modes with higher $|m|\geq 1$ in the saturated state, using Eq. (\ref{b_i_(r,m,k_z)}) we compute $b^2(r, m)=\sum_{k_z} (|\bar{b}_r(r,m,k_z)|^2+|\bar{b}_{\phi}(r,m,k_z)|^2+|\bar{b}_z(r,m,k_z)|^2)$, which in fact gives the $z$-integrated magnetic energy as a function of radius $r$ and $m$. Figure \ref{fig:Mode_energy_vs_r} shows $b^2(r, m)$ in the saturated state at high $Re=6\times 10^4$ and $4\times 10^4$, respectively, for points B and C from Fig. \ref{fig:compare_m0_m1_mu}.  
It is seen that in both cases, in the saturated state, the non-axisymmetric modes are generally concentrated near the inner cylinder, being more attached to the wall the higher $m$ is. This indicates an important role of boundaries in the dynamics of these modes. 

In Paper II, we demonstrated that the saturation of the axisymmetric SMRI mode occurs via magnetic reconnection and results in the modification of the mean azimuthal flow profile from a standard TC profile. This modified flow profile involves steep velocity gradients at the cylinder walls, which lead to the development of a specific boundary layer. At high $Re$, these boundary layers get thinner and the radial gradient of azimuthal velocity -- the local shear parameter $q=- \partial ~{\rm ln}\,\Omega/\partial~{\rm ln}\,r$ therein steeper [see Fig. 8(c) of Paper II]. As a result, the boundary layers become Rayleigh-unstable with high-$q$ values (much more than the Rayleigh-stability limit $q=2$) and eventually break down into turbulence, which develops first near the inner cylinder [see the distribution of $q$ in Figs. \ref{fig:r_z_slices}(e), \ref{fig:r_z_slices}(j) and \ref{fig:r_z_slices}(o) below]. Thus, different $m$-modes, whose total energy evolution at various $Re$ has been analyzed above in Fig. \ref{fig:Mag_energyRm35Lu678_m20} and then separately for each $m$ in Fig. \ref{fig:Mode_energyRm35Lu678_m20}, emerge and grow primarily near the inner cylinder wall. In other words, the boundary layers formed as a result of the saturation of the axisymmetric SMRI mode are prone to magnetically-modified hydrodynamic instability of (small-scale) non-axisymmetric modes, which render this layer turbulent. The non-axisymmetric modes are mainly localized in this layer and maintained via energy extraction from high shear therein as well as mutual nonlinear interactions. This also implies that the saturation mechanisms of axisymmetric and non-axisymmetric modes are different: the axisymmetric mode saturates via magnetic reconnnection, as analyzed in detail in Paper II, and more or less keeps its overall laminar structure except for boundary layers (see also Fig. \ref{fig:r_z_slices}), whereas the non-axisymmetric modes are mostly confined near walls, saturating in boundary layer turbulence.

\subsection{Structure of the nonlinear states} \label{subsection_rz_and_meridional slices}

To better illustrate the evolution of non-axisymmetric modes in the boundary layers described in the above subsection, in Fig. \ref{fig:r_z_slices} we show the radial and axial velocity, ($u_r, \, u_z$), and magnetic field, ($b_r, \, b_z$), structures for non-axisymmetric $|m|\geq 1$ and for all $m$ modes in the ($r,z$)-plane for $Lu=6.78$, $Rm=35$ and $Re=4\times10^4$ at different evolutionary moments marked with points F, G and H in Fig. \ref{fig:Mag_energyRm35Lu678_m20}(c) \footnote{Point E in Fig. \ref{fig:Mag_energyRm35Lu678_m20}(c) marks only the exponential growth phase of non-axisymmetric modes in the linear regime when their overall $(r,z)$-structure is nearly similar to that obtained from the linear stability analysis of the dominant $|m|=1$ mode (Fig. \ref{fig:eigen_linearcode}), so we make this comparison in Appendix \ref{appendix_r_z_structures_growth_phase}}.

Figures \ref{fig:r_z_slices}(a)-\ref{fig:r_z_slices}(d) show the structures composed of all non-axisymmetric $|m|\geq 1$ modes at the beginning of their rapid growth phase [moment F in Fig. \ref{fig:Mag_energyRm35Lu678_m20}(c)]. It can be clearly seen that, in stark contrast to the eigenfunction structure (Fig. \ref{fig:eigen_linearcode}), at this time, the strong non-axisymmetric perturbations of velocity and magnetic field gradually emerge near the inner cylinder boundary. A closer look at the local shear parameter $q$ in Fig. \ref{fig:r_z_slices}(e) shows the disruption of the laminar boundary layers due to localized non-axisymmetric perturbations -- turbulent spots, indicating that these boundary layers are gradually becoming turbulent [see a zoomed segment in Fig. \ref{fig:r_z_slices}(e)]. Figures \ref{fig:r_z_slices}(f)-\ref{fig:r_z_slices}(i) show the velocity and magnetic field structures for these non-axisymmetric modes at later times in the saturated state [moment G in Fig. \ref{fig:Mag_energyRm35Lu678_m20}(c)] when they are well developed near the inner wall and are clearly visible. The boundary layers  are  consequently fully turbulent [Fig. \ref{fig:r_z_slices}(j)]. 

Figures \ref{fig:r_z_slices}(k)-\ref{fig:r_z_slices}(n) show the structure of velocity and magnetic fields for all $m$ modes in the saturated state [moment H in Fig. \ref{fig:Mag_energyRm35Lu678_m20}(c)]. Evidently, they are dominated by the axisymmetric $m=0$ mode and therefore are similar to those analyzed in Paper II, exhibiting the typical radial jet [red areas in Fig. \ref{fig:r_z_slices}(k)] and the corresponding thin recoonnection (current sheet) layer seen in the magnetic field map  where the radial fields with opposite directions meet resulting in high azimuthal current density there [Fig. \ref{fig:r_z_slices}(m)]. These are the main players in the saturation dynamics of the axisymmetric SMRI investigated in detail in Paper II.

\begin{figure}
\centering
\includegraphics[width=0.85\columnwidth]{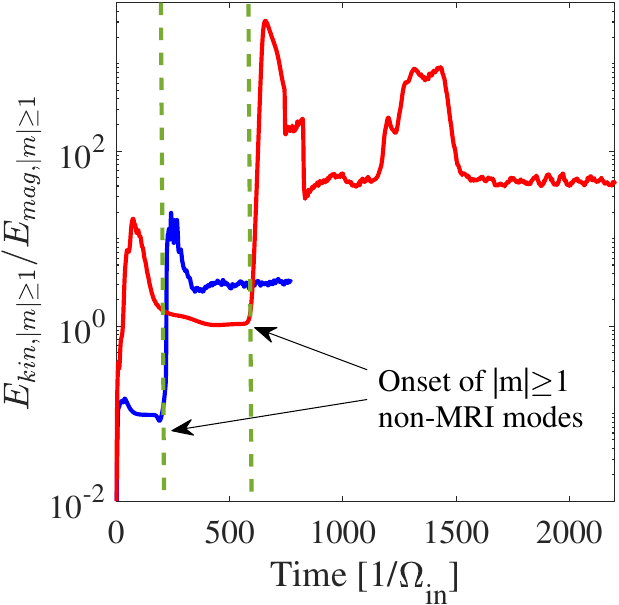}
\caption{Evolution of the ratio of the kinetic to magnetic energies of non-axisymmetric modes, $E_{kin, |m|\geq1}/E_{mag, |m|\geq1}$ for $Lu=5$, $Rm=20$, $Re=6\times 10^4$ (red) and $Lu=6.78$, $Rm=35$, $Re=4\times 10^4$ (blue) corresponding to Figs. \ref{fig:Mag_energyRm20Lu5_m20} and \ref{fig:Mag_energyRm35Lu678_m20}, respectively. Dashed green lines mark the onset moment of the non-axisymmetric nonmagnetic (non-MRI) modes.} \label{fig:KE_to_ME_ratio}
\end{figure}

From the above analysis, we conclude that these small-scale non-axisymmetric modes, which emerge and grow after the saturation of the dominant axisymmetric SMRI mode and survive in the saturated state, are \textit{not} SMRI modes. Instead, these are the non-axisymmetric modes excited in the Rayleigh-unstable (turbulent) boundary layers, which, as mentioned above, arise as a result of the saturation process of the axisymmetric SMRI mode. The passive magnetic field perturbations in these modes arise as a result of velocity perturbations advecting (dragging) the imposed axial magnetic field. Figure \ref{fig:KE_to_ME_ratio} gives an additional confirmation of the nonmagnetic nature of these modes. It shows the evolution of the ratio of the total kinetic to magnetic energies of the non-axisymmetric modes, $E_{kin, |m|\geq 1}/E_{mag, |m|\geq1}$, for the parameters used in Figs. \ref{fig:Mag_energyRm20Lu5_m20} and \ref{fig:Mag_energyRm35Lu678_m20}, respectively.  In the initial exponential growth phase, the non-axisymmetric modes are SMRI modes, so this ratio is almost constant, as is typical of the normal mode, and the kinetic energy is smaller or comparable to the magnetic energy. However, once the axisymmetric SMRI mode starts to saturate and modify the mean azimuthal flow profile, the kinetic energy shoots up orders of magnitude higher over the magnetic energy (indicated with green dashed lines), signalling the onset of nonmagnetic, or as we refer to them non-MRI, non-axisymmetric modes in the flow. During the subsequent evolution their kinetic energy remains always larger than the magnetic energy by a factor which is higher at higher $Re$. Since these modes are nonmagnetic in nature, they can, in principle, appear subcritically for large enough $Re$ in the presence of original TC flow at those $Lu$ and $Rm$ which are linearly stable against non-axisymmetric SMRI but unstable for axisymmetric SMRI (Fig. \ref{fig:Mag_energyRm20Lu5_m20}).

\begin{figure}
\centering
\includegraphics[width=\columnwidth]{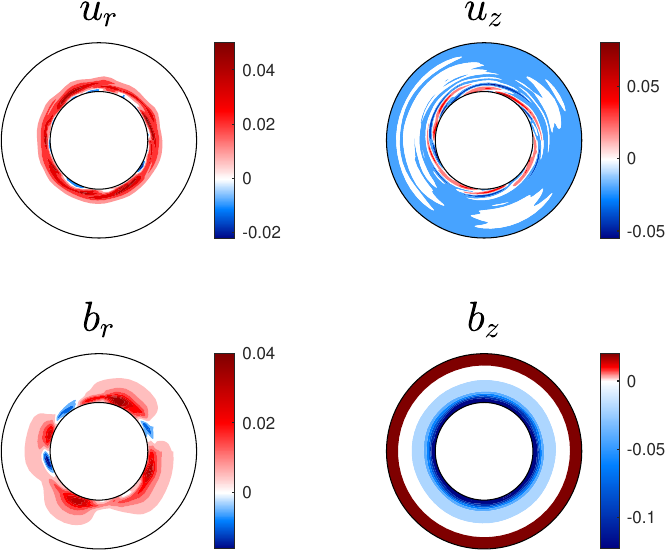}
\caption{Radial and axial velocity ($u_r, \, u_z$) and magnetic field ($b_r, \, b_z$) structures in the $(r, \phi)$-plane in the saturated state at $z=4.16$ shown by orange dashed line in the bottom row of Fig. \ref{fig:r_z_slices} for $Lu=6.78$, $Rm=35$ and $Re=4\times10^4$. }\label{fig:r_theta_plane_z600}
\end{figure}

\begin{figure}
    \centering
    \includegraphics[width=\columnwidth]{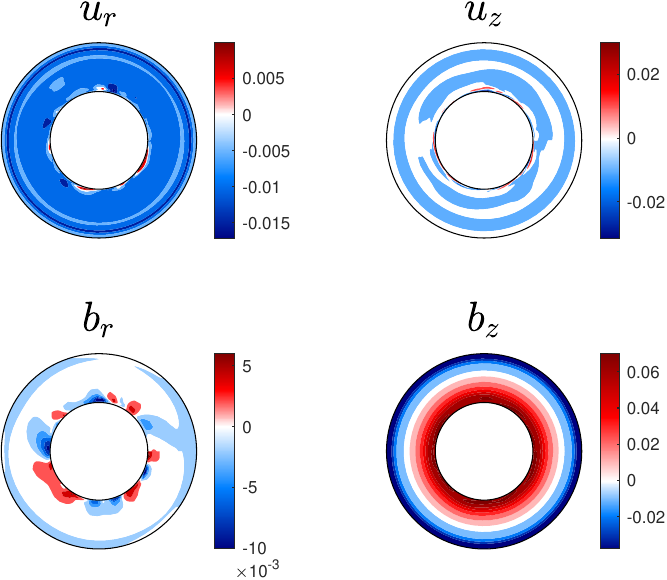}
    \caption{Same as Fig. \ref{fig:r_theta_plane_z600} but at $z=2.52$ shown by the green dashed line in the bottom row of Fig. \ref{fig:r_z_slices}}
    \label{fig:r_theta_plane_z365}
\end{figure}

In Fig. \ref{fig:r_theta_plane_z600}, we plot the perturbed total (including all $m$) radial and axial velocity ($u_r, \, u_z$) and magnetic field ($b_r, \, b_z$) in the $(r, \phi)$-plane at $z=4.16$ marked by orange dashed line in Fig. \ref{fig:r_z_slices} for the same parameters. The axial coordinate $z=4.16$ is chosen such that the extent of velocity and  magnetic field at the site of radial jet and magnetic reconnection can be analysed qualitatively in the meridional plane. The radial velocity $u_r$ is strongly concentrated along the inner cylinder. Since it is positive at the inner boundary, the jet travels from the inner cylinder towards outer one with an effective penetration radius $r \approx 1.3$ and is negligible elsewhere. The radial magnetic field $b_r$ is concentrated predominantly near the inner cylinder and is nearly zero elsewhere. However, the axial velocity $u_z$ is sheared (or spiralled) axially along inner cylinder with up-down motion very close to inner cylinder while bulk of the flow is either stationary or moving slowly downward. The perturbed axial magnetic field $b_z$ is mostly concentrated along the inner and outer cylinders and directed oppositely, while in the bulk of the flow it is nearly zero.

Figure \ref{fig:r_theta_plane_z365} shows the radial and axial velocity and magnetic field distribution in the $(r, \phi)$ plane for the same parameters as in Fig.  \ref{fig:r_theta_plane_z600} but at $z=2.52$ marked as green dashed line in Fig. \ref{fig:r_z_slices}, which is chosen such that the effect of the radial velocity jet (or magnetic reconnection) is minimal and is representative of the bulk flow between successive jets. In this case, the large-scale radial velocity, corresponding to the motion of the bulk flow from outer to inner cylinder, dominates over that of the small-scale non-axisymmetric modes very close to the inner cylinder. A similar trend is seen for the axial velocity. Perturbed radial and axial magnetic field are mostly concentrated near the inner cylinder with very small perturbations of magnetic field in the bulk flow. 

\begin{figure*}
\centering
\includegraphics[width=0.31\textwidth]{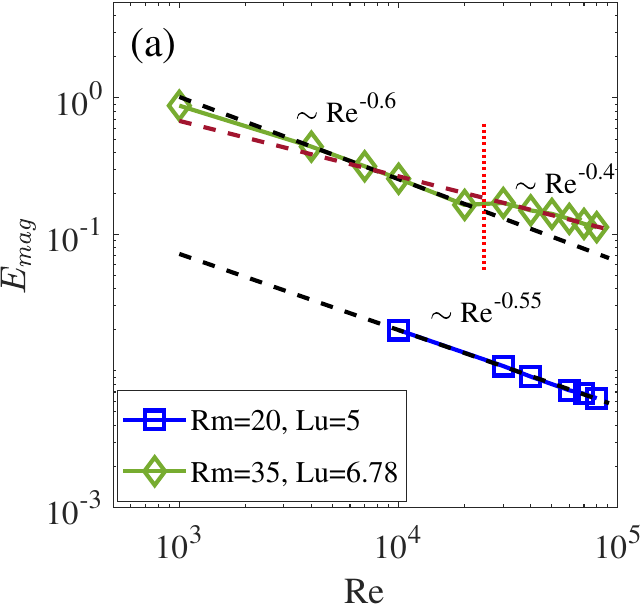}
\hspace{1em}
\includegraphics[width=0.30\textwidth]{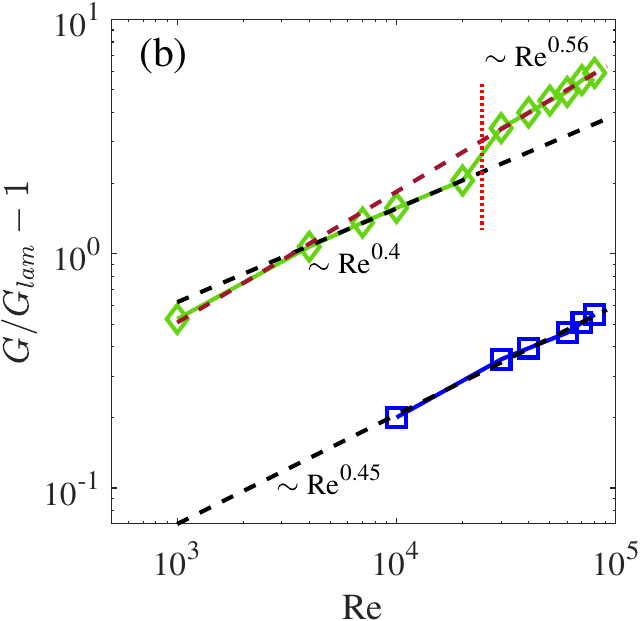}
\hspace{1em}
\includegraphics[width=0.31\textwidth]{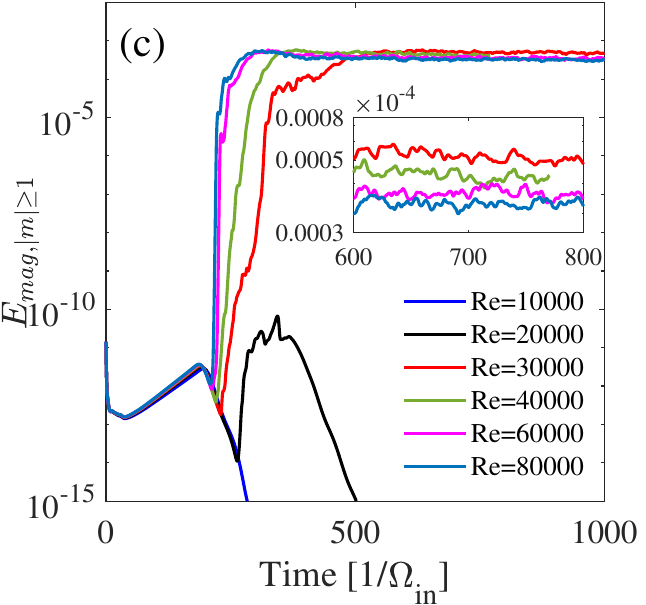}
\caption{(a) Total magnetic energy of all $m$- and $k_z$-modes, $E_{mag}=\sum_m\mathcal{E}_{mag}$, and (b) torque $G/G_{lam}-1$ in the saturated state as a function of $Re$ at point A with $Lu=5$, $Rm=20$ (blue lines) and point C with $Lu=6.78$, $Rm=35$ (green lines). Dashed lines are the power-law fits. Vertical red dotted line indicates the change in the scaling behavior of these quantities. (c) Evolution of the magnetic energy of all the non-axisymmetric $|m|\geq 1$ modes, $E_{mag,|m|\geq 1}$, for point C and different $Re=\{1, 2, 3, 4, 6, 8\}\times 10^4$.}
    \label{fig:scaling}
\end{figure*}

\subsection{Scalings of the magnetic energy and torque with $Re$}

The emergence of non-axisymmetric non-MRI modes during the saturation process of the axisymmetric SMRI mode naturally gives rise to the question whether the $Re$-scalings of saturated state magnetic energy and torque reported in Paper II still hold in the general 3D case including non-axisymmetric modes. Of particular interest is the influence of non-axisymmetric modes on the empirical scaling relation between the saturated magnetic energy and normalized perturbation torque, $E_{mag}^{-1}(G/G_{lam}-1) \sim  Re$, where as defined above $E_{mag}=\sum_{m}\mathcal{E}_{mag}$ is the total (i.e., summed over all $m$, $k_z$ and integrated in $r$) magnetic energy of perturbations (which is equivalent to the volume-integrated magnetic energy used in Paper II) and $G$ is the torque at the cylinders, which in the quasi-steady state is on average the same at the inner and outer cylinder walls and is given by \cite{Mamatsashvili_etal2018, Mishra_etal_nonlinear_2022}
\begin{equation}\label{torque_at_cylinders}
G= -\frac{r_{in, out}^3}{Re}\int_0^{2\pi}\int_0^{L_z} \frac{d}{dr}\Big(\frac{u_\phi}{r} \Big)|_{r=r_{in, out}} d\phi dz.
\end{equation}
For the basic TC flow (\ref{TC_flow}), the torque in the laminar state $G_{lam}=-(2\pi L_z/Re)r_{in}^3d\Omega/dr|_{r=r_{in}}$ and is used here to normalize the total viscous torque, $G/G_{lam}$, in order to characterize the effective angular momentum transport in the nonlinear state.

To verify the scaling relations in the presence of non-axisymmetric modes, in Fig. \ref{fig:scaling}(a) we plot $E_{mag}$ as a function of $Re$ for $Lu=5$, $Rm=20$ (point A) and $Lu=6.78$, $Rm=35$ (point C) in the saturated state. For $Re \lesssim 2\times10^4$, the scaling of the magnetic energy follows $Re^{-0.55}$ in the case A and the scaling $Re^{-0.6}$ in the case C, which are similar to those reported for purely axisymmetric SMRI in Paper II. Interestingly, for higher $Re \gtrsim 2\times10^4$ in the case C [marked by the vertical red dotted line in Fig. \ref{fig:scaling}(a)], the scaling of $E_{mag}$ becomes less steep, $Re^{-0.4}$, which is due to the stronger non-axisymmetric modes at such high $Re$, $Lu$ and  [Figs. \ref{fig:Mag_energyRm35Lu678_m20}(c) and \ref{fig:scaling}(c)], so that their nonlinear back-reaction on the dominant axisymmetric mode somewhat modifies the scaling behavior of the latter. On the other hand, the scaling for lower $Lu$ and $Rm$ in the case A remains the same across all Reynolds numbers [blue line in Fig. \ref{fig:scaling}(a)]. This is because the saturated energy of non-axisymmetric modes is several orders smaller than that for higher $Lu$ and $Rm$ [green curve in Fig. \ref{fig:scaling}(a), see also Appendix \ref{appendix_scaling_Rm40}].

Let us now look at the behavior of the total torque with Reynolds number. Figure \ref{fig:scaling}(b) shows the normalized torque due to perturbations or simply torque $G/G_{lam}-1$ (Paper II) for the same parameters. At $Re \lesssim 2\times10^4$ in the case A, the scaling of the torque follows $Re^{0.45}$, while in the case C it follows $Re^{0.4}$ in agreement with the scaling in Paper II. At higher $Re \gtrsim 2\times10^4$ [marked by the vertical red dotted line in Fig. \ref{fig:scaling}(b)], like for $E_{mag}$, only the scaling in the case C changes to steeper $Re^{0.56}$. Nevertheless, we emphasize that in both cases the scaling relation $E_{mag}^{-1}(G/G_{lam}-1) \sim  Re$ holds. This underscores the robustness of this relation and is consistent with the conclusions drawn from the detailed analysis of axisymmetric simulations in Paper II. In the present simulations, we observed that having a high $Re$ alone is not sufficient to modify the above-obtained scaling laws due to the non-axisymmetric instabilities in the flow, but $Lu$ and $Rm$ should also be high enough, as seen in the scalings of $E_{mag}$ and $G/G_{lam}-1$ in Figs. \ref{fig:scaling}(a) and \ref{fig:scaling}(b). 

We demonstrated in Paper II that the saturation of axisymmetric SMRI and corresponding scalings of the magnetic energy and torque are determined by the interplay between the magnetic reconnection and boundary layer dynamics, where the former is characterized by $Lu$, $Rm$ and the latter by $Re$. As we have seen above the scaling exponents derived in that study assuming stable (non-turbulent) boundary layer well carry over in the 3D case at lower $Re \lesssim 2\times10^4$, where non-axisymmetric modes decay. On the other hand, at higher $Re \gtrsim 2\times10^4$ small-scale non-axisymmetric, non-MRI modes develop in the inner boundary layer and render it turbulent, leading to different $Re$-scaling properties than that in the laminar case. This in turn affects the scaling behavior of the saturated SMRI state in the 3D case, causing a slight deviation of the scaling exponents at high $Re$ from those of the axisymmetric SMRI obtained in Paper II, as observed above. However, it is seen from Fig. \ref{fig:scaling} that this effect due to the non-axisymetric modes appears to depend not only on $Re$, but also on $Lu$ and $Rm$, being more appreciable at higher values of these numbers.   

It is now interesting to examine whether the non-axisymmetric modes themselves obey any specific scaling behavior. Figure \ref{fig:scaling}(c) shows the evolution of the magnetic energy of all the non-axisymmetric $|m|\geq 1$ modes for point C with $Lu=6.78$, $Rm=35$ and $Re= \{1, 2, 3, 4, 6, 8\}\times 10^4$. Although for larger $Re$ an early onset, rapid growth and saturation of the non-axisymmetric non-MRI modes happens, the saturation levels for all these modes for different $Re$ are actually quite close to each other. However, on taking a closer look at the saturated state [inset in Fig. \ref{fig:scaling}(c)], it is seen that the magnetic energy does decrease as $Re$ increases, indicating a certain systematic dependence on $Re$. In this example, this dependence is approximately $Re^{-0.45}$, which closely resembles the scaling of the magnetic energy for the axisymmetric MRI mode, suggesting a possible influence of the latter mode on the former. 

In conclusion, the scaling laws shown in Fig. \ref{fig:scaling} for typical values of $Lu$ and $Rm$ in the DRESDYN-MRI experiments (Table \ref{Table1:non_dimensional}) certainly strengthen our finding that the bulk of the flow is dominated by the axisymmetric SMRI mode and validate the robustness of the scalings of the energy and torque with respect to $Re$ and the relationship  $E_{mag}^{-1}(G/G_{lam}-1) \sim  Re$. Higher $Rm \geq 35$, at which deviation of the scaling laws has been observed at $Re \gtrsim 2\times10^4$, still represent upper limits on $Rm$ reached when operating the DRESDYN-TC device at full capacity and hence may not be typically used in these experiments. These scaling laws are quite useful and important since they allow us to extrapolate the characteristic quantities (energies, torques) in the saturated state to much higher $Re$ (which are numerically demanding) and thus to optimize the parameters in the DRESDYN-MRI experiments. In this way, we can minimize the effects of non-axisymmetric modes for an unambiguous detection of SMRI. For this purpose, high-resolution 3D simulations are needed to: 1. quantify the amplitude of non-axisymmetric modes for a broader area of the $(Lu,Rm)$-plane and thereby 2. validate the scaling behavior of the energy and torque at lower $Rm \lesssim 35$ but higher, experimentally relevant $Re \gtrsim  10^6$, which has been obtained so far at $Re \leq 10^5$ in this paper and in Paper II. This is outside the scope of the present paper and will be addressed elsewhere.

\section{Conclusion} \label{conclusion}
	
In this sequel paper to our previous analysis of axisymmetric SMRI in a magnetized TC flow of infinitely long cylinders for the DRESDYN-MRI experiments (Papers I and II), we investigated in a similar TC flow setup the linear and nonlinear dynamics of non-axisymmetric modes. First, using linear stability analysis, we explored the parameter regime of onset of non-axisymmetric SMRI and other nonmagnetic instability modes. We showed that in the parameter regime achievable in the experiment the dominant unstable non-axisymmetric mode has an azimuthal wavenumber $|m|=1$ with the growth rate, however, much smaller than that of the axisymmetric SMRI mode, while higher $|m|\geq 2$ modes are stable. In the DRESDYN-MRI experiment, the achievable values of Lundquist and magnetic Reynolds numbers, $Lu\leq 10$ and $Rm \leq 40$,  are large enough  for the onset of both axisymmetric and non-axisymmetric SMRI.  Hence,  this study is important for determining the parameter regime over which these modes can be uniquely and unambiguously identified. We showed that for a fixed ratio of the cylinders' angular velocities, $\mu$, the critical $Lu_c$ and $Rm_c$ for the $m=0$ SMRI is several times smaller than those for the $|m|=1$ SMRI mode. This difference between the critical values of the onset of the non-axiymmetric instability will allow a definitive and unambiguous detection of SMRI in laboratory, which thus should be its axisymmetric mode first.

After linear analysis, we conducted nonlinear analysis of non-axisymmetric modes with azimuthal wavenumbers $|m|\in[0, 20]$ and found that at small $Re \lesssim 4\times 10^4$ the non-axisymmetric SMRI modes do not saturate and decay. However, at large enough $Re \gtrsim 4\times 10^4$ non-axisymmetric nonmagnetic, or so-called non-MRI modes of hydrodynamic origin are excited in the TC flow preferably near the inner cylinder wall and create the turbulent boundary there. The turbulence in this boundary layer arises due to the deviation of the mean azimuthal flow profile from classical TC profile as a result of the saturation of the axisymmetric SMRI mode. At large enough $Re$, this modified flow profile introduces steep velocity gradients at the boundaries, which becoming Rayleigh-unstable, give rise to rapid growth and saturation of the non-axisymmetric non-MRI modes preferably concentrated in the inner boundary layer. On the other hand, the unstable non-axisymmetric SMRI modes growing initially in the original classical TC flow, decay once the flow profile is modified by the saturated axisymmetric SMRI mode. This indicates that the axisymmetric SMRI mode is essentially responsible for the growth of small-scale non-axisymmetric non-MRI modes within the inner boundary layer of the given magnetized TC flow. The saturation of these modes occur at the levels few orders lower than that of the large-scale axisymmetic MRI mode, which still remains a prevalent mode in the flow. 

Finally, we also explored the scaling behavior of the total magnetic energy and torque with respect to $Re$ at several $Lu$ and $Rm$ in the present 3D study and compared with that of saturated axisymmetric SMRI. We showed that the main conclusion of Paper II regarding the scaling of these two quantities holds true in the fully 3D case too, because of the dominant role of the axisymetric SMRI mode. A slight deviation from these scalings has been observed at higher $Rm\gtrsim 35$, due to larger amplitude of non-axisymmetric non-MRI modes, but such $Rm$ are still somewhat higher than usual values reached in the DRESDYN-MRI experiments. These scalings are quite important since they allow us to extrapolate the key characteristic quantities of the saturated state (energy, torque) to numerically demanding but experimentally relevant higher $Re\gtrsim 10^6$ than those considered here in order to find the saturated values of magnetic fields and velocity perturbations expected in the upcoming DRESDYN-MRI experiment.   However,  it is necessary to further investigate the nonlinear (turbulent) regimes of SMRI in a magnetized TC flow at such large $Re$ using 3D higher resolution simulations to rigorously validate these scalings and better understand the properties of the boundary layer turbulence near the cylinder walls found in the present study.

\begin{acknowledgments}
We thank Prof.  R.  Hollerbach for providing the linear 1D code and and Anna Guseva for the nonlinear  code.  AM acknowledges useful discussions with Vivaswat Kumar and Paolo Personnettaz.  We also thank anonymous Referees for useful suggestions. This work received funding from the European Union's Horizon 2020 research and innovation program under the ERC Advanced Grant Agreement No. 787544 and from Shota Rustaveli National Science Foundation of Georgia (SRNSFG) [grant number FR-23-1277].
\end{acknowledgments}

\bibliography{ref.bib}

\begin{thebibliography}{49}%
\makeatletter
\providecommand \@ifxundefined [1]{%
 \@ifx{#1\undefined}
}%
\providecommand \@ifnum [1]{%
 \ifnum #1\expandafter \@firstoftwo
 \else \expandafter \@secondoftwo
 \fi
}%
\providecommand \@ifx [1]{%
 \ifx #1\expandafter \@firstoftwo
 \else \expandafter \@secondoftwo
 \fi
}%
\providecommand \natexlab [1]{#1}%
\providecommand \enquote  [1]{``#1''}%
\providecommand \bibnamefont  [1]{#1}%
\providecommand \bibfnamefont [1]{#1}%
\providecommand \citenamefont [1]{#1}%
\providecommand \href@noop [0]{\@secondoftwo}%
\providecommand \href [0]{\begingroup \@sanitize@url \@href}%
\providecommand \@href[1]{\@@startlink{#1}\@@href}%
\providecommand \@@href[1]{\endgroup#1\@@endlink}%
\providecommand \@sanitize@url [0]{\catcode `\\12\catcode `\$12\catcode
  `\&12\catcode `\#12\catcode `\^12\catcode `\_12\catcode `\%12\relax}%
\providecommand \@@startlink[1]{}%
\providecommand \@@endlink[0]{}%
\providecommand \url  [0]{\begingroup\@sanitize@url \@url }%
\providecommand \@url [1]{\endgroup\@href {#1}{\urlprefix }}%
\providecommand \urlprefix  [0]{URL }%
\providecommand \Eprint [0]{\href }%
\providecommand \doibase [0]{https://doi.org/}%
\providecommand \selectlanguage [0]{\@gobble}%
\providecommand \bibinfo  [0]{\@secondoftwo}%
\providecommand \bibfield  [0]{\@secondoftwo}%
\providecommand \translation [1]{[#1]}%
\providecommand \BibitemOpen [0]{}%
\providecommand \bibitemStop [0]{}%
\providecommand \bibitemNoStop [0]{.\EOS\space}%
\providecommand \EOS [0]{\spacefactor3000\relax}%
\providecommand \BibitemShut  [1]{\csname bibitem#1\endcsname}%
\let\auto@bib@innerbib\@empty
\bibitem [{\citenamefont {{Velikhov}}(1959)}]{Velikhov_1959}%
  \BibitemOpen
  \bibfield  {author} {\bibinfo {author} {\bibfnamefont {E.}~\bibnamefont
  {{Velikhov}}},\ }\bibfield  {title} {\bibinfo {title} {{Stability of an
  ideally conducting liquid flowing between rotating cylinders in a magnetic
  field}},\ }\href@noop {} {\bibfield  {journal} {\bibinfo  {journal} {Zh.
  Eksp. Teor. Fiz.}\ }\textbf {\bibinfo {volume} {36}},\ \bibinfo {pages}
  {1398} (\bibinfo {year} {1959})}\BibitemShut {NoStop}%
\bibitem [{\citenamefont {{Balbus}}\ and\ \citenamefont
  {{Hawley}}(1991)}]{Balbus_Hawley_1991}%
  \BibitemOpen
  \bibfield  {author} {\bibinfo {author} {\bibfnamefont {S.~A.}\ \bibnamefont
  {{Balbus}}}\ and\ \bibinfo {author} {\bibfnamefont {J.~F.}\ \bibnamefont
  {{Hawley}}},\ }\bibfield  {title} {\bibinfo {title} {{A Powerful Local Shear
  Instability in Weakly Magnetized Disks. I. Linear Analysis}},\ }\href
  {https://doi.org/10.1086/170270} {\bibfield  {journal} {\bibinfo  {journal}
  {Astrophys. J.}\ }\textbf {\bibinfo {volume} {376}},\ \bibinfo {pages} {214}
  (\bibinfo {year} {1991})}\BibitemShut {NoStop}%
\bibitem [{\citenamefont {{Sisan}}\ \emph {et~al.}(2004)\citenamefont
  {{Sisan}}, \citenamefont {{Mujica}}, \citenamefont {{Tillotson}},
  \citenamefont {{Huang}}, \citenamefont {{Dorland}}, \citenamefont {{Hassam}},
  \citenamefont {{Antonsen}},\ and\ \citenamefont
  {{Lathrop}}}]{Sisan_etal2004}%
  \BibitemOpen
  \bibfield  {author} {\bibinfo {author} {\bibfnamefont {D.~R.}\ \bibnamefont
  {{Sisan}}}, \bibinfo {author} {\bibfnamefont {N.}~\bibnamefont {{Mujica}}},
  \bibinfo {author} {\bibfnamefont {W.~A.}\ \bibnamefont {{Tillotson}}},
  \bibinfo {author} {\bibfnamefont {Y.-M.}\ \bibnamefont {{Huang}}}, \bibinfo
  {author} {\bibfnamefont {W.}~\bibnamefont {{Dorland}}}, \bibinfo {author}
  {\bibfnamefont {A.~B.}\ \bibnamefont {{Hassam}}}, \bibinfo {author}
  {\bibfnamefont {T.~M.}\ \bibnamefont {{Antonsen}}},\ and\ \bibinfo {author}
  {\bibfnamefont {D.~P.}\ \bibnamefont {{Lathrop}}},\ }\bibfield  {title}
  {\bibinfo {title} {{Experimental Observation and Characterization of the
  Magnetorotational Instability}},\ }\href
  {https://doi.org/10.1103/PhysRevLett.93.114502} {\bibfield  {journal}
  {\bibinfo  {journal} {Phys. Rev. Lett.}\ }\textbf {\bibinfo {volume} {93}},\
  \bibinfo {eid} {114502} (\bibinfo {year} {2004})}\BibitemShut {NoStop}%
\bibitem [{\citenamefont {Nornberg}\ \emph {et~al.}(2010)\citenamefont
  {Nornberg}, \citenamefont {Ji}, \citenamefont {Schartman}, \citenamefont
  {Roach},\ and\ \citenamefont {Goodman}}]{Nornberg_Ji_etal_2010_PhysRevLett}%
  \BibitemOpen
  \bibfield  {author} {\bibinfo {author} {\bibfnamefont {M.~D.}\ \bibnamefont
  {Nornberg}}, \bibinfo {author} {\bibfnamefont {H.}~\bibnamefont {Ji}},
  \bibinfo {author} {\bibfnamefont {E.}~\bibnamefont {Schartman}}, \bibinfo
  {author} {\bibfnamefont {A.}~\bibnamefont {Roach}},\ and\ \bibinfo {author}
  {\bibfnamefont {J.}~\bibnamefont {Goodman}},\ }\bibfield  {title} {\bibinfo
  {title} {Observation of magnetocoriolis waves in a liquid metal
  taylor-couette experiment},\ }\href
  {https://doi.org/10.1103/PhysRevLett.104.074501} {\bibfield  {journal}
  {\bibinfo  {journal} {Phys. Rev. Lett.}\ }\textbf {\bibinfo {volume} {104}},\
  \bibinfo {pages} {074501} (\bibinfo {year} {2010})}\BibitemShut {NoStop}%
\bibitem [{\citenamefont {{Roach}}\ \emph {et~al.}(2012)\citenamefont
  {{Roach}}, \citenamefont {{Spence}}, \citenamefont {{Gissinger}},
  \citenamefont {{Edlund}}, \citenamefont {{Sloboda}}, \citenamefont
  {{Goodman}},\ and\ \citenamefont {{Ji}}}]{Roach_etal2012PhRvL}%
  \BibitemOpen
  \bibfield  {author} {\bibinfo {author} {\bibfnamefont {A.~H.}\ \bibnamefont
  {{Roach}}}, \bibinfo {author} {\bibfnamefont {E.~J.}\ \bibnamefont
  {{Spence}}}, \bibinfo {author} {\bibfnamefont {C.}~\bibnamefont
  {{Gissinger}}}, \bibinfo {author} {\bibfnamefont {E.~M.}\ \bibnamefont
  {{Edlund}}}, \bibinfo {author} {\bibfnamefont {P.}~\bibnamefont {{Sloboda}}},
  \bibinfo {author} {\bibfnamefont {J.}~\bibnamefont {{Goodman}}},\ and\
  \bibinfo {author} {\bibfnamefont {H.}~\bibnamefont {{Ji}}},\ }\bibfield
  {title} {\bibinfo {title} {{Observation of a Free-Shercliff-Layer Instability
  in Cylindrical Geometry}},\ }\href
  {https://doi.org/10.1103/PhysRevLett.108.154502} {\bibfield  {journal}
  {\bibinfo  {journal} {Phys. Rev. Lett.}\ }\textbf {\bibinfo {volume} {108}},\
  \bibinfo {eid} {154502} (\bibinfo {year} {2012})}\BibitemShut {NoStop}%
\bibitem [{\citenamefont {{Hung}}\ \emph {et~al.}(2019)\citenamefont {{Hung}},
  \citenamefont {{Blackman}}, \citenamefont {{Caspary}}, \citenamefont
  {{Gilson}},\ and\ \citenamefont {{Ji}}}]{Hung_etal2019CmPhy}%
  \BibitemOpen
  \bibfield  {author} {\bibinfo {author} {\bibfnamefont {D.~M.~H.}\
  \bibnamefont {{Hung}}}, \bibinfo {author} {\bibfnamefont {E.~G.}\
  \bibnamefont {{Blackman}}}, \bibinfo {author} {\bibfnamefont {K.~J.}\
  \bibnamefont {{Caspary}}}, \bibinfo {author} {\bibfnamefont {E.~P.}\
  \bibnamefont {{Gilson}}},\ and\ \bibinfo {author} {\bibfnamefont
  {H.}~\bibnamefont {{Ji}}},\ }\bibfield  {title} {\bibinfo {title}
  {{Experimental confirmation of the standard magnetorotational instability
  mechanism with a spring-mass analogue}},\ }\href
  {https://doi.org/10.1038/s42005-018-0103-7} {\bibfield  {journal} {\bibinfo
  {journal} {Communications Physics}\ }\textbf {\bibinfo {volume} {2}},\
  \bibinfo {eid} {7} (\bibinfo {year} {2019})}\BibitemShut {NoStop}%
\bibitem [{\citenamefont {{Wang}}\ \emph
  {et~al.}(2022{\natexlab{a}})\citenamefont {{Wang}}, \citenamefont {{Gilson}},
  \citenamefont {{Ebrahimi}}, \citenamefont {{Goodman}},\ and\ \citenamefont
  {{Ji}}}]{Wang_etal2022b_PRL}%
  \BibitemOpen
  \bibfield  {author} {\bibinfo {author} {\bibfnamefont {Y.}~\bibnamefont
  {{Wang}}}, \bibinfo {author} {\bibfnamefont {E.~P.}\ \bibnamefont
  {{Gilson}}}, \bibinfo {author} {\bibfnamefont {F.}~\bibnamefont
  {{Ebrahimi}}}, \bibinfo {author} {\bibfnamefont {J.}~\bibnamefont
  {{Goodman}}},\ and\ \bibinfo {author} {\bibfnamefont {H.}~\bibnamefont
  {{Ji}}},\ }\bibfield  {title} {\bibinfo {title} {{Observation of Axisymmetric
  Standard Magnetorotational Instability in the Laboratory}},\ }\href
  {https://doi.org/10.1103/PhysRevLett.129.115001} {\bibfield  {journal}
  {\bibinfo  {journal} {Phys. Rev. Lett.}\ }\textbf {\bibinfo {volume} {129}},\
  \bibinfo {eid} {115001} (\bibinfo {year} {2022}{\natexlab{a}})}\BibitemShut
  {NoStop}%
\bibitem [{\citenamefont {{R{\"u}diger}}\ \emph {et~al.}(2018)\citenamefont
  {{R{\"u}diger}}, \citenamefont {{Gellert}}, \citenamefont {{Hollerbach}},
  \citenamefont {{Schultz}},\ and\ \citenamefont
  {{Stefani}}}]{Ruediger_etal_2018_PhysRepo}%
  \BibitemOpen
  \bibfield  {author} {\bibinfo {author} {\bibfnamefont {G.}~\bibnamefont
  {{R{\"u}diger}}}, \bibinfo {author} {\bibfnamefont {M.}~\bibnamefont
  {{Gellert}}}, \bibinfo {author} {\bibfnamefont {R.}~\bibnamefont
  {{Hollerbach}}}, \bibinfo {author} {\bibfnamefont {M.}~\bibnamefont
  {{Schultz}}},\ and\ \bibinfo {author} {\bibfnamefont {F.}~\bibnamefont
  {{Stefani}}},\ }\bibfield  {title} {\bibinfo {title} {Stability and
  instability of hydromagnetic taylor-couette flows},\ }\href@noop {}
  {\bibfield  {journal} {\bibinfo  {journal} {Phys. Rep.}\ }\textbf {\bibinfo
  {volume} {741}},\ \bibinfo {pages} {1} (\bibinfo {year} {2018})}\BibitemShut
  {NoStop}%
\bibitem [{\citenamefont {{Ji}}\ and\ \citenamefont
  {{Goodman}}(2023)}]{Ji_Goodman2023}%
  \BibitemOpen
  \bibfield  {author} {\bibinfo {author} {\bibfnamefont {H.}~\bibnamefont
  {{Ji}}}\ and\ \bibinfo {author} {\bibfnamefont {J.}~\bibnamefont
  {{Goodman}}},\ }\bibfield  {title} {\bibinfo {title} {{Taylor-Couette flow
  for astrophysical purposes}},\ }\href
  {https://doi.org/10.1098/rsta.2022.0119} {\bibfield  {journal} {\bibinfo
  {journal} {Philosophical Transactions of the Royal Society of London Series
  A}\ }\textbf {\bibinfo {volume} {381}},\ \bibinfo {eid} {20220119} (\bibinfo
  {year} {2023})}\BibitemShut {NoStop}%
\bibitem [{\citenamefont {{Wang}}\ \emph
  {et~al.}(2022{\natexlab{b}})\citenamefont {{Wang}}, \citenamefont {{Gilson}},
  \citenamefont {{Ebrahimi}}, \citenamefont {{Goodman}}, \citenamefont
  {{Caspary}}, \citenamefont {{Winarto}},\ and\ \citenamefont
  {{Ji}}}]{Wang_etal2022a_NatCom}%
  \BibitemOpen
  \bibfield  {author} {\bibinfo {author} {\bibfnamefont {Y.}~\bibnamefont
  {{Wang}}}, \bibinfo {author} {\bibfnamefont {E.~P.}\ \bibnamefont
  {{Gilson}}}, \bibinfo {author} {\bibfnamefont {F.}~\bibnamefont
  {{Ebrahimi}}}, \bibinfo {author} {\bibfnamefont {J.}~\bibnamefont
  {{Goodman}}}, \bibinfo {author} {\bibfnamefont {K.~J.}\ \bibnamefont
  {{Caspary}}}, \bibinfo {author} {\bibfnamefont {H.~W.}\ \bibnamefont
  {{Winarto}}},\ and\ \bibinfo {author} {\bibfnamefont {H.}~\bibnamefont
  {{Ji}}},\ }\bibfield  {title} {\bibinfo {title} {{Identification of a
  non-axisymmetric mode in laboratory experiments searching for standard
  magnetorotational instability}},\ }\href
  {https://doi.org/10.1038/s41467-022-32278-0} {\bibfield  {journal} {\bibinfo
  {journal} {Nature Communications}\ }\textbf {\bibinfo {volume} {13}},\
  \bibinfo {eid} {4679} (\bibinfo {year} {2022}{\natexlab{b}})}\BibitemShut
  {NoStop}%
\bibitem [{\citenamefont {Rüdiger}\ and\ \citenamefont
  {Schultz}(2024)}]{Ruediger2023}%
  \BibitemOpen
  \bibfield  {author} {\bibinfo {author} {\bibfnamefont {G.}~\bibnamefont
  {Rüdiger}}\ and\ \bibinfo {author} {\bibfnamefont {M.}~\bibnamefont
  {Schultz}},\ }\bibfield  {title} {\bibinfo {title} {The gap-size influence on
  the excitation of magnetorotational instability in cylindrictaylor–couette
  flows},\ }\href {https://doi.org/10.1017/S0022377823001356} {\bibfield
  {journal} {\bibinfo  {journal} {Journal of Plasma Physics}\ }\textbf
  {\bibinfo {volume} {90}},\ \bibinfo {pages} {905900105} (\bibinfo {year}
  {2024})}\BibitemShut {NoStop}%
\bibitem [{\citenamefont {Hollerbach}\ and\ \citenamefont
  {R\"udiger}(2005)}]{Hollerbach_Rudiger_2005}%
  \BibitemOpen
  \bibfield  {author} {\bibinfo {author} {\bibfnamefont {R.}~\bibnamefont
  {Hollerbach}}\ and\ \bibinfo {author} {\bibfnamefont {G.}~\bibnamefont
  {R\"udiger}},\ }\bibfield  {title} {\bibinfo {title} {New type of
  magnetorotational instability in cylindrical {\uppercase
  {t}}aylor-{\uppercase {c}}ouette flow},\ }\href
  {https://doi.org/10.1103/PhysRevLett.95.124501} {\bibfield  {journal}
  {\bibinfo  {journal} {Phys. Rev. Lett.}\ }\textbf {\bibinfo {volume} {95}},\
  \bibinfo {pages} {124501} (\bibinfo {year} {2005})}\BibitemShut {NoStop}%
\bibitem [{\citenamefont {Hollerbach}\ \emph {et~al.}(2010)\citenamefont
  {Hollerbach}, \citenamefont {Teeluck},\ and\ \citenamefont
  {R\"udiger}}]{Hollerbach_Rudiger_2010}%
  \BibitemOpen
  \bibfield  {author} {\bibinfo {author} {\bibfnamefont {R.}~\bibnamefont
  {Hollerbach}}, \bibinfo {author} {\bibfnamefont {V.}~\bibnamefont
  {Teeluck}},\ and\ \bibinfo {author} {\bibfnamefont {G.}~\bibnamefont
  {R\"udiger}},\ }\bibfield  {title} {\bibinfo {title} {Nonaxisymmetric
  magnetorotational instabilities in cylindrical taylor-couette flow},\ }\href
  {https://doi.org/10.1103/PhysRevLett.104.044502} {\bibfield  {journal}
  {\bibinfo  {journal} {Phys. Rev. Lett.}\ }\textbf {\bibinfo {volume} {104}},\
  \bibinfo {pages} {044502} (\bibinfo {year} {2010})}\BibitemShut {NoStop}%
\bibitem [{\citenamefont {{Stefani}}\ \emph {et~al.}(2006)\citenamefont
  {{Stefani}}, \citenamefont {{Gundrum}}, \citenamefont {{Gerbeth}},
  \citenamefont {{R{\"u}diger}}, \citenamefont {{Schultz}}, \citenamefont
  {{Szklarski}},\ and\ \citenamefont
  {{Hollerbach}}}]{Stefani_Gundrum_Gerbeth_etal_2006PhRvL}%
  \BibitemOpen
  \bibfield  {author} {\bibinfo {author} {\bibfnamefont {F.}~\bibnamefont
  {{Stefani}}}, \bibinfo {author} {\bibfnamefont {T.}~\bibnamefont
  {{Gundrum}}}, \bibinfo {author} {\bibfnamefont {G.}~\bibnamefont
  {{Gerbeth}}}, \bibinfo {author} {\bibfnamefont {G.}~\bibnamefont
  {{R{\"u}diger}}}, \bibinfo {author} {\bibfnamefont {M.}~\bibnamefont
  {{Schultz}}}, \bibinfo {author} {\bibfnamefont {J.}~\bibnamefont
  {{Szklarski}}},\ and\ \bibinfo {author} {\bibfnamefont {R.}~\bibnamefont
  {{Hollerbach}}},\ }\bibfield  {title} {\bibinfo {title} {{Experimental
  Evidence for Magnetorotational Instability in a Taylor-Couette Flow under the
  Influence of a Helical Magnetic Field}},\ }\href
  {https://doi.org/10.1103/PhysRevLett.97.184502} {\bibfield  {journal}
  {\bibinfo  {journal} {Phys. Rev. Lett.}\ }\textbf {\bibinfo {volume} {97}},\
  \bibinfo {eid} {184502} (\bibinfo {year} {2006})}\BibitemShut {NoStop}%
\bibitem [{\citenamefont {Stefani}\ \emph {et~al.}(2009)\citenamefont
  {Stefani}, \citenamefont {Gerbeth}, \citenamefont {Gundrum}, \citenamefont
  {Hollerbach}, \citenamefont {Priede}, \citenamefont {R\"udiger},\ and\
  \citenamefont {Szklarski}}]{Stefani_Gerbeth_Gundrum_Etal_2009PhysRevE}%
  \BibitemOpen
  \bibfield  {author} {\bibinfo {author} {\bibfnamefont {F.}~\bibnamefont
  {Stefani}}, \bibinfo {author} {\bibfnamefont {G.}~\bibnamefont {Gerbeth}},
  \bibinfo {author} {\bibfnamefont {T.}~\bibnamefont {Gundrum}}, \bibinfo
  {author} {\bibfnamefont {R.}~\bibnamefont {Hollerbach}}, \bibinfo {author}
  {\bibfnamefont {J.~b.~a.}\ \bibnamefont {Priede}}, \bibinfo {author}
  {\bibfnamefont {G.}~\bibnamefont {R\"udiger}},\ and\ \bibinfo {author}
  {\bibfnamefont {J.}~\bibnamefont {Szklarski}},\ }\bibfield  {title} {\bibinfo
  {title} {Helical magnetorotational instability in a {\uppercase
  {t}}aylor-{\uppercase {c}}ouette flow with strongly reduced {\uppercase
  {e}}kman pumping},\ }\href {https://doi.org/10.1103/PhysRevE.80.066303}
  {\bibfield  {journal} {\bibinfo  {journal} {Phys. Rev. E}\ }\textbf {\bibinfo
  {volume} {80}},\ \bibinfo {pages} {066303} (\bibinfo {year}
  {2009})}\BibitemShut {NoStop}%
\bibitem [{\citenamefont {{Seilmayer}}\ \emph {et~al.}(2014)\citenamefont
  {{Seilmayer}}, \citenamefont {{Galindo}}, \citenamefont {{Gerbeth}},
  \citenamefont {{Gundrum}}, \citenamefont {{Stefani}}, \citenamefont
  {{Gellert}}, \citenamefont {{R{\"u}diger}}, \citenamefont {{Schultz}},\ and\
  \citenamefont {{Hollerbach}}}]{Seilmayer_etal2014}%
  \BibitemOpen
  \bibfield  {author} {\bibinfo {author} {\bibfnamefont {M.}~\bibnamefont
  {{Seilmayer}}}, \bibinfo {author} {\bibfnamefont {V.}~\bibnamefont
  {{Galindo}}}, \bibinfo {author} {\bibfnamefont {G.}~\bibnamefont
  {{Gerbeth}}}, \bibinfo {author} {\bibfnamefont {T.}~\bibnamefont
  {{Gundrum}}}, \bibinfo {author} {\bibfnamefont {F.}~\bibnamefont
  {{Stefani}}}, \bibinfo {author} {\bibfnamefont {M.}~\bibnamefont
  {{Gellert}}}, \bibinfo {author} {\bibfnamefont {G.}~\bibnamefont
  {{R{\"u}diger}}}, \bibinfo {author} {\bibfnamefont {M.}~\bibnamefont
  {{Schultz}}},\ and\ \bibinfo {author} {\bibfnamefont {R.}~\bibnamefont
  {{Hollerbach}}},\ }\bibfield  {title} {\bibinfo {title} {{Experimental
  Evidence for Nonaxisymmetric Magnetorotational Instability in a Rotating
  Liquid Metal Exposed to an Azimuthal Magnetic Field}},\ }\href
  {https://doi.org/10.1103/PhysRevLett.113.024505} {\bibfield  {journal}
  {\bibinfo  {journal} {Phys. Rev. Lett.}\ }\textbf {\bibinfo {volume} {113}},\
  \bibinfo {eid} {024505} (\bibinfo {year} {2014})}\BibitemShut {NoStop}%
\bibitem [{\citenamefont {Mishra}\ \emph {et~al.}(2021)\citenamefont {Mishra},
  \citenamefont {Mamatsashvili}, \citenamefont {Galindo},\ and\ \citenamefont
  {Stefani}}]{Mishra_etal2021}%
  \BibitemOpen
  \bibfield  {author} {\bibinfo {author} {\bibfnamefont {A.}~\bibnamefont
  {Mishra}}, \bibinfo {author} {\bibfnamefont {G.}~\bibnamefont
  {Mamatsashvili}}, \bibinfo {author} {\bibfnamefont {V.}~\bibnamefont
  {Galindo}},\ and\ \bibinfo {author} {\bibfnamefont {F.}~\bibnamefont
  {Stefani}},\ }\bibfield  {title} {\bibinfo {title} {Convective, absolute and
  global azimuthal magnetorotational instabilities},\ }\href
  {https://doi.org/10.1017/jfm.2021.548} {\bibfield  {journal} {\bibinfo
  {journal} {J. Fluid Mech.}\ }\textbf {\bibinfo {volume} {922}},\ \bibinfo
  {pages} {R4} (\bibinfo {year} {2021})}\BibitemShut {NoStop}%
\bibitem [{\citenamefont {{Stefani}}\ \emph {et~al.}(2019)\citenamefont
  {{Stefani}}, \citenamefont {{Gailitis}}, \citenamefont {{Gerbeth}},
  \citenamefont {{Giesecke}}, \citenamefont {{Gundrum}}, \citenamefont
  {{R{\"u}diger}}, \citenamefont {{Seilmayer}},\ and\ \citenamefont
  {{Vogt}}}]{Stefani_etal2019}%
  \BibitemOpen
  \bibfield  {author} {\bibinfo {author} {\bibfnamefont {F.}~\bibnamefont
  {{Stefani}}}, \bibinfo {author} {\bibfnamefont {A.}~\bibnamefont
  {{Gailitis}}}, \bibinfo {author} {\bibfnamefont {G.}~\bibnamefont
  {{Gerbeth}}}, \bibinfo {author} {\bibfnamefont {A.}~\bibnamefont
  {{Giesecke}}}, \bibinfo {author} {\bibfnamefont {T.}~\bibnamefont
  {{Gundrum}}}, \bibinfo {author} {\bibfnamefont {G.}~\bibnamefont
  {{R{\"u}diger}}}, \bibinfo {author} {\bibfnamefont {M.}~\bibnamefont
  {{Seilmayer}}},\ and\ \bibinfo {author} {\bibfnamefont {T.}~\bibnamefont
  {{Vogt}}},\ }\bibfield  {title} {\bibinfo {title} {{The DRESDYN project:
  liquid metal experiments on dynamo action and magnetorotational
  instability}},\ }\href@noop {} {\bibfield  {journal} {\bibinfo  {journal}
  {Geophys. Astrophys. Fluid Dyn.}\ }\textbf {\bibinfo {volume} {113}},\
  \bibinfo {pages} {51} (\bibinfo {year} {2019})}\BibitemShut {NoStop}%
\bibitem [{\citenamefont {{Mishra}}\ \emph {et~al.}(2022)\citenamefont
  {{Mishra}}, \citenamefont {{Mamatsashvili}},\ and\ \citenamefont
  {{Stefani}}}]{Mishra_etal_2022PhRvF}%
  \BibitemOpen
  \bibfield  {author} {\bibinfo {author} {\bibfnamefont {A.}~\bibnamefont
  {{Mishra}}}, \bibinfo {author} {\bibfnamefont {G.}~\bibnamefont
  {{Mamatsashvili}}},\ and\ \bibinfo {author} {\bibfnamefont {F.}~\bibnamefont
  {{Stefani}}},\ }\bibfield  {title} {\bibinfo {title} {{From helical to
  standard magnetorotational instability: Predictions for upcoming liquid
  sodium experiments}},\ }\href
  {https://doi.org/10.1103/PhysRevFluids.7.064802} {\bibfield  {journal}
  {\bibinfo  {journal} {Phys. Rev. Fluids}\ }\textbf {\bibinfo {volume} {7}},\
  \bibinfo {eid} {064802} (\bibinfo {year} {2022})}\BibitemShut {NoStop}%
\bibitem [{\citenamefont {{Seilmayer}}\ \emph {et~al.}(2012)\citenamefont
  {{Seilmayer}}, \citenamefont {{Stefani}}, \citenamefont {{Gundrum}},
  \citenamefont {{Weier}}, \citenamefont {{Gerbeth}}, \citenamefont
  {{Gellert}},\ and\ \citenamefont {{R{\"u}diger}}}]{Seilmayer_etal2012PhRvL}%
  \BibitemOpen
  \bibfield  {author} {\bibinfo {author} {\bibfnamefont {M.}~\bibnamefont
  {{Seilmayer}}}, \bibinfo {author} {\bibfnamefont {F.}~\bibnamefont
  {{Stefani}}}, \bibinfo {author} {\bibfnamefont {T.}~\bibnamefont
  {{Gundrum}}}, \bibinfo {author} {\bibfnamefont {T.}~\bibnamefont {{Weier}}},
  \bibinfo {author} {\bibfnamefont {G.}~\bibnamefont {{Gerbeth}}}, \bibinfo
  {author} {\bibfnamefont {M.}~\bibnamefont {{Gellert}}},\ and\ \bibinfo
  {author} {\bibfnamefont {G.}~\bibnamefont {{R{\"u}diger}}},\ }\bibfield
  {title} {\bibinfo {title} {{Experimental Evidence for a Transient Tayler
  Instability in a Cylindrical Liquid-Metal Column}},\ }\href
  {https://doi.org/10.1103/PhysRevLett.108.244501} {\bibfield  {journal}
  {\bibinfo  {journal} {Phys. Rev. Lett.}\ }\textbf {\bibinfo {volume} {108}},\
  \bibinfo {eid} {244501} (\bibinfo {year} {2012})}\BibitemShut {NoStop}%
\bibitem [{\citenamefont {Mamatsashvili}\ \emph {et~al.}(2019)\citenamefont
  {Mamatsashvili}, \citenamefont {Stefani}, \citenamefont {Hollerbach},\ and\
  \citenamefont {R\"udiger}}]{Mamatsashvili_Stefani_Hollerbach_Rudiger_2019}%
  \BibitemOpen
  \bibfield  {author} {\bibinfo {author} {\bibfnamefont {G.}~\bibnamefont
  {Mamatsashvili}}, \bibinfo {author} {\bibfnamefont {F.}~\bibnamefont
  {Stefani}}, \bibinfo {author} {\bibfnamefont {R.}~\bibnamefont
  {Hollerbach}},\ and\ \bibinfo {author} {\bibfnamefont {G.}~\bibnamefont
  {R\"udiger}},\ }\bibfield  {title} {\bibinfo {title} {Two types of
  axisymmetric helical magnetorotational instability in rotating flows with
  positive shear},\ }\href {https://doi.org/10.1103/PhysRevFluids.4.103905}
  {\bibfield  {journal} {\bibinfo  {journal} {Phys. Rev. Fluids}\ }\textbf
  {\bibinfo {volume} {4}},\ \bibinfo {pages} {103905} (\bibinfo {year}
  {2019})}\BibitemShut {NoStop}%
\bibitem [{\citenamefont {{Hawley}}\ and\ \citenamefont
  {{Balbus}}(1991)}]{Hawley_Balbus_1991ApJ_nonlinear}%
  \BibitemOpen
  \bibfield  {author} {\bibinfo {author} {\bibfnamefont {J.~F.}\ \bibnamefont
  {{Hawley}}}\ and\ \bibinfo {author} {\bibfnamefont {S.~A.}\ \bibnamefont
  {{Balbus}}},\ }\bibfield  {title} {\bibinfo {title} {{A Powerful Local Shear
  Instability in Weakly Magnetized Disks. II. Nonlinear Evolution}},\ }\href
  {https://doi.org/10.1086/170271} {\bibfield  {journal} {\bibinfo  {journal}
  {Astrophys. J.}\ }\textbf {\bibinfo {volume} {376}},\ \bibinfo {pages} {223}
  (\bibinfo {year} {1991})}\BibitemShut {NoStop}%
\bibitem [{\citenamefont {{Balbus}}\ and\ \citenamefont
  {{Hawley}}(1998)}]{Balbus_Hawley_1998}%
  \BibitemOpen
  \bibfield  {author} {\bibinfo {author} {\bibfnamefont {S.~A.}\ \bibnamefont
  {{Balbus}}}\ and\ \bibinfo {author} {\bibfnamefont {J.~F.}\ \bibnamefont
  {{Hawley}}},\ }\bibfield  {title} {\bibinfo {title} {{Instability,
  turbulence, and enhanced transport in accretion disks}},\ }\href
  {https://doi.org/10.1103/RevModPhys.70.1} {\bibfield  {journal} {\bibinfo
  {journal} {Rev. Mod. Phys.}\ }\textbf {\bibinfo {volume} {70}},\ \bibinfo
  {pages} {1} (\bibinfo {year} {1998})}\BibitemShut {NoStop}%
\bibitem [{\citenamefont {{Balbus}}(2003)}]{Balbus2003}%
  \BibitemOpen
  \bibfield  {author} {\bibinfo {author} {\bibfnamefont {S.~A.}\ \bibnamefont
  {{Balbus}}},\ }\bibfield  {title} {\bibinfo {title} {{Enhanced Angular
  Momentum Transport in Accretion Disks}},\ }\href
  {https://doi.org/10.1146/annurev.astro.41.081401.155207} {\bibfield
  {journal} {\bibinfo  {journal} {Annu. Rev. Astron. Astrophys.}\ }\textbf
  {\bibinfo {volume} {41}},\ \bibinfo {pages} {555} (\bibinfo {year}
  {2003})}\BibitemShut {NoStop}%
\bibitem [{\citenamefont {{Knobloch}}\ and\ \citenamefont
  {{Julien}}(2005)}]{Knobloch_Julien_2005PhFl}%
  \BibitemOpen
  \bibfield  {author} {\bibinfo {author} {\bibfnamefont {E.}~\bibnamefont
  {{Knobloch}}}\ and\ \bibinfo {author} {\bibfnamefont {K.}~\bibnamefont
  {{Julien}}},\ }\bibfield  {title} {\bibinfo {title} {{Saturation of the
  magnetorotational instability}},\ }\href {https://doi.org/10.1063/1.2047592}
  {\bibfield  {journal} {\bibinfo  {journal} {Phys. Fluids}\ }\textbf {\bibinfo
  {volume} {17}},\ \bibinfo {eid} {094106-094106-6} (\bibinfo {year}
  {2005})}\BibitemShut {NoStop}%
\bibitem [{\citenamefont {{Liu}}\ \emph {et~al.}(2006)\citenamefont {{Liu}},
  \citenamefont {{Goodman}},\ and\ \citenamefont
  {{Ji}}}]{Liu_Goodman_Ji_NonlinearMRI_2006ApJ}%
  \BibitemOpen
  \bibfield  {author} {\bibinfo {author} {\bibfnamefont {W.}~\bibnamefont
  {{Liu}}}, \bibinfo {author} {\bibfnamefont {J.}~\bibnamefont {{Goodman}}},\
  and\ \bibinfo {author} {\bibfnamefont {H.}~\bibnamefont {{Ji}}},\ }\bibfield
  {title} {\bibinfo {title} {{Simulations of Magnetorotational Instability in a
  Magnetized Couette Flow}},\ }\href {https://doi.org/10.1086/501495}
  {\bibfield  {journal} {\bibinfo  {journal} {Astrophys. J.}\ }\textbf
  {\bibinfo {volume} {643}},\ \bibinfo {pages} {306} (\bibinfo {year}
  {2006})}\BibitemShut {NoStop}%
\bibitem [{\citenamefont {{Gellert}}\ \emph {et~al.}(2012)\citenamefont
  {{Gellert}}, \citenamefont {{R{\"u}diger}},\ and\ \citenamefont
  {{Schultz}}}]{Gellert_etal2012}%
  \BibitemOpen
  \bibfield  {author} {\bibinfo {author} {\bibfnamefont {M.}~\bibnamefont
  {{Gellert}}}, \bibinfo {author} {\bibfnamefont {G.}~\bibnamefont
  {{R{\"u}diger}}},\ and\ \bibinfo {author} {\bibfnamefont {M.}~\bibnamefont
  {{Schultz}}},\ }\bibfield  {title} {\bibinfo {title} {{The angular momentum
  transport by standard MRI in quasi-Kepler cylindrical Taylor-Couette
  flows}},\ }\href {https://doi.org/10.1051/0004-6361/201117892} {\bibfield
  {journal} {\bibinfo  {journal} {Astron. Astrophys.}\ }\textbf {\bibinfo
  {volume} {541}},\ \bibinfo {eid} {A124} (\bibinfo {year} {2012})}\BibitemShut
  {NoStop}%
\bibitem [{\citenamefont {Gissinger}\ \emph {et~al.}(2012)\citenamefont
  {Gissinger}, \citenamefont {Goodman},\ and\ \citenamefont
  {Ji}}]{Gissinger_Goodman_Ji_2012PhFl}%
  \BibitemOpen
  \bibfield  {author} {\bibinfo {author} {\bibfnamefont {C.}~\bibnamefont
  {Gissinger}}, \bibinfo {author} {\bibfnamefont {J.}~\bibnamefont {Goodman}},\
  and\ \bibinfo {author} {\bibfnamefont {H.}~\bibnamefont {Ji}},\ }\bibfield
  {title} {\bibinfo {title} {The role of boundaries in the magnetorotational
  instability},\ }\href {https://doi.org/10.1063/1.4737657} {\bibfield
  {journal} {\bibinfo  {journal} {Phys. Fluids}\ }\textbf {\bibinfo {volume}
  {24}},\ \bibinfo {pages} {074109} (\bibinfo {year} {2012})}\BibitemShut
  {NoStop}%
\bibitem [{\citenamefont {{Wei}}\ \emph {et~al.}(2016)\citenamefont {{Wei}},
  \citenamefont {{Ji}}, \citenamefont {{Goodman}}, \citenamefont {{Ebrahimi}},
  \citenamefont {{Gilson}}, \citenamefont {{Jenko}},\ and\ \citenamefont
  {{Lackner}}}]{Wei_etal2016}%
  \BibitemOpen
  \bibfield  {author} {\bibinfo {author} {\bibfnamefont {X.}~\bibnamefont
  {{Wei}}}, \bibinfo {author} {\bibfnamefont {H.}~\bibnamefont {{Ji}}},
  \bibinfo {author} {\bibfnamefont {J.}~\bibnamefont {{Goodman}}}, \bibinfo
  {author} {\bibfnamefont {F.}~\bibnamefont {{Ebrahimi}}}, \bibinfo {author}
  {\bibfnamefont {E.}~\bibnamefont {{Gilson}}}, \bibinfo {author}
  {\bibfnamefont {F.}~\bibnamefont {{Jenko}}},\ and\ \bibinfo {author}
  {\bibfnamefont {K.}~\bibnamefont {{Lackner}}},\ }\bibfield  {title} {\bibinfo
  {title} {{Numerical simulations of the Princeton magnetorotational
  instability experiment with conducting axial boundaries}},\ }\href
  {https://doi.org/10.1103/PhysRevE.94.063107} {\bibfield  {journal} {\bibinfo
  {journal} {Phys. Rev. E}\ }\textbf {\bibinfo {volume} {94}},\ \bibinfo {eid}
  {063107} (\bibinfo {year} {2016})}\BibitemShut {NoStop}%
\bibitem [{\citenamefont {{Choi}}\ \emph {et~al.}(2019)\citenamefont {{Choi}},
  \citenamefont {{Ebrahimi}}, \citenamefont {{Caspary}}, \citenamefont
  {{Gilson}}, \citenamefont {{Goodman}},\ and\ \citenamefont
  {{Ji}}}]{Choi_etal2019}%
  \BibitemOpen
  \bibfield  {author} {\bibinfo {author} {\bibfnamefont {D.}~\bibnamefont
  {{Choi}}}, \bibinfo {author} {\bibfnamefont {F.}~\bibnamefont {{Ebrahimi}}},
  \bibinfo {author} {\bibfnamefont {K.~J.}\ \bibnamefont {{Caspary}}}, \bibinfo
  {author} {\bibfnamefont {E.~P.}\ \bibnamefont {{Gilson}}}, \bibinfo {author}
  {\bibfnamefont {J.}~\bibnamefont {{Goodman}}},\ and\ \bibinfo {author}
  {\bibfnamefont {H.}~\bibnamefont {{Ji}}},\ }\bibfield  {title} {\bibinfo
  {title} {{Nonaxisymmetric simulations of the Princeton magnetorotational
  instability experiment with insulating and conducting axial boundaries}},\
  }\href {https://doi.org/10.1103/PhysRevE.100.033116} {\bibfield  {journal}
  {\bibinfo  {journal} {Phys. Rev. E}\ }\textbf {\bibinfo {volume} {100}},\
  \bibinfo {eid} {033116} (\bibinfo {year} {2019})}\BibitemShut {NoStop}%
\bibitem [{\citenamefont {{Winarto}}\ \emph {et~al.}(2020)\citenamefont
  {{Winarto}}, \citenamefont {{Ji}}, \citenamefont {{Goodman}}, \citenamefont
  {{Ebrahimi}}, \citenamefont {{Gilson}},\ and\ \citenamefont
  {{Wang}}}]{Winarto_etal2020}%
  \BibitemOpen
  \bibfield  {author} {\bibinfo {author} {\bibfnamefont {H.~W.}\ \bibnamefont
  {{Winarto}}}, \bibinfo {author} {\bibfnamefont {H.}~\bibnamefont {{Ji}}},
  \bibinfo {author} {\bibfnamefont {J.}~\bibnamefont {{Goodman}}}, \bibinfo
  {author} {\bibfnamefont {F.}~\bibnamefont {{Ebrahimi}}}, \bibinfo {author}
  {\bibfnamefont {E.~P.}\ \bibnamefont {{Gilson}}},\ and\ \bibinfo {author}
  {\bibfnamefont {Y.}~\bibnamefont {{Wang}}},\ }\bibfield  {title} {\bibinfo
  {title} {{Parameter space mapping of the Princeton magnetorotational
  instability experiment}},\ }\href
  {https://doi.org/10.1103/PhysRevE.102.023113} {\bibfield  {journal} {\bibinfo
   {journal} {Phys. Rev. E}\ }\textbf {\bibinfo {volume} {102}},\ \bibinfo
  {eid} {023113} (\bibinfo {year} {2020})}\BibitemShut {NoStop}%
\bibitem [{\citenamefont {{Umurhan}}\ \emph
  {et~al.}(2007{\natexlab{a}})\citenamefont {{Umurhan}}, \citenamefont
  {{Menou}},\ and\ \citenamefont {{Regev}}}]{Umurhan_Menou_Regev_2007PhRvL}%
  \BibitemOpen
  \bibfield  {author} {\bibinfo {author} {\bibfnamefont {O.~M.}\ \bibnamefont
  {{Umurhan}}}, \bibinfo {author} {\bibfnamefont {K.}~\bibnamefont {{Menou}}},\
  and\ \bibinfo {author} {\bibfnamefont {O.}~\bibnamefont {{Regev}}},\
  }\bibfield  {title} {\bibinfo {title} {{Weakly Nonlinear Analysis of the
  Magnetorotational Instability in a Model Channel Flow}},\ }\href
  {https://doi.org/10.1103/PhysRevLett.98.034501} {\bibfield  {journal}
  {\bibinfo  {journal} {Phys. Rev. Lett.}\ }\textbf {\bibinfo {volume} {98}},\
  \bibinfo {eid} {034501} (\bibinfo {year} {2007}{\natexlab{a}})}\BibitemShut
  {NoStop}%
\bibitem [{\citenamefont {{Umurhan}}\ \emph
  {et~al.}(2007{\natexlab{b}})\citenamefont {{Umurhan}}, \citenamefont
  {{Regev}},\ and\ \citenamefont {{Menou}}}]{Umurhan_Regev_Menou_2007PhRvE}%
  \BibitemOpen
  \bibfield  {author} {\bibinfo {author} {\bibfnamefont {O.~M.}\ \bibnamefont
  {{Umurhan}}}, \bibinfo {author} {\bibfnamefont {O.}~\bibnamefont {{Regev}}},\
  and\ \bibinfo {author} {\bibfnamefont {K.}~\bibnamefont {{Menou}}},\
  }\bibfield  {title} {\bibinfo {title} {{Nonlinear saturation of the
  magnetorotational instability near threshold in a thin-gap Taylor-Couette
  setup}},\ }\href {https://doi.org/10.1103/PhysRevE.76.036310} {\bibfield
  {journal} {\bibinfo  {journal} {Phys. Rev. E}\ }\textbf {\bibinfo {volume}
  {76}},\ \bibinfo {eid} {036310} (\bibinfo {year}
  {2007}{\natexlab{b}})}\BibitemShut {NoStop}%
\bibitem [{\citenamefont {{Clark}}\ and\ \citenamefont
  {{Oishi}}(2017{\natexlab{a}})}]{Clark_Oishi_2017ApJ_local_geometry}%
  \BibitemOpen
  \bibfield  {author} {\bibinfo {author} {\bibfnamefont {S.~E.}\ \bibnamefont
  {{Clark}}}\ and\ \bibinfo {author} {\bibfnamefont {J.~S.}\ \bibnamefont
  {{Oishi}}},\ }\bibfield  {title} {\bibinfo {title} {{The Weakly Nonlinear
  Magnetorotational Instability in a Local Geometry}},\ }\href
  {https://doi.org/10.3847/1538-4357/aa6ff1} {\bibfield  {journal} {\bibinfo
  {journal} {Astrophys. J.}\ }\textbf {\bibinfo {volume} {841}},\ \bibinfo
  {eid} {1} (\bibinfo {year} {2017}{\natexlab{a}})}\BibitemShut {NoStop}%
\bibitem [{\citenamefont {{Clark}}\ and\ \citenamefont
  {{Oishi}}(2017{\natexlab{b}})}]{Clark_Oishi_2017ApJ_TC_geometry}%
  \BibitemOpen
  \bibfield  {author} {\bibinfo {author} {\bibfnamefont {S.~E.}\ \bibnamefont
  {{Clark}}}\ and\ \bibinfo {author} {\bibfnamefont {J.~S.}\ \bibnamefont
  {{Oishi}}},\ }\bibfield  {title} {\bibinfo {title} {{The Weakly Nonlinear
  Magnetorotational Instability in a Global, Cylindrical Taylor-Couette
  Flow}},\ }\href {https://doi.org/10.3847/1538-4357/aa6ff6} {\bibfield
  {journal} {\bibinfo  {journal} {Astrophys. J.}\ }\textbf {\bibinfo {volume}
  {841}},\ \bibinfo {eid} {2} (\bibinfo {year}
  {2017}{\natexlab{b}})}\BibitemShut {NoStop}%
\bibitem [{\citenamefont {{Mishra}}\ \emph {et~al.}(2023)\citenamefont
  {{Mishra}}, \citenamefont {{Mamatsashvili}},\ and\ \citenamefont
  {{Stefani}}}]{Mishra_etal_nonlinear_2022}%
  \BibitemOpen
  \bibfield  {author} {\bibinfo {author} {\bibfnamefont {A.}~\bibnamefont
  {{Mishra}}}, \bibinfo {author} {\bibfnamefont {G.}~\bibnamefont
  {{Mamatsashvili}}},\ and\ \bibinfo {author} {\bibfnamefont {F.}~\bibnamefont
  {{Stefani}}},\ }\bibfield  {title} {\bibinfo {title} {{Nonlinear evolution of
  magnetorotational instability in a magnetized Taylor-Couette flow: Scaling
  properties and relation to upcoming DRESDYN-MRI experiment}},\ }\href
  {https://doi.org/10.1103/PhysRevFluids.8.083902} {\bibfield  {journal}
  {\bibinfo  {journal} {Phys. Rev. Fluids}\ }\textbf {\bibinfo {volume} {8}},\
  \bibinfo {eid} {083902} (\bibinfo {year} {2023})}\BibitemShut {NoStop}%
\bibitem [{\citenamefont {{Ogilvie}}\ and\ \citenamefont
  {{Pringle}}(1996)}]{Ogilvie_Pringle1996}%
  \BibitemOpen
  \bibfield  {author} {\bibinfo {author} {\bibfnamefont {G.~I.}\ \bibnamefont
  {{Ogilvie}}}\ and\ \bibinfo {author} {\bibfnamefont {J.~E.}\ \bibnamefont
  {{Pringle}}},\ }\bibfield  {title} {\bibinfo {title} {{The non-axisymmetric
  instability of a cylindrical shear flow containing an azimuthal magnetic
  field}},\ }\href {https://doi.org/10.1093/mnras/279.1.152} {\bibfield
  {journal} {\bibinfo  {journal} {Mon. Not. R. Astron. Soc.}\ }\textbf
  {\bibinfo {volume} {279}},\ \bibinfo {pages} {152} (\bibinfo {year}
  {1996})}\BibitemShut {NoStop}%
\bibitem [{\citenamefont {{Khalzov}}\ \emph {et~al.}(2006)\citenamefont
  {{Khalzov}}, \citenamefont {{Ilgisonis}}, \citenamefont {{Smolyakov}},\ and\
  \citenamefont {{Velikhov}}}]{Khalzov2006}%
  \BibitemOpen
  \bibfield  {author} {\bibinfo {author} {\bibfnamefont {I.~V.}\ \bibnamefont
  {{Khalzov}}}, \bibinfo {author} {\bibfnamefont {V.~I.}\ \bibnamefont
  {{Ilgisonis}}}, \bibinfo {author} {\bibfnamefont {A.~I.}\ \bibnamefont
  {{Smolyakov}}},\ and\ \bibinfo {author} {\bibfnamefont {E.~P.}\ \bibnamefont
  {{Velikhov}}},\ }\bibfield  {title} {\bibinfo {title} {{Magnetorotational
  instability in electrically driven flow of liquid metal: Spectral analysis of
  global modes}},\ }\href {https://doi.org/10.1063/1.2408513} {\bibfield
  {journal} {\bibinfo  {journal} {Phys. Fluids}\ }\textbf {\bibinfo {volume}
  {18}},\ \bibinfo {pages} {124107} (\bibinfo {year} {2006})}\BibitemShut
  {NoStop}%
\bibitem [{\citenamefont {{Goedbloed}}\ and\ \citenamefont
  {{Keppens}}(2022)}]{Goedbloed_Keppens2022}%
  \BibitemOpen
  \bibfield  {author} {\bibinfo {author} {\bibfnamefont {H.}~\bibnamefont
  {{Goedbloed}}}\ and\ \bibinfo {author} {\bibfnamefont {R.}~\bibnamefont
  {{Keppens}}},\ }\bibfield  {title} {\bibinfo {title} {{The Super-Alfv{\'e}nic
  Rotational Instability in Accretion Disks about Black Holes}},\ }\href
  {https://doi.org/10.3847/1538-4365/ac573c} {\bibfield  {journal} {\bibinfo
  {journal} {Astrophys. J., Suppl. Ser.}\ }\textbf {\bibinfo {volume} {259}},\
  \bibinfo {eid} {65} (\bibinfo {year} {2022})}\BibitemShut {NoStop}%
\bibitem [{\citenamefont {{Ebrahimi}}\ and\ \citenamefont
  {{Pharr}}(2022)}]{Ebrahimi2022}%
  \BibitemOpen
  \bibfield  {author} {\bibinfo {author} {\bibfnamefont {F.}~\bibnamefont
  {{Ebrahimi}}}\ and\ \bibinfo {author} {\bibfnamefont {M.}~\bibnamefont
  {{Pharr}}},\ }\bibfield  {title} {\bibinfo {title} {{A Nonlocal
  Magneto-curvature Instability in a Differentially Rotating Disk}},\ }\href
  {https://doi.org/10.3847/1538-4357/ac892d} {\bibfield  {journal} {\bibinfo
  {journal} {Astrophys. J.}\ }\textbf {\bibinfo {volume} {936}},\ \bibinfo
  {eid} {145} (\bibinfo {year} {2022})}\BibitemShut {NoStop}%
\bibitem [{\citenamefont {{R{\"u}diger}}\ \emph {et~al.}(2003)\citenamefont
  {{R{\"u}diger}}, \citenamefont {{Schultz}},\ and\ \citenamefont
  {{Shalybkov}}}]{Rudiger_Schultz_Shalybkov_2003PhRvE}%
  \BibitemOpen
  \bibfield  {author} {\bibinfo {author} {\bibfnamefont {G.}~\bibnamefont
  {{R{\"u}diger}}}, \bibinfo {author} {\bibfnamefont {M.}~\bibnamefont
  {{Schultz}}},\ and\ \bibinfo {author} {\bibfnamefont {D.}~\bibnamefont
  {{Shalybkov}}},\ }\bibfield  {title} {\bibinfo {title} {{Linear
  magnetohydrodynamic Taylor-Couette instability for liquid sodium}},\ }\href
  {https://doi.org/10.1103/PhysRevE.67.046312} {\bibfield  {journal} {\bibinfo
  {journal} {Phys. Rev. E}\ }\textbf {\bibinfo {volume} {67}},\ \bibinfo {eid}
  {046312} (\bibinfo {year} {2003})}\BibitemShut {NoStop}%
\bibitem [{\citenamefont {{Shalybkov}}\ \emph {et~al.}(2002)\citenamefont
  {{Shalybkov}}, \citenamefont {{R{\"u}diger}},\ and\ \citenamefont
  {{Schultz}}}]{Shalybkov_Rudiger_Schultz_2002A&A_nonaxisMRI}%
  \BibitemOpen
  \bibfield  {author} {\bibinfo {author} {\bibfnamefont {D.~A.}\ \bibnamefont
  {{Shalybkov}}}, \bibinfo {author} {\bibfnamefont {G.}~\bibnamefont
  {{R{\"u}diger}}},\ and\ \bibinfo {author} {\bibfnamefont {M.}~\bibnamefont
  {{Schultz}}},\ }\bibfield  {title} {\bibinfo {title} {{Nonaxisymmetric
  patterns in the linear theory of MHD Taylor-Couette instability}},\ }\href
  {https://doi.org/10.1051/0004-6361:20021284} {\bibfield  {journal} {\bibinfo
  {journal} {Astron. Astrophys.}\ }\textbf {\bibinfo {volume} {395}},\ \bibinfo
  {pages} {339} (\bibinfo {year} {2002})}\BibitemShut {NoStop}%
\bibitem [{\citenamefont {{Hollerbach}}\ and\ \citenamefont
  {{Fournier}}(2004)}]{Hollerbach_Fournier2004}%
  \BibitemOpen
  \bibfield  {author} {\bibinfo {author} {\bibfnamefont {R.}~\bibnamefont
  {{Hollerbach}}}\ and\ \bibinfo {author} {\bibfnamefont {A.}~\bibnamefont
  {{Fournier}}},\ }\bibfield  {title} {\bibinfo {title} {{End-effects in
  rapidly rotating cylindrical Taylor-Couette flow}},\ }in\ \href
  {https://doi.org/10.1063/1.1832141} {\emph {\bibinfo {booktitle} {MHD Couette
  Flows: Experiments and Models}}},\ \bibinfo {series} {American Institute of
  Physics Conference Series}, Vol.\ \bibinfo {volume} {733},\ \bibinfo {editor}
  {edited by\ \bibinfo {editor} {\bibfnamefont {R.}~\bibnamefont {{Rosner}}},
  \bibinfo {editor} {\bibfnamefont {G.}~\bibnamefont {{R{\"u}diger}}},\ and\
  \bibinfo {editor} {\bibfnamefont {A.}~\bibnamefont {{Bonanno}}}}\ (\bibinfo
  {year} {2004})\ pp.\ \bibinfo {pages} {114--121}\BibitemShut {NoStop}%
\bibitem [{\citenamefont {{Szklarski}}(2007)}]{Szklarski2007}%
  \BibitemOpen
  \bibfield  {author} {\bibinfo {author} {\bibfnamefont {J.}~\bibnamefont
  {{Szklarski}}},\ }\bibfield  {title} {\bibinfo {title} {{Reduction of
  boundary effects in the spiral MRI experiment PROMISE}},\ }\href
  {https://doi.org/10.1002/asna.200710774} {\bibfield  {journal} {\bibinfo
  {journal} {Astron. Nachr.}\ }\textbf {\bibinfo {volume} {328}},\ \bibinfo
  {pages} {499} (\bibinfo {year} {2007})}\BibitemShut {NoStop}%
\bibitem [{Note1()}]{Note1}%
  \BibitemOpen
  \bibinfo {note} {Because of symmetry, the results for the $m<0$ modes are the
  same as those for the $m>0$ modes, so without loss of generality everywhere
  below we use the absolute value $|m|$ when describing non-axisymmetric mode
  dynamics.}\BibitemShut {Stop}%
\bibitem [{Note2()}]{Note2}%
  \BibitemOpen
  \bibinfo {note} {In the present case with resistivity, the meaning of the
  Alfv\'en resonance points is, however, less clear.}\BibitemShut {Stop}%
\bibitem [{\citenamefont {{Guseva}}\ \emph {et~al.}(2015)\citenamefont
  {{Guseva}}, \citenamefont {{Willis}}, \citenamefont {{Hollerbach}},\ and\
  \citenamefont {{Avila}}}]{Guseva_etal_2015NJPh}%
  \BibitemOpen
  \bibfield  {author} {\bibinfo {author} {\bibfnamefont {A.}~\bibnamefont
  {{Guseva}}}, \bibinfo {author} {\bibfnamefont {A.~P.}\ \bibnamefont
  {{Willis}}}, \bibinfo {author} {\bibfnamefont {R.}~\bibnamefont
  {{Hollerbach}}},\ and\ \bibinfo {author} {\bibfnamefont {M.}~\bibnamefont
  {{Avila}}},\ }\bibfield  {title} {\bibinfo {title} {{Transition to
  magnetorotational turbulence in Taylor-Couette flow with imposed azimuthal
  magnetic field}},\ }\href {https://doi.org/10.1088/1367-2630/17/9/093018}
  {\bibfield  {journal} {\bibinfo  {journal} {New Journal of Physics}\ }\textbf
  {\bibinfo {volume} {17}},\ \bibinfo {eid} {093018} (\bibinfo {year}
  {2015})}\BibitemShut {NoStop}%
\bibitem [{\citenamefont {{Mamatsashvili}}\ \emph {et~al.}(2018)\citenamefont
  {{Mamatsashvili}}, \citenamefont {{Stefani}}, \citenamefont {{Guseva}},\ and\
  \citenamefont {{Avila}}}]{Mamatsashvili_etal2018}%
  \BibitemOpen
  \bibfield  {author} {\bibinfo {author} {\bibfnamefont {G.}~\bibnamefont
  {{Mamatsashvili}}}, \bibinfo {author} {\bibfnamefont {F.}~\bibnamefont
  {{Stefani}}}, \bibinfo {author} {\bibfnamefont {A.}~\bibnamefont
  {{Guseva}}},\ and\ \bibinfo {author} {\bibfnamefont {M.}~\bibnamefont
  {{Avila}}},\ }\bibfield  {title} {\bibinfo {title} {{Quasi-two-dimensional
  nonlinear evolution of helical magnetorotational instability in a magnetized
  Taylor-Couette flow}},\ }\href {https://doi.org/10.1088/1367-2630/aa9d65}
  {\bibfield  {journal} {\bibinfo  {journal} {New Journal of Physics}\ }\textbf
  {\bibinfo {volume} {20}},\ \bibinfo {eid} {013012} (\bibinfo {year}
  {2018})}\BibitemShut {NoStop}%
\bibitem [{Note3()}]{Note3}%
  \BibitemOpen
  \bibinfo {note} {Point E in Fig. \ref {fig:Mag_energyRm35Lu678_m20}(c) marks
  only the exponential growth phase of non-axisymmetric modes in the linear
  regime when their overall $(r,z)$-structure is nearly similar to that
  obtained from the linear stability analysis of the dominant $|m|=1$ mode
  (Fig. \ref {fig:eigen_linearcode}), so we make this comparison in Appendix
  \ref {appendix_r_z_structures_growth_phase}}\BibitemShut {NoStop}%
\end{thebibliography}%

\appendix*
\section{} \label{appendix}

\begin{figure}
\centering
\includegraphics[width=0.4\textwidth]{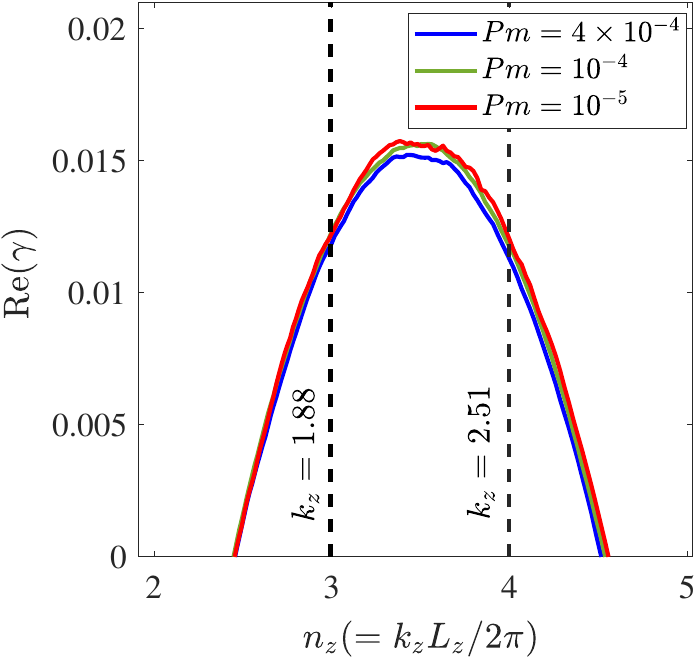}
\caption{Growth rate, ${\rm Re}(\gamma)$,  versus the axial mode number $n_z=k_zL_z/2\pi$ for the $|m|=1$ mode of SMRI at $Lu=6.78$, $Rm=35$, $\mu=0.35$ and different $Pm=10^{-5}, 10^{-4}$, $4\times 10^{-4}$.}
\label{fig:diff_Pm}
\end{figure}
\begin{figure*}
\centering
\includegraphics[width=\textwidth]{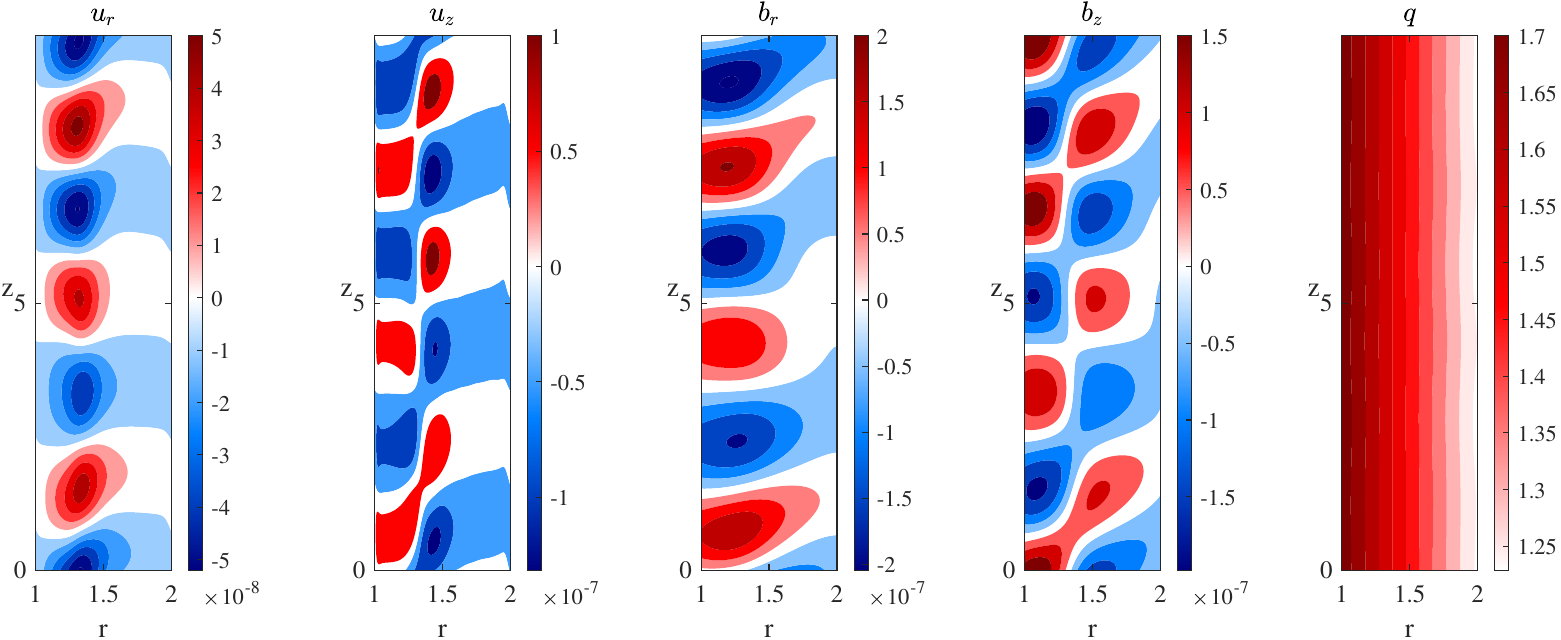}
\caption{Radial and axial velocity ($u_r, \, u_z$) and magnetic ($b_r, \, b_z$) field structures of the non-axisymmetric modes in the ($r,z$)-plane during the early exponential growth stage [moment E in Fig. \ref{fig:Mag_energyRm35Lu678_m20}(c)] obtained from the nonlinear simulations, which are dominated by the $|m|=1$ SMRI mode. The parameters are: $\mu=0.35$, $Lu=6.78$, $Rm=35$ (point C in Fig. \ref{fig:compare_m0_m1_mu}) and $Re=4\times 10^4$.}
\label{fig:r_z_structure_nonaxis_growth}
\end{figure*}

\subsection{Independence of the linear SMRI from $Pm$}\label{Pm_dependence}

To demonstrate that the linear dynamics of SMRI is essentially insensitive to $Pm$ when $Pm \ll 1$, in Fig. \ref{fig:diff_Pm} we show the linear growth rate of the $|m|=1$ mode as a function of the axial mode number $n_z=k_zL_z/2\pi$ at $\mu=0.35$, $Lu=6.78$, $Rm=35$ (point C in Fig. \ref{fig:compare_m0_m1_mu}) and different $Pm=10^{-5}, 10^{-4}$ and $4\times 10^{-4}$. It is evident that the dispersion curves almost coincide despite an order of magnitude increase in $Pm$.  Since in the nonlinear analysis the axial wavenumber $k_z$ is discrete and hence the mode number $n_z$ is integer, the only unstable modes in the flow domain for these $Lu$ and $Rm$ are $n_z=3$ ($k_z=1.88$) and 4 ($k_z=2.51$), as seen in Fig. \ref{fig:diff_Pm}, with the former having the largest growth rate.

\begin{figure*}
\centering
\includegraphics[width=0.25\textwidth]{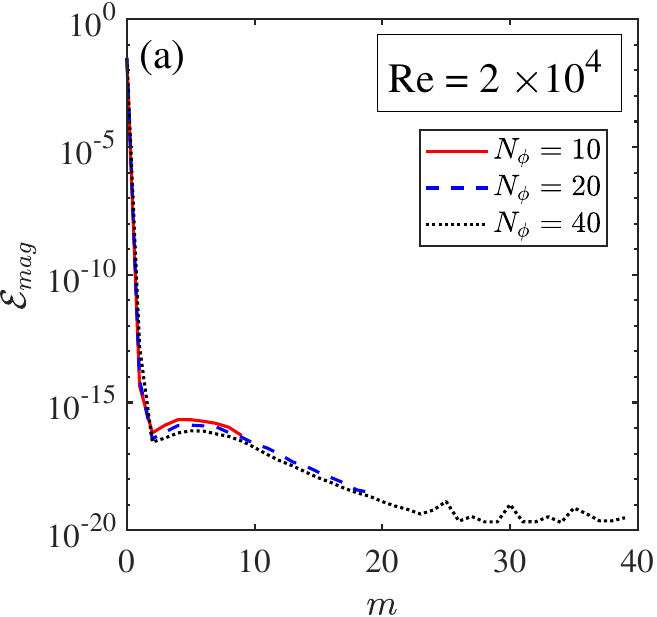}
\hspace{0.25em}
\includegraphics[width=0.23\textwidth]{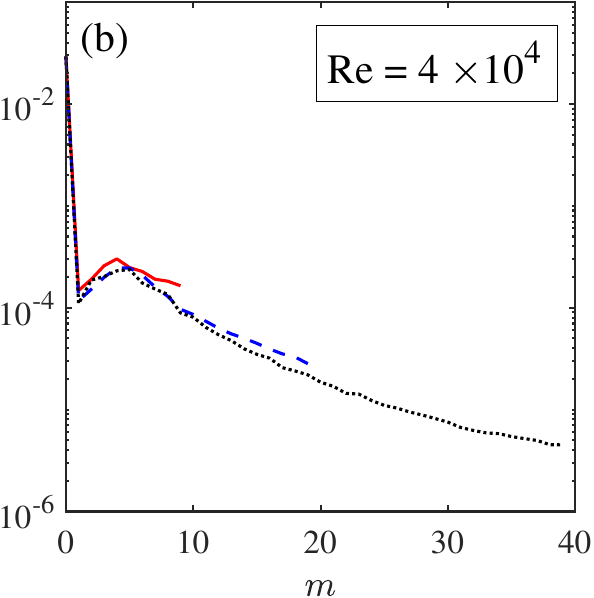}
\hspace{0.25em}
\includegraphics[width=0.23\textwidth]{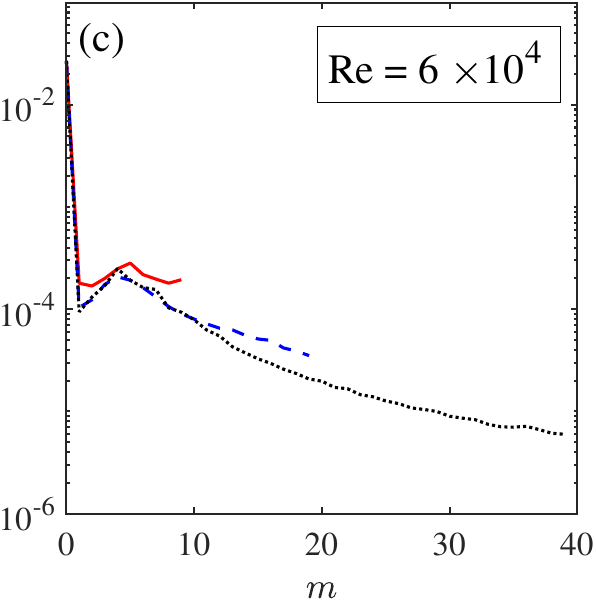}
\hspace{0.25em}
\includegraphics[width=0.23\textwidth]{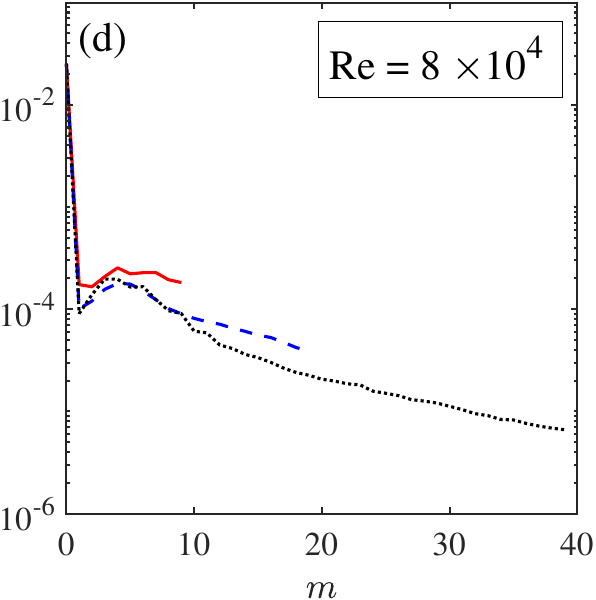}
\caption{Azimuthal spectra of the magnetic energy density, $\mathcal{E}_{mag}$,  versus $m$ for different resolutions $N_{\phi}=\{10, 20, 40\}$ in the azimuthal and a fixed resolution $N_z=480$ in the axial directions at $Lu=6.78$, $Rm=35$ (point C in Fig. \ref{fig:compare_m0_m1_mu}) and  different (a) $Re=2\times10^4$, (b) $4 \times 10^4$, (c) $6 \times 10^4$ and (d) $8 \times 10^4$.  It is seen that the most energy-containing non-axisymmetric modes with $|m| \lesssim 10$ are well resolved from $N_{\phi}=20$, exhibiting the convergence of the spectra at these high $Re$.} \label{fig:mode_spectra_Rm35Lu678}
\end{figure*}
\begin{figure*}
	\includegraphics[width=0.24\textwidth]{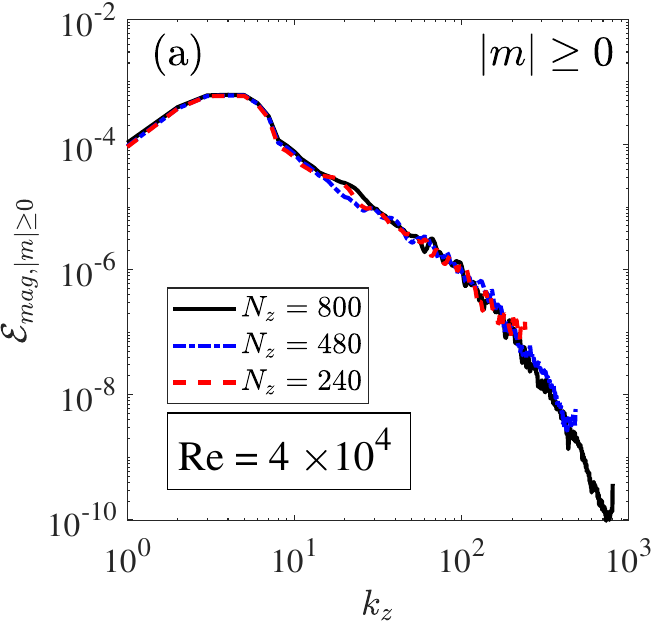}
	\includegraphics[width=0.24\textwidth]{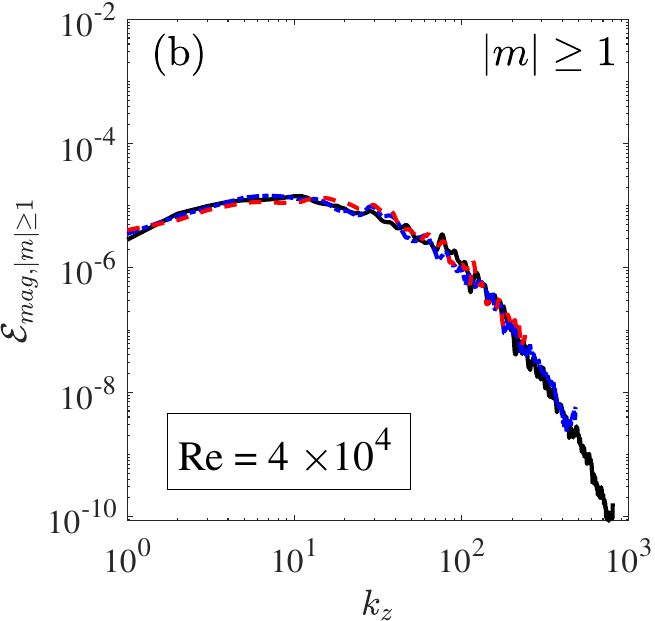}
	\includegraphics[width=0.24\textwidth]{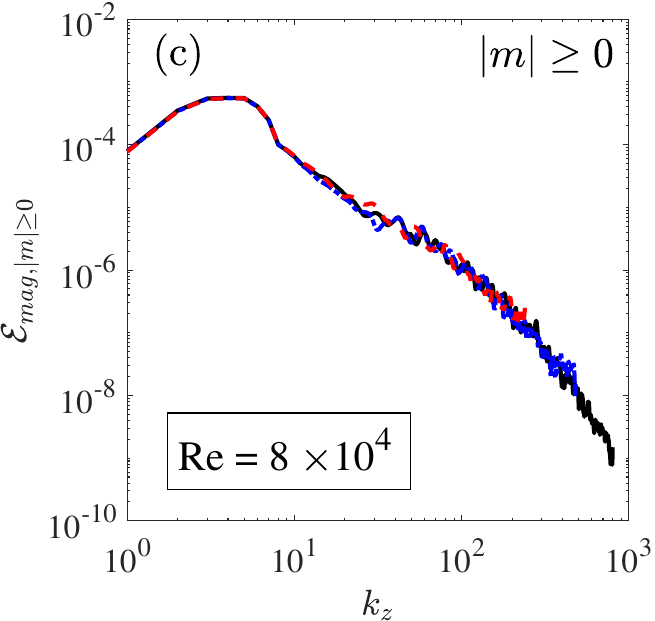}
	\includegraphics[width=0.24\textwidth]{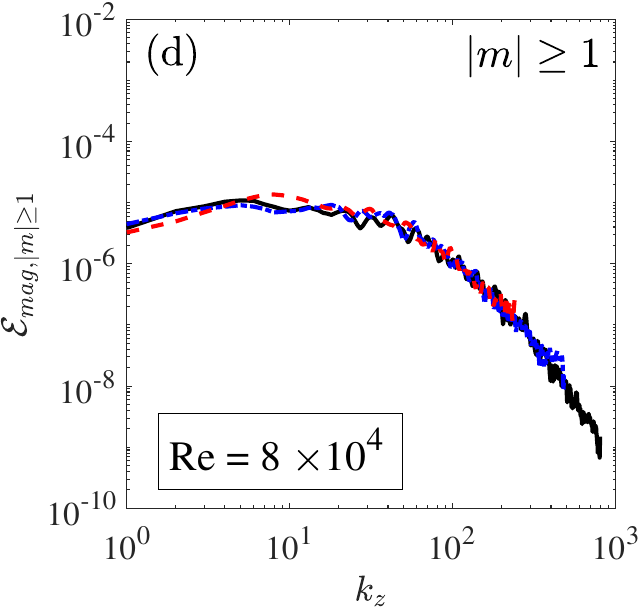}
	\caption{Axial spectra of the magnetic energy density for  [(a),(c)] all $m$-modes, $\mathcal{E}_{mag,|m|\geq 0}$ and [(b),(d)] for non-axisymmetric $|m|\geq 1$ modes,  $\mathcal{E}_{mag,|m|\geq 1}$ versus $k_z$ at different resolutions $N_z=\{240, 480, 800\}$ in the axial and a fixed resolution $N_{\phi}=20$ in the azimuthal directions at the same $Lu=6.78$ and $Rm=35$ as in Fig. \ref{fig:mode_spectra_Rm35Lu678} and different [(a),(b)]$Re=4\times10^4$ and [(c),(d)] $8 \times 10^4$.  It is seen that both axial spectra well converge with resolution.} \label{fig:axial_resolution}
\end{figure*}

\subsection{$(r,z)$-structure of the non-axisymmetric modes in the linear regime}\label{appendix_r_z_structures_growth_phase}

Figure \ref{fig:r_z_structure_nonaxis_growth} shows the radial and axial velocity, ($u_r, \, u_z$), and magnetic field, ($b_r, \, b_z$), structures for all non-axisymmetric modes in the ($r,z$)-plane for $Lu=6.78$, $Rm=35$ (point C) as in Fig. \ref{fig:eigen_linearcode} and $Re=4\times10^4$ during the early exponential growth phase [moment E in Fig. \ref{fig:Mag_energyRm35Lu678_m20}(c)]. It is seen from Fig. \ref{fig:Mode_energyRm35Lu678_m20} that in this case only non-axisymmetric $|m|=1$ mode is unstable and dominates the velocity and magnetic field structures. Hence, the $(r,z)$-maps shown here from the simulations look similar to those obtained for this mode using the linear stability analysis (Fig. \ref{fig:eigen_linearcode}), having the same wavelength, or the number of axial modes $n_z=3$, which corresponds to the largest growth rate in the domain (Fig. \ref{fig:diff_Pm}). Note that the orientation of the $(r,z)$-structures is reversed compared to that in Fig. \ref{fig:eigen_linearcode}. Since $m=1$ and $m=-1$ modes have generally equal weights, incidentally the nonlinear simulation shows somewhat more preference for the $m=-1$ mode, which could be due to initial conditions. 

\begin{figure*}
\centering
\includegraphics[width=0.24\textwidth]{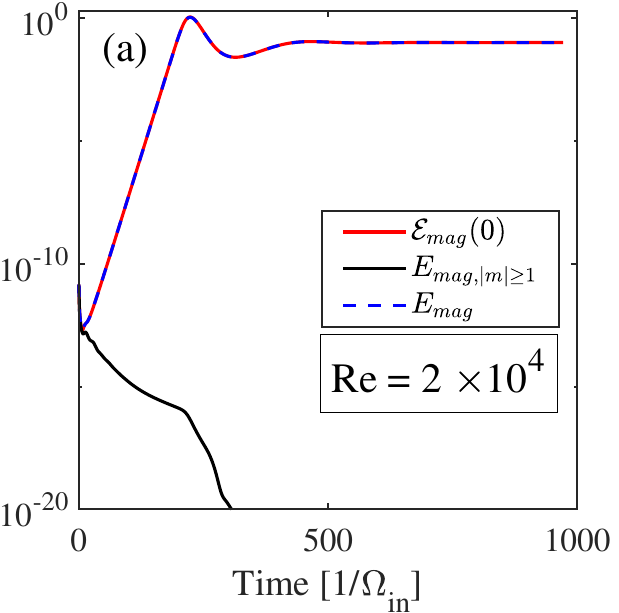}
\includegraphics[width=0.24\textwidth]{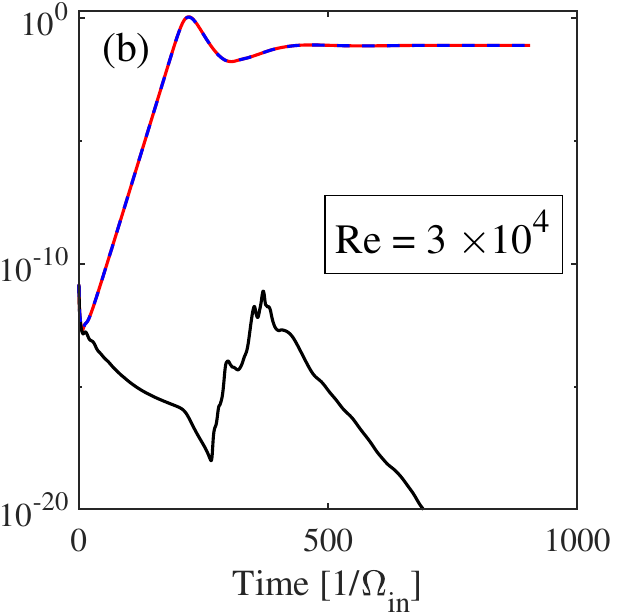}
\includegraphics[width=0.24\textwidth]{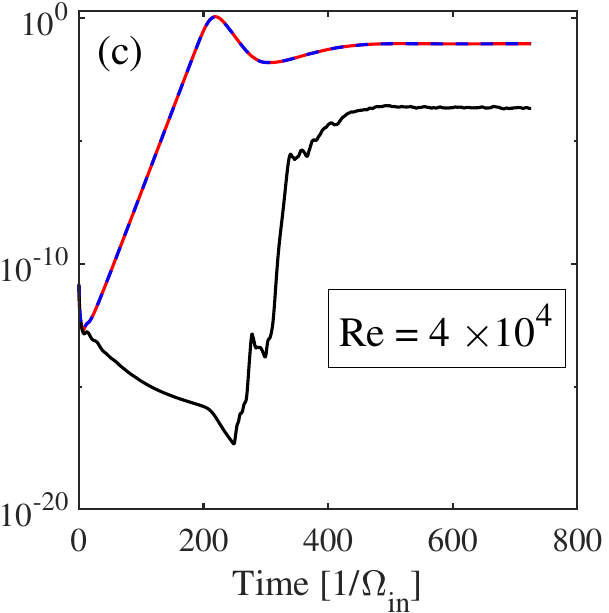}
\includegraphics[width=0.24\textwidth]{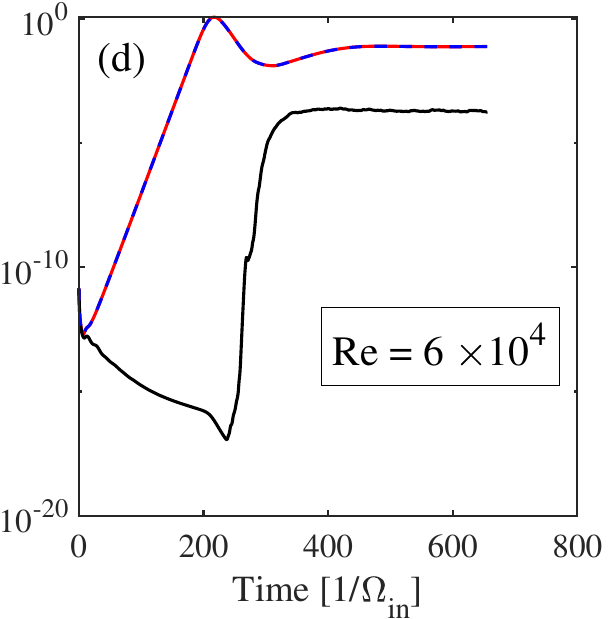}
\caption{Same as in Fig. \ref{fig:Mag_energyRm20Lu5_m20} but at $Lu=6$ and $Rm=30$ (point B in Fig. \ref{fig:compare_m0_m1_mu}) at different (a) $Re = 2\times10^4$, (b) $3\times 10^4$, (c) $4\times 10^4$ and (d) $6\times 10^4$.}
	\label{fig:Mag_energyRm30Lu6_m20}
\end{figure*}
\begin{figure*}
\centering
\includegraphics[width=\textwidth]{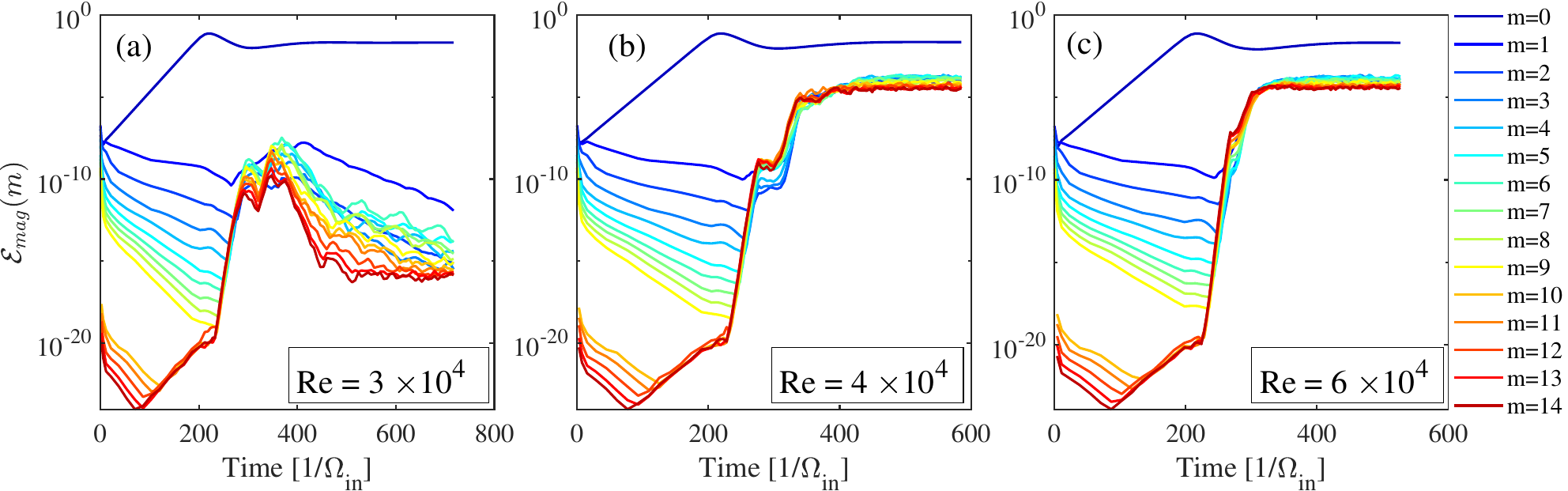}
\caption{Evolution of the azimuthal spectrum of the magnetic energy density, $\mathcal{E}_{mag}(m)$, for different $m \in [0,14]$ and $Lu=6$, $Rm=30$ and (a) $Re = 3\times 10^4$, (b) $4\times 10^4$ and (c) $6\times 10^4$.}
\label{fig:Mode_energyRm30Lu6_m20}
\end{figure*}
\begin{figure*}
\centering
\includegraphics[width=0.4\textwidth]{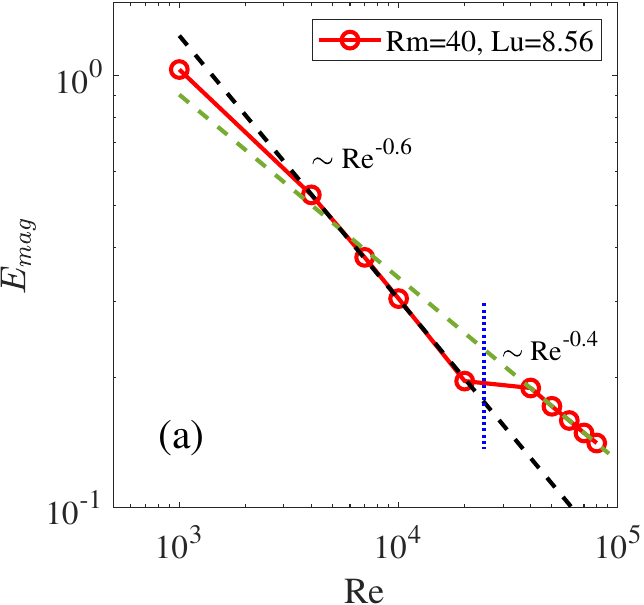}
\hspace{2em}
\includegraphics[width=0.39\textwidth]{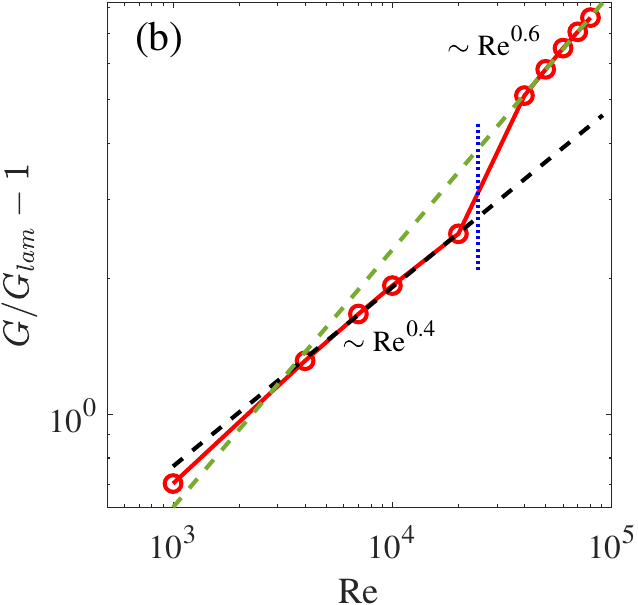}
\caption{Same as Figs. \ref{fig:scaling}(a) and \ref{fig:scaling}(b) but for higher $Lu=8.56$ and $Rm=40$. Blue dotted line shows the change in the scaling behavior for the saturated magnetic energy $E_{mag}$ and torque $G/G_{lam}-1$ due to the presence of stronger non-axisymmetric non-MRI modes at these relatively high $Lu$ and $Rm$.}
\label{fig:scaling_Rm40}
\end{figure*}

\subsection{Resolution test}\label{appendix_resolution_test}

To justify the resolution used in the main study, here we present resolution tests both in the $\phi-$azimuthal and $z-$axial directions. Figure \ref{fig:mode_spectra_Rm35Lu678} shows the azimuthal $m$-spectra of the magnetic energy, $\mathcal{E}_{mag}$, in the saturated state at three different azimuthal resolutions $|m|\leq N_{\phi}=\{10, 20, 40\}$ and an axial resolution $N_z=480$ (i.e.,  $|k_z|\leq 2\pi N_z/L_z$) used in Sec. \ref{nonlin_results} for $Lu=6.78$, $Rm=35$ and $Re = \{2, 4, 6, 8\} \times 10^4$. It is seen that the spectra at $N_{\phi}=20$ and 40 converge with a little deviation at higher $|m| \gtrsim 10$ for $Re=6\times 10^4$ and $8\times 10^4$. Thus,  $N_{\phi}=20$ used in the main analysis appears to be adequate, capturing a significant portion of the most energy-containing non-axisymmetric modes with $|m|\lesssim 10$. 

Figure \ref{fig:axial_resolution} shows the axial $k_z$-spectra of the magnetic energy in the saturated state summed over all $m$-modes, $\mathcal{E}_{mag,|m|\geq 0}(k_z)=\sum_{m} \bar{\mathcal{E}}_{mag}(m,k_z)$, as well as over only non-axisymmetric $|m| \geq 1$ modes,  $\mathcal{E}_{mag,|m|\geq 1}(k_z)=\sum_{|m|\geq 1} \bar{\mathcal{E}}_{mag}(m,k_z)$ at three different axial resolutions $N_z=\{240, 480, 800\}$ and a fixed azimuthal resolution $N_{\phi}=20$.  $Lu$ and $Rm$ are the same as above and $Re = \{4, 8\} \times 10^4$.  It is seen that both axial spectra of all the modes (which are dominated by the axisymmetric one) and only smaller-scale non-axisymmetric modes exhibit a very good convergence with resolution at all $k_z$. Therefore,  $N_z=480$ used in Sec. \ref{nonlin_results} is sufficient for resolving the small-scale turbulent structurs of the flow near the cylinder walls along the $z$-axis.
\\
\subsection{Magnetic energy evolution at point B of Fig. \ref{fig:compare_m0_m1_mu}}\label{appendix_Lu6_Rm30_erergy_growth}

Figure \ref{fig:Mag_energyRm30Lu6_m20} shows the evolution of the magnetic energy of axisymmetric, $\mathcal{E}_{mag}(0)$, and all the non-axisymmetric, $E_{mag,|m|\geq 1}$, modes for $Lu=6$, $Rm=30$ (point B in Fig. \ref{fig:compare_m0_m1_mu}) and $Re = \{2, 3, 4, 6\}\times10^4$. Since point B is just outside but near the marginal stability curve of the $|m|=1$ mode, this and higher-$m$ non-axisymmetric modes decay for all $Re$ in the linear growth phase, where the basic flow profile is still the original TC flow (\ref{TC_flow}). Noticeable rapid growth of non-axisymmetric modes occur at large $Re \gtrsim 3\times 10^4$ as seen in panels (b), (c) and (d), which is triggered by the change in the TC flow profile as the axisymmetric SMRI mode saturates. This growth is only transient and eventually decays for $Re = 3\times 10^4$, whereas for $Re = 4\times 10^4$ and $6\times 10^4$ it saturates to a finite value, which is still orders of magnitude smaller than that of the axisymmetric mode.

Figure \ref{fig:Mode_energyRm30Lu6_m20} shows the evolution of the azimuthal spectral magnetic energy density $\mathcal{E}_{mag}$ for different $m \in [0,14]$, the same $Lu=6$, $Rm=30$ as above and $Re = \{3,\, 4, \, 6\}\times 10^4$. Note that in the early phase, before the full saturation of the dominant axisymmetric mode, non-axisymmetric modes with smaller $|m|<10$ decay for all $Re$ while the modes with larger $|m|\geq 10$ first decrease and then start to grow. The reason for this behavior could be that in the middle of the exponential growth of the axisymmetric SMRI mode the boundary layers with high shear start to form and, although the shear therein may not be yet strong enough, still appears sufficient to trigger the growth of these high-$m$ modes. Later, once the axisymmetric mode has reached a saturation point, a much steeper increase in the magnetic energy of all $|m| \geq 1$ non-axisymmetric  modes is seen, which is, however, not sustained for Re= $3\times 10^4$, while it saturates for $Re= 4\times 10^4$ and $6\times 10^4$ at orders of magnitude higher levels. 
\\

\subsection{Scaling relations at Rm=40, Lu=8.56}\label{appendix_scaling_Rm40}

Figure \ref{fig:scaling_Rm40} shows the scaling behavior for the saturated total magnetic energy $E_{mag}$ and torque $G/G_{lam}-1$ as a function of $Re$ for higher $Lu=8.56$ and $Rm=40$ that those used in the main study. Similar to the observation in Figs. \ref{fig:scaling}(a) and \ref{fig:scaling}(b), the scalings of $E_{mag}$ and $G/G_{lam}-1$ change after $ Re \gtrsim 2\times10^4$ due to the presence of stronger non-axisymmetric non-MRI modes. Still, the scaling relation $E_{mag}^{-1}(G/G_{lam}-1) \sim Re$ is satisfied.   

\end{document}